\documentclass[11pt,twoside]{article}
\usepackage[latin9]{inputenc}
\usepackage[a4paper]{geometry}
\usepackage{amsmath}
\usepackage{amssymb}
\usepackage{graphicx}

\makeatletter
\@ifundefined{date}{}{\date{}}
\usepackage{amsfonts}
\providecommand{\U}[1]{\protect\rule{.1in}{.1in}}

\makeatother

\begin{document}
\title{Electromagnetic Formalism of the Propagation and Amplification of
Light}
\author{\textbf{F. Javier Fraile-Peláez}\\
 Dept. Teoría de la Señal y Comunicaciones, Universidad de Vigo.\\
 ETSI de Telecomunicación, Campus Universitario.\\
 E-36310 Vigo (Spain). \and \textbf{Andrés Macho}\\
 Nanophotonics Technology Center, Universidad Politécnica de Valencia.\\
 Edificio 8F, Planta 2, Camino de Vera, s/n.\\
 E-46022 Valencia (Spain). }
\maketitle
\begin{abstract}
\noindent In this work, we present a simplified but comprehensive
derivation of all the key concepts and main results concerning light
pulse propagation in dielectric media, including a brief extension
to the case of active media and laser oscillation. Clarifications
of the concepts of slow light and ``superluminality''\ are provided,
and a detailed discussion on the concept of transform-limited pulses
is also included in the Appendix. 
\end{abstract}
\tableofcontents{}

\pdfoutput=1

\section{Introduction}

In this article we present a route that starts from basic electromagnetism
concepts and leads to the development of the essential theory of various
basic devices and subsystems widely employed in telecommunication
photonics.

Of course, such derivations can be found by the hundreds in the literature,
but, in our opinion, they are rarely self-contained and are often
hasty, in some sense ``improvised'', with the result that it is
difficult for the reader to recognize the key ideas beyond what is
simply accessory. Sometimes, the routine and careless use of certain
formalisms (which, on some occasions, are not even really well-understood)
further obscures the conceptualization process.

Here we present the key concepts in the simplest possible way, but
in a rigorous fashion as to the logical order of the development and
the connections between the successive results, both from the field
of electromagnetism in dielectrics and from signal theory. Most of
all, we want to avoid the ``scattered''\ nature of the fragmented
presentations referred to above.

The required electromagnetic theory is presented in Section \ref{J02sctEMOW}.
We can, and will, limit ourselves to the simplest case: the propagation
of plane waves in \emph{homogeneous}, infinite dielectrics. Additionally,
the dielectrics will be, at this time, \emph{linear} and \emph{isotropic}.
With this background, Section \ref{J02sctPS} studies the transmission
of signals in dielectrics without any mention to waveguiding structures;
in our presentation, all the key ideas will be introduced and clarified
in the simple frame of non-guided, plane wave propagation. The great
advantage of this approach is that the fundamental concepts thus acquired
can be very easily generalized, at a later stage, to more realistic
situations.\footnote{We think, for example, that introducing the concept of chromatic dispersion
directly by the formal study of pulse propagation along a waveguide,
as is often found in textbooks, is not a good idea from the didactic
point of view.}

\section[Electromagnetism of optical waves]{Electromagnetism of optical waves \sectionmark{Electromagnetism
of optical waves}}

\label{J02sctEMOW}

In this section we will mainly be concerned with the derivation of
the wave equation in linear dispersive dielectrics. A few general,
simple results directly following from it are also presented.

\subsection{Linear response of an (isotropic) dielectric}

\label{J02sbsctLRID}

In their \emph{fundamental} form, the Maxwell equations read as follows:
\begin{align}
\ \nabla\cdot\mbox{\boldmath\ensuremath{\mathcal{E}}}(\boldsymbol{r},t) & =\frac{\rho(\boldsymbol{r},t)}{\epsilon_{0}}\label{J02max1}\\
\nabla\cdot\mbox{\boldmath\ensuremath{\mathcal{B}}}(\boldsymbol{r},t) & =0\label{J02max2}\\
\nabla\times\mbox{\boldmath\ensuremath{\mathcal{E}}}(\boldsymbol{r},t) & =-\frac{\partial}{\partial\,t}\mbox{\boldmath\ensuremath{\mathcal{B}}}(\boldsymbol{r},t)\label{J02max3}\\
c^{2}\nabla\times\mbox{\boldmath\ensuremath{\mathcal{B}}}(\boldsymbol{r},t) & =\frac{\mbox{\boldmath\ensuremath{\mathcal{J}}}(\boldsymbol{r},t)}{\epsilon_{0}}+\frac{\partial}{\partial\,t}\mbox{\boldmath\ensuremath{\mathcal{E}}}(\boldsymbol{r},t).\label{J02max4}
\end{align}

The adjective ``fundamental''\ means two things here:

\textbf{(a)} Only two fields, $\mbox{\boldmath\ensuremath{\mathcal{E}}}$
(electric field strength) and $\mbox{\boldmath\ensuremath{\mathcal{B}}}$
(magnetic induction), appear in the equations. These are considered
to be the genuine fields which describe the electromagnetism in nature.
The \emph{electric displacement} field $\mbox{\boldmath\ensuremath{\mathcal{D}}}$,
often included in the set of Maxwell equations on the same footing
as $\mbox{\boldmath\ensuremath{\mathcal{E}}}$, is an ``artificial''\ field
introduced as an aid to describe the macroscopic response of the dielectric
materials to the electromagnetic (EM) fields. In many cases, the material
response leads to a surprisingly simple (albeit approximate) linear
relation between both fields: $\mbox{\boldmath\ensuremath{\mathcal{D}}}(t)=\varepsilon\mbox{\boldmath\ensuremath{\mathcal{E}}}(t)$,
as remarked in the Introduction. Under these circumstances, all the
effects of the material response can be trivially accounted for, with
very good accuracy, by a mere constant $\varepsilon$ (the \emph{dielectric
permittivity }or \emph{dielectric constant} of the material). However,
this will frequently \emph{not} be the case in Photonics. On the other
hand, since the dielectrics materials we will be concerned with are
diamagnetic, a simple linear relation between $\mbox{\boldmath\ensuremath{\mathcal{B}}}$
and the so-called \emph{magnetic induction}, $\mbox{\boldmath\ensuremath{\mathcal{H}}},$
will be generally valid in the situations studied here: 
\begin{equation}
\mbox{\boldmath\ensuremath{\mathcal{B}}}(t)=\mu\mbox{\boldmath\ensuremath{\mathcal{H}}}(t),\label{J02mag}
\end{equation}
with $\mu$ the \emph{magnetic permeability}. We will thus not need
to dwell on any magnetic properties of matter, and either $\mbox{\boldmath\ensuremath{\mathcal{B}}}$
or $\mbox{\boldmath\ensuremath{\mathcal{H}}}$ will be used at our
convenience. Moreover, almost all materials of interest in Photonics
are diamagnetic or paramagnetic, so the magnetic permeability can
be taken as that of the vacuum in all cases: $\mu\simeq\mu_{0}=4\pi\times10^{-7}$
N$\times$A$^{-2}.$

\textbf{(b)} $\rho(\boldsymbol{r},t)$ in (\ref{J02max1}) is the
\emph{total} change density in space, including not only the ``free''\ charges
(which give rise to the electric conduction current), but also the
bound charges of the atoms which make up the dielectric. This is,
\begin{equation}
\rho(\boldsymbol{r},t)=\rho_{d}(\boldsymbol{r},t)+\rho_{f}(\boldsymbol{r},t),\label{J02rorro}
\end{equation}
where $d$ and $f$ stand for ``dielectric''\ and ``free'', respectively.
Likewise, $\mbox{\boldmath\ensuremath{\mathcal{J}}}$ in (\ref{J02max4})
is given by 
\begin{equation}
\mbox{\boldmath\ensuremath{\mathcal{J}}}(\boldsymbol{r},t)=\mbox{\boldmath\ensuremath{\mathcal{J}}}{}_{d}(\boldsymbol{r},t)+\mbox{\boldmath\ensuremath{\mathcal{J}}}{}_{f}(\boldsymbol{r},t),\label{J02jj}
\end{equation}
where $\mbox{\boldmath\ensuremath{\mathcal{J}}}{}_{d},$ the ``dielectric
current''\ originated from the temporal variation of the dielectric
charge density, adds to the free conduction current $\mbox{\boldmath\ensuremath{\mathcal{J}}}{}_{f}.$The
key issue will be to obtain the form of $\rho_{d}$ (and $\mbox{\boldmath\ensuremath{\mathcal{J}}}{}_{d}$),
but, as long as a specific model for the dielectric interaction has
not been chosen, eqs. (\ref{J02max1})--(\ref{J02max4}) remain general.

The \emph{wave equation} is the equation for one field only, whether
$\mbox{\boldmath\ensuremath{\mathcal{E}}}$ or $\mbox{\boldmath\ensuremath{\mathcal{B}}}.$
In our context, the electric field is almost always chosen and the
derivation of the wave equation is straightforward. Taking the curl
of (\ref{J02max3}), replacing $\nabla\times\mbox{\boldmath\ensuremath{\mathcal{B}}}$
in favour of $\mbox{\boldmath\ensuremath{\mathcal{E}}}$ by means
of (\ref{J02max4}), applying the general vector relation $\nabla\times\lbrack\nabla\times\boldsymbol{F}]=\nabla(\nabla\cdot\boldsymbol{F})-\nabla^{2}\boldsymbol{F,}$
and making use of eq. (\ref{J02max1}), the following equation is
obtained: 
\begin{align}
\nabla^{2}\mbox{\boldmath\ensuremath{\mathcal{E}}}(\boldsymbol{r},t)-\dfrac{1}{c^{2}}\frac{\partial^{2}}{\partial\,t^{2}}\mbox{\boldmath\ensuremath{\mathcal{E}}}(\boldsymbol{r},t) & =\dfrac{1}{\varepsilon_{0}}\nabla\rho_{d}(\boldsymbol{r},t)+\dfrac{1}{\varepsilon_{0}c^{2}}\frac{\partial}{\partial\,t}\mbox{\boldmath\ensuremath{\mathcal{J}}}{}_{d}(\boldsymbol{r},t)\label{J02weq1}\\
 & \qquad\qquad\qquad\qquad\qquad\qquad\qquad\text{\textsf{{\small\{no free charges\}}}}\emph{.}\nonumber 
\end{align}
It has also been assumed in (\ref{J02weq1}) that the dielectric is
a perfect isolator;\footnote{This is the case, for example, of optical fibers and many other photonic
devices. In other cases, the equations may have to be reformulated.} therefore $\rho_{f}=0$ and $\mbox{\boldmath\ensuremath{\mathcal{J}}}{}_{f}=\mathbf{0}.$

We will first consider the vacuum as the medium where the propagation
takes place. Then, $\rho_{d}=0$ and $\mbox{\boldmath\ensuremath{\mathcal{J}}}{}_{d}=\mathbf{0},$
and (\ref{J02weq1}) simplifies further to 
\begin{equation}
\nabla^{2}\mbox{\boldmath\ensuremath{\mathcal{E}}}(\boldsymbol{r},t)-\dfrac{1}{c^{2}}\frac{\partial^{2}}{\partial\,t^{2}}\mbox{\boldmath\ensuremath{\mathcal{E}}}(\boldsymbol{r},t)=\mathbf{0}\qquad\text{\textsf{{\small\{vacuum\}}}}\emph{.}\label{J02vacio}
\end{equation}
Assuming a linearly-polarized plane wave propagating in the $z$ direction,\footnote{$\boldsymbol{\hat{u}}$ is the unitary vector defining the polarization
of the electric field. As is explained in any elementary textbook
in electromagnetism, the electric and magnetic fields of such plane
waves are transverse and mutually orthogonal (this follows trivially
from forcing the fields to fullfill the individual Maxwell's equations).
The direction of $\boldsymbol{\hat{u}}$ is thus arbitrary, but within
the $xy$ plane.} $\mbox{\boldmath\ensuremath{\mathcal{E}}}=\boldsymbol{\hat{u}~}f(z,t),$
it can be readily verified that \emph{any} function shape $f$ satisfies
eq. (\ref{J02vacio}) (in which $\nabla^{2}$ now reduces to $\partial^{2}/\partial z^{2}$)
\emph{as long as} the spatial and temporal arguments are tied to each
other in this specific form: $f(z,t)=f(z\mp ct).$ Any such function
obviously describes an electric field with an initial (say, at $t=0$)
spatial distribution $f(z,0),$ which propagates --- unaltered ---
in the $\pm z$ direction, precisely at speed $c.$

Let us now turn to a dielectric medium. In this case, there is charge
density due to the protons and electrons of the atoms of the dielectric.
However, the function $\rho_{d}(\boldsymbol{r},t)$ cannot be formulated
\emph{microscopically}. If an electron, for example, is considered
to be a point charge --- which is almost the universal choice in
both classical and quantum electrodynamics\footnote{Bizarre as it may appear, a mathematical point in $
\mathbb{R}
^{3}$, lacking any ``size''\ by definition (its volume is not even differential,
but strictly \emph{zero}!), is supposed to house charge, mass, spin...
It is thus not surprising that the point charge model leads to unsolved
theoretical problems at the fundamental level.} ---, its associated charge density can only be $\rho_{d}(\boldsymbol{r},t)=-q\delta(\boldsymbol{r}-\boldsymbol{r}_{e}(t)),$
where $\boldsymbol{r}_{e}(t)$ is the instantaneous position of the
electron (which, by the way, is uncertain because of thermal fluctuations
and quantum uncertainty). Even if $\boldsymbol{r}_{e}(t)$ were known,
we would have to conclude that $\rho_{d}(\boldsymbol{r},t)$ varies
wildly at an atomic scale. On the contrary, in Maxwell's equations
the electric charge is idealized as being some sort of macroscopic
``fluid''. This means that a spatial charge averaging is carried
out in the following fashion:

Take a sphere centered at $\boldsymbol{r}$ with a very small radius
at a \emph{macroscopic scale} (but large enough at a microscopic scale
so as to embrace many atoms or molecules\footnote{For example, a cube with a 10 nm side can accomodate roughly $\sim40,000$
SiO$_{2}$ molecules inside.}). We can then think of the macroscopic charge density ``at''\ the
point $\boldsymbol{r}$ as: $\rho_{d}(\boldsymbol{r},t)=(\sum_{i}q_{i})/V_{\boldsymbol{r}}$
(C/m$^{3}$), with $\{q_{i}\}$ being all the charges inside the small
volume $V_{\boldsymbol{r}}$ of the sphere located at $\boldsymbol{r}.$
In this way $\rho_{d}(\boldsymbol{r},t)$ becomes a continuous and
differentiable function (with the possible exception at the sharp
boundaries between different materials), suitable for the Maxwell
equations. Expressed colloquially,\ we have mashed the peas (the
discrete point charges) and filled the space with the resulting purï¿œe
($\rho_{d}$).

\begin{figure}[h]
\centering{}\includegraphics[width=1.8928in,height=1.5629in]{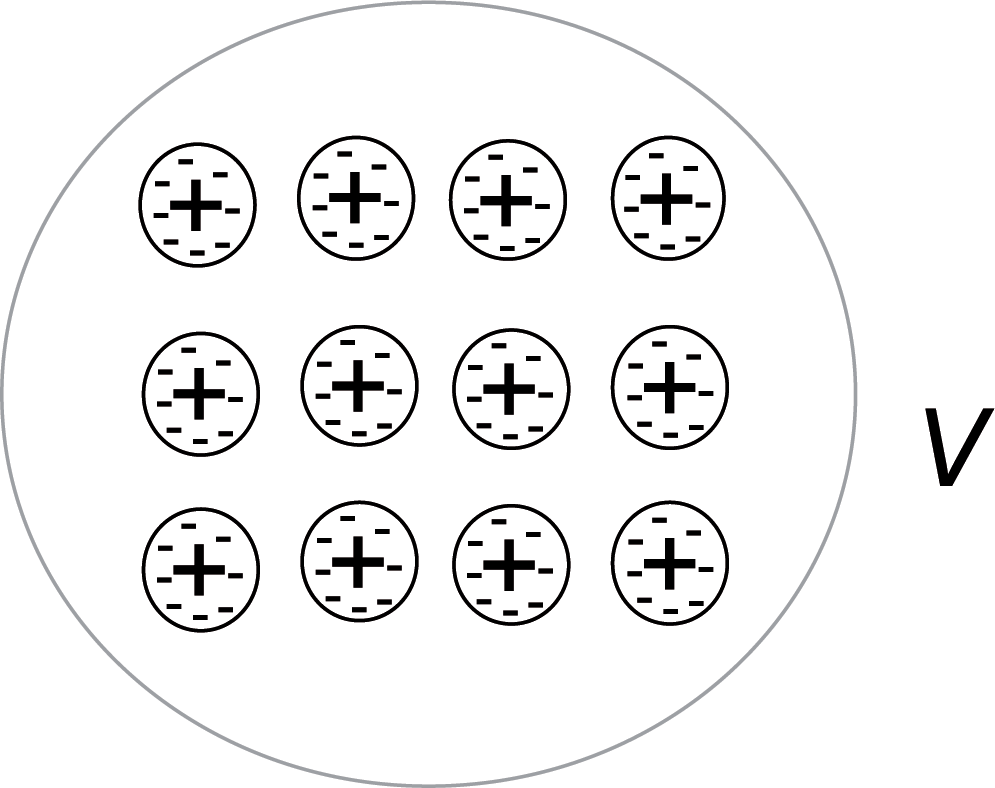}\caption{Apparently, the macroscopic (microscopically averaged) net charge
at any location in a dielectric should always be zero.}
\label{J02fneutro} 
\end{figure}

\begin{figure}[h]
\centering{}\includegraphics[width=5.8284in,height=1.7757in,keepaspectratio]{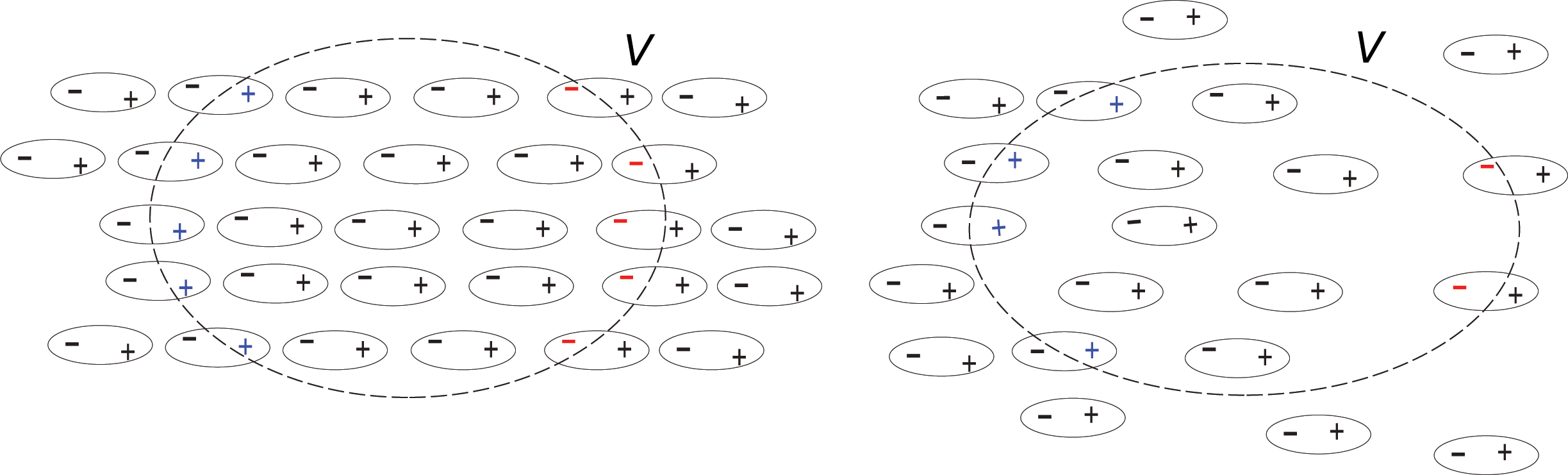}\caption{These rough schemes are only intended to illustrate how a ``net macroscopic
charge''\ can arise inside a dielectric. \emph{Left.} Although the
propagated electric field polarizes the molecules of the dielectric,
if the dipolar moment distribution is spatially homogeneous --- more
accurately, if it does not diverge ---, the net dielectric charge
is zero at all ``macroscopic''\ points.This is grafically motivated
by assuming an horizontal (left-to-right) electric field and computing
the net charge within a small ``macroscopic''\ volume\ $V$ centered
at some point. All dipoles within the volme contribute with zero net
charge, while those ``crossing''\ the left border (in this somewhat
naive view) are also compensated by those crossing the right border.
\emph{Right.} When the dipolar moments are inhomogeneously distributed
(namely, when $\nabla\cdot\mbox{\boldmath\ensuremath{\mathcal{P}}}=0$),
nonzero dielectric charge arises. In this case, for example, the density
of the material could be thought of as decreasing to the right, thus
yielding a non-zero locally-averaged charge distribution.}
\label{J02fdipolos} 
\end{figure}

In view of the discussion above, any ordinary material appears to
have $\rho_{d}(\boldsymbol{r},t)=0$ and $\mbox{\boldmath\ensuremath{\mathcal{J}}}{}_{d}(\boldsymbol{r},t)=\mathbf{0}$
at all (``macroscopic'') points $\boldsymbol{r}$, since every atom
has the same the number of electrons and protons, and any charge averaging
will seemingly yield a zero value, as illustrated in Fig. \ref{J02fneutro}.
So, apparently, the presence of any dielectric material whatsoever
would not make any difference in the form of the Maxwell equations
with respect to the vacuum! Actually, a difference does arise when
an externally generated electric field (such as that of an electromagnetic
wave) is present in the dielectric. Such an electric field alters
the original equilibrium charge distribution of the atoms or molecules
--- which tends to be non-polar --- inducing atomic or molecular
dipole moments, as sketched in Fig. \ref{J02fdipolos} (left). But
this is not enough, since the average net charge would still be zero
in the depicted example. Thus, we conclude that the appearance of
net dielectric charge not only requires the presence of induced polarization,
but also that it be spatially \emph{inhomogeneous}, as illustrated
in Fig. \ref{J02fdipolos} (right). Actually, the following result
can be obtained:\footnote{With different levels of detail and rigour, the derivation of (\ref{J02rhod})
and other related results is presented in innumerable textbooks on
electromagnetism, optics and solid state physics (see for example
\cite{ashcroft_solid_1976}). While clear and (relatively) simple
accounts can be found in some classical references, deeper and more
comprehensive analyses do exist \cite{cho_reconstruction_2010}. The
subject is far more complicated than it would first appear, and it
cannot be considered to be completely settled.} 
\begin{equation}
\rho_{d}(\boldsymbol{r},t)=-\nabla\cdot\mbox{\boldmath\ensuremath{\mathcal{P}}}(\boldsymbol{r},t).\label{J02rhod}
\end{equation}
$\mbox{\boldmath\ensuremath{\mathcal{P}}}(\boldsymbol{r},t)$ is the
\textit{electric polarization,} and basically measures the quantity
and strength of the dipoles induced by the field. Idealizing the material
at a (macroscopic) point as a collection of very many tiny dipoles
having a dipole moment $\boldsymbol{p}=q\boldsymbol{s}$ (corresponding
to two point charges $+q$ and $-q$ separated by a distance $s$),
we can write 
\begin{equation}
\mbox{\boldmath\ensuremath{\mathcal{P}}}(\boldsymbol{r},t)=N(\boldsymbol{r})\boldsymbol{p}(\boldsymbol{r},t)=qN(\boldsymbol{r})\boldsymbol{s}(\boldsymbol{r},t)\qquad\text{[C/m}^{2}\text{],}\label{J02Np}
\end{equation}
where $N(\boldsymbol{r})$ is the volumetric density of dipoles (in
m$^{-3}$). $N$ will in general depend on the (macroscopic) position
$\boldsymbol{r}$ within the material if the dielectric is not spatially
homogeneous (its chemical composition and /or density change with
position). The dipolar distance $\boldsymbol{s}$ may be position-dependent
for the same reason, but also because the strength of the electric
field --- the agent inducing the dipoles --- will in general be
a function of $\boldsymbol{r}.$

The problem would be solved if the functional form of $\boldsymbol{p}$
with the applied field $\mbox{\boldmath\ensuremath{\mathcal{E}}}$
were known. It appears reasonable to think that the dipole moments
$\boldsymbol{p}$ somehow follow $\mbox{\boldmath\ensuremath{\mathcal{E}}}$
since, if the electric force $\mbox{\boldmath\ensuremath{\mathcal{F}}}{}_{e}=-q\mbox{\boldmath\ensuremath{\mathcal{E}}}$)
increases, the negative charge cloud will be displaced farther away
from the (fixed) positive charge; this means that\ $\boldsymbol{s}$
will be increased, and so will $\boldsymbol{p}$ (and $\mbox{\boldmath\ensuremath{\mathcal{P}}}$).
Actually, for ``moderate''\ field intensities, the relation between
$\boldsymbol{p}$ and $\mbox{\boldmath\ensuremath{\mathcal{E}}}$
can be taken as linear... but not necessarily instantaneous, as we
will see next. Note, finally, that the result (\ref{J02rhod}) is
consistent with the qualitative reasoning illustrated in Fig. \ref{J02fdipolos}:
only if the polarization vector diverges --- which accounts for an
inhomogeneous dipolar distribution ---, can an effective dielectric
charge exist.

\begin{figure}[h]
\centering{}\includegraphics[width=2.543in,height=0.6768in,keepaspectratio]{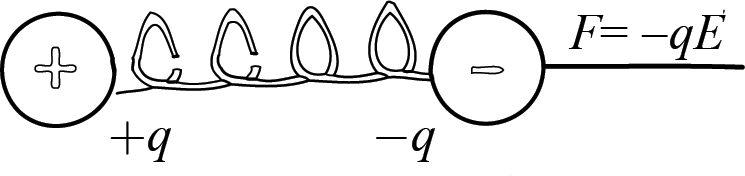}\caption{For small displacements around its equilibrium position ($s=0$),
an active electron is bound to the nucleus by an elastic force ($F_{\text{elast}}=ks)$\ symbolized
by the spring. Isotropy is assumed in this model, i.e., $k$\ is
the same regardless of the spatial direction of $s.$}
\label{J02fmuelle} 
\end{figure}

The theoretical framework to calculate the form of $\mbox{\boldmath\ensuremath{\mathcal{P}}}$
is the Quantum theory, but that is well beyond the scope of this lecture,
so we will use the naive (yet fortunate) classical model of Lorentz,
in which an electron responding to the electric field is envisioned
as being bound to an atomic nucleus by an elastic force. The model
is sketched in Fig. \ref{J02fmuelle}. We choose to call $x$ the
polarization direction of the $\mbox{\boldmath\ensuremath{\mathcal{E}}};$
\emph{i.e.}, $\mbox{\boldmath\ensuremath{\mathcal{E}}}=\boldsymbol{\hat{x}}\mathcal{E}$.
Then also, it can be accepted that $\boldsymbol{s}$ $=\boldsymbol{\hat{x}}s,$
as shown in the figure (this in indeed the case in \emph{isotropic}
dielectrics). Needless to say, the spring in Fig. \ref{J02fmuelle}
is fictitious and intended solely as a reminder that a force of elastic
type binds the electron to the nucleus. Thus for a \emph{static} electric
field $\mbox{\boldmath\ensuremath{\mathcal{E}}}=-\hat{x}\mathcal{E}$
at the electron location, the Coulomb force must equal the elastic
force, which, for small elongations, is proportional to the displacement
of the electron from its equilibrium position: 
\begin{equation}
ks=q\mathcal{E}\qquad\text{\textsf{{\small\{static electric field\}}}}\emph{.}\label{J02Hook estatica}
\end{equation}
In the purely mechanical interpretation, (\ref{J02Hook estatica})
derives from Hooke's law, which states that $F_{\text{elastic}}=ks$
for small elongations of the spring, with $k$ the \emph{elastic constant}
of the latter. Is is assumed here that the much more massive nucleus
is hardly affected by the electric field and remains static at $x=0$.
Relation (\ref{J02Hook estatica}) shows that $s\propto\mathcal{E},$
so that, according to (\ref{J02Np}), $P(\boldsymbol{r})\propto\mathcal{E}$
$(\boldsymbol{r}),$ as anticipated. But the electric field of a wave
will always be time-varying, so the electron will experience accelerations,
and the following dynamic equation must be applied rather than (\ref{J02Hook estatica}):
\begin{equation}
m\,\ddot{s}(t)=q\mathcal{E}(t)-k\,s(t)-\zeta\dot{s}(t).\label{J02Hook dinamica}
\end{equation}
The overdots denote time derivative, $d/dt.$ Relation (\ref{J02Hook dinamica})
is the Abraham--Lorentz equation, and simply expresses Newton's second
law for the electron: its mass times its acceleration must equal the
net force upon it. The forces involved are the Coulomb force\footnote{Since the electron moves at a velocity $v(t)=\dot{s}(t),$ one might
wonder why the magnetic part of the Lorentz force, $q\boldsymbol{v}\times
\mbox{\boldmath\ensuremath{\mathcal{B}}}
$, has not been included in the dynamical equation. Actually, it can
be shown that the magnetic contribution is negligible, both in the
ficticious mechanical model (as long as $|v|<<c$) and in the ``real''\ one.} (in the $+x$ direction), the elastic force (in the $-x$ direction),
and a ``dissipative''\ force, introduced \emph{ad hoc}, which accounts
for the unavoidable energy losses. In the mechanical model, it is
attributed to the ``friction''\ of the spring system, which opposes
to the electron movement with a force proportional to its velocity
$\dot{s}(t)$, $\zeta$ being the proportionality constant. In reality,
losses are caused by different physical processes to be discussed
in Subsection \ref{J02sbsctTFDRI}. 


Only for a static situation, $d/dt=0$ and equation (\ref{J02Hook dinamica})
obviously reduces to (\ref{J02Hook estatica}). Otherwise we see that
the polarization does not follow the electric field instantaneously,
because of the time derivatives. However, the differential equation
(\ref{J02Hook dinamica}) is linear with constant coefficients, so
it describes a linear system which can be most easily understood in
the frequency domain. Taking the Fourier transform (FT) of (\ref{J02Hook dinamica})
(see Appendix), a \emph{linear, frequency-dependent} relationship
is obtained between the \emph{FTs} of $\mathcal{E}(t)$ and $s(t),$
$E(\omega)$ and $S(\omega),$ respectively: 
\begin{equation}
S(\omega)=\frac{q/m}{k/m-\omega^{2}+i\omega(\zeta/m)}E(\omega)\equiv K_{\text{diel}}(\omega)E(\omega).\label{J02x}
\end{equation}

If the input field is purely monochromatic, the \emph{phasor} version
(\ref{J02x}) can be used --- see Subsection \ref{JApAfourierseries}

The simplicity of (\ref{J02x}) compared with (\ref{J02Hook dinamica})
suggests that we carry on the derivations in the frequency domain
if possible. Thus taking the FT of (\ref{J02Np}) and using (\ref{J02x}),
we obtain (omitting the variable $\boldsymbol{r}$ for clarity): 
\begin{align}
\boldsymbol{P}(\omega) & =qN\boldsymbol{S}(\omega)=qNK_{\text{diel}}(\omega)\boldsymbol{E}(\omega)\nonumber \\
 & \equiv\varepsilon_{0}\chi(\omega)\boldsymbol{E}(\omega)\qquad\text{\textsf{{\small\{linear and isotropic\}}}}\emph{.}\label{J02PEm}
\end{align}
where $\chi(\omega)$ is the (linear) \emph{dielectric susceptibility,}
defined as the frequency-dependent polarization response of the dielectric
to an electric field. For the moment, even if the appearance of $qNK_{\text{diel}}(\omega)$
arises from a fictitious model, we will keep to the equality $\chi(\omega)=qNK_{\text{diel}}(\omega)/\varepsilon_{0}$
implied in (\ref{J02PEm}) as the functional form of the susceptibility.
The unimportant prefactor $\varepsilon_{0}$ has been introduced for
later convenience. Then,\footnote{Our simple argument has overlooked an important difficulty. $\mathcal{E}$
in (\ref{J02Hook dinamica}) or in (\ref{J02Hook estatica}) has to
be the actual electric field at the electron site, that is, the \emph{local}
field. But the local field does not coincide with the ``external''\ field
of the propagating wave (which is the macroscopic electric field appearing
in the Maxwell equations): The neighbouring atoms, which become polarized
as well, generate additional contributions to the field at the location
considered. In non-dense media such as a rarified gas, the effect
of the relatively distant surrounding dipoles can be neglected, but
this is not the case in a solid. Generally, it can be shown that the
correction is such that the macroscopic polarization is, again, proportional
to the macroscopic field, so that equation (\ref{J02PEm}) remains
valid.} 
\begin{equation}
\chi(\omega)=\dfrac{qN}{\varepsilon_{0}}\frac{q/m}{k/m-\omega^{2}+i\omega(\zeta/m)}\qquad\text{(Lorentz's model).}\label{J02chiL}
\end{equation}

A non-differential relationship can be established in the time domain
between $\mbox{\boldmath\ensuremath{\mathcal{P}}}$ and $\mbox{\boldmath\ensuremath{\mathcal{E}}}$
through a \emph{convolution} (Subsection \ref{JApAconvolution}).
Taking the FT$^{-1}$ of (\ref{J02PEm}), we have 
\begin{equation}
\mbox{\boldmath\ensuremath{\mathcal{P}}}(t)=\varepsilon_{0}\breve{\chi}(t)\ast\mbox{\boldmath\ensuremath{\mathcal{E}}}(t)=\varepsilon_{0}\int_{-\infty}^{\infty}\breve{\chi}(\tau)\mbox{\boldmath\ensuremath{\mathcal{E}}}(t-\tau)d\tau=\varepsilon_{0}\int_{0}^{\infty}\breve{\chi}(\tau)\mbox{\boldmath\ensuremath{\mathcal{E}}}(t-\tau)d\tau.\label{J02conv}
\end{equation}
Thus $\breve{\chi}(t)=\,$FT$^{-1}[\chi(\omega)]$ is the ``impulse
response''\ of the dielectric (at the particular point $\boldsymbol{r}$
considered), using the terminology of linear systems theory. The last
equality in (\ref{J02conv}) follows from the \emph{causality} of
the response, that is, from the presumption that the polarization
of the material at time $t$ cannot depend on the \emph{on the future}
electric field, at $\tau>t.$ This requirement can only be ensured
in (\ref{J02conv}) if $\breve{\chi}(\tau)=0$ for $\tau<0$, a condition
to which we indeed adhere. Then, the response of the dielectric at
time $t$ depends on all the \emph{past} history, $\tau\leq t,$ of
the electric field. We then say that the dielectric response is\emph{\ dispersive}
in time.\footnote{In contrast, $
\mbox{\boldmath\ensuremath{\mathcal{P}}}
(\boldsymbol{r})$ only depends on $
\mbox{\boldmath\ensuremath{\mathcal{E}}}
(\boldsymbol{r})$ in our model. But there are cases --- not to be considered in our
discussion --- with \emph{spatial dispersion}, in which $
\mbox{\boldmath\ensuremath{\mathcal{P}}}
(\boldsymbol{r})$ is significantly contributed by $
\mbox{\boldmath\ensuremath{\mathcal{E}}}
(\boldsymbol{r})$ at the neighbouring points also. This happens when there is energy
transport by other mechanisms besides the electromagnetic field.}

$\mbox{\boldmath\ensuremath{\mathcal{J}}}{}_{d}(\boldsymbol{r},t),$
still unknown, is easily related to $\rho_{d}(\boldsymbol{r},t)$
through the law of conservation of charge:\footnote{This law is contained in the Maxwell equations, its derivation being
straighforward. Physically it states that, if charge happens to be
accumulating inside (disappearing from) some differential volume of
space, then necessarily a net electric current is entering (exiting)
the volume. In other words: (a) charge is conserved, and (b) it does
not ``magically''\ jump between distant locations.} $\nabla\cdot\mbox{\boldmath\ensuremath{\mathcal{J}}}{}_{d}=-\partial\mbox{\boldmath\ensuremath{\mathcal{P}}}/\partial t.$
Using (\ref{J02rhod}), it follows that\footnote{As a matter of fact, the solution is $
\mbox{\boldmath\ensuremath{\mathcal{J}}}
_{d}=\partial\,
\mbox{\boldmath\ensuremath{\mathcal{P}}}
/\partial\,t+
\mbox{\boldmath\ensuremath{\mathcal{V}}}
,$ with \emph{any} $
\mbox{\boldmath\ensuremath{\mathcal{V}}}
$ such that $\nabla\cdot
\mbox{\boldmath\ensuremath{\mathcal{V}}}
=0;$ but the most economical possibility, $
\mbox{\boldmath\ensuremath{\mathcal{V}}}
=0,$ yields consistent results.} 
\begin{equation}
\mbox{\boldmath\ensuremath{\mathcal{J}}}_{d}(\boldsymbol{r},t)=\dfrac{\partial\,\rho_{d}(\boldsymbol{r},t)}{\partial\,t}.\label{J02Jd}
\end{equation}
Therefore, if the dielectric polarization varies with time, a dielectric
current will be generated.\footnote{This is not a conduction current due to freely moving charges, as
in a metallic conductor. Each electron remains bound to its nucleus
and only shifts back and forth around its fixed equilibrium position.
However, since, according to (\ref{J02rhod}), there \emph{is} a resulting
dielectric charge, any time variation of its spatial distribution
may naturally result in an effect of ``dielectric''\ current. Think,
for example, of the ``moving''\ patterns of light generated by
synchronized on-and-off switching of the fixed light bulbs on a marquee.}

\subsection{Wave equations in time and in frequency}

\label{J02sbsctWETF}

Replacing (\ref{J02rhod}) and (\ref{J02Jd}) in (\ref{J02weq1}),
a time wave equation is obtained expressed in terms of the polarization
(the variables $\boldsymbol{r},t$ are omitted): 
\begin{equation}
\fbox{\ensuremath{\nabla^{2}\mbox{\boldmath\ensuremath{\mathcal{E}}}-\dfrac{1}{\varepsilon_{0}c^{2}}\dfrac{\partial^{2}}{\partial\,t^{2}}\left[\varepsilon_{0}\mbox{\boldmath\ensuremath{\mathcal{E}}}+\mbox{\boldmath\ensuremath{\mathcal{P}}}\right]=\left\{ \begin{tabular}{l}
 \ensuremath{=-\dfrac{1}{\varepsilon_{0}}\nabla(\nabla\cdot\mbox{\boldmath\ensuremath{\mathcal{P}}})}\\
 \ensuremath{=\text{ }0\text{\quad(or }\simeq0\text{)}\quad\text{often}} 
\end{tabular}\ \right.\qquad\text{\textsf{{\small\{general\}}}}\quad}}\label{J02weq2}
\end{equation}
By \emph{general} we mean that the equation also \emph{applies to
nonlinear and/or anisotropic dielectrics}, since the form of $\mbox{\boldmath\ensuremath{\mathcal{P}}}$
is unspecified and not necessarily that of (\ref{J02conv}). Actually,
(\ref{J02weq2}) should be the starting point in the study of such
cases, but here we will limit ourselves to (linear) isotropic, homogeneous
dielectrics. While isotropy is already implicit in relation (\ref{J02PEm}),
homogeneity means $\boldsymbol{r}$-independence: $\chi(\boldsymbol{r},\omega)=\chi(\omega).$
Thus in this case the RHS of (\ref{J02weq2}) is identically zero
for linear response.\footnote{This is proved most easily in the frequency domain. Substituting the
FT of relation (\ref{J02rhod}) in the FT of (\ref{J02max1}), and
using (\ref{J02PEm}), we obtain: $\nabla\cdot\boldsymbol{E}(\boldsymbol{r},\omega)=-\varepsilon_{0}^{-1}\nabla\cdot\boldsymbol{P}(\boldsymbol{r},\omega)=\allowbreak-\chi(\omega)\nabla\cdot\boldsymbol{E}(\boldsymbol{r},\omega).$
Since $\chi(\omega)\neq0,$ then necessarily $\nabla\cdot\boldsymbol{P}(\boldsymbol{r},\omega)=0$
for all $\omega$; therefore, 
\[
\nabla\cdot
\mbox{\boldmath\ensuremath{\mathcal{P}}}
(\boldsymbol{r},t)=0.
\]
}

The term in square brackets on the LHS of (\ref{J02weq2}) is defined
as the \emph{electric displacement} vector: 
\begin{equation}
\mbox{\boldmath\ensuremath{\mathcal{E}}}(\boldsymbol{r},t)+\mbox{\boldmath\ensuremath{\mathcal{P}}}(\boldsymbol{r},t).\label{J02DEP}
\end{equation}

In order to obtain the \emph{homogeneous} wave equation in a linear
dielectric, we apply the FT to (\ref{J02weq2}) and make use of (\ref{J02PEm}),
which leads to 
\begin{equation}
\fbox{\ensuremath{\nabla^{2}\boldsymbol{E}+\dfrac{\omega^{2}}{c^{2}}n^{2}(\omega)\boldsymbol{E}=\mathbf{0}\qquad\left\{ \begin{array}{l}
\text{\textsf{{\small linear, isotropic}}}\\
\text{\textsf{{\small\& homogeneous}}}
\end{array}\right\} }}\label{J02weq3}
\end{equation}
where 
\begin{equation}
n(\omega)\equiv\sqrt{\varepsilon^{\prime}(\omega)}\label{J02epspri}
\end{equation}
is the (frequency-dependent) \emph{index of refraction}, or \emph{refractive
index}, defined as the square root of the \emph{relative dielectric
permittivity}, which is in turn defined as 
\begin{equation}
\varepsilon^{\prime}(\omega)\equiv1+\chi(\omega).
\end{equation}
Naturally, $\varepsilon^{\prime}$ and $n$ are $\boldsymbol{r}$-dependent
if $\chi$ is. Note that, even if $\boldsymbol{E}=\boldsymbol{E}(\boldsymbol{r},\omega),$
no differential operator acts on $\omega,$ so the frequency is a
parameter, rather than a variable, in the Fourier-transformed wave
equation.

In the isotropic, linear (but not necessarily homogeneous) case, the
FT of (\ref{J02DEP}) yields 
\begin{equation}
\boldsymbol{D}(\boldsymbol{r},\omega)=\varepsilon(\boldsymbol{r},\omega)\boldsymbol{E}(\boldsymbol{r},\omega)\qquad\text{\textsf{{\small\{linear and isotropic\}}}\emph{{\small,}}}
\end{equation}
with $\varepsilon(\boldsymbol{r},\omega)\equiv\varepsilon_{0}\varepsilon^{\prime}(\boldsymbol{r},\omega),$
the (absolute) \emph{dielectric permittivity}.\footnote{It is also straightforwad to see that, when Fourier-transformed, Maxwell's
equation (\ref{J02max1}) becomes

\begin{equation}
\nabla\cdot\lbrack\varepsilon(\boldsymbol{r},\omega)\boldsymbol{E}(\boldsymbol{r},\omega)]=0\qquad\text{\textsf{{\small\{linear and isotropic\}}}\emph{{\small;}}}\label{note11}
\end{equation}
or equivalently, $\nabla\cdot\boldsymbol{D}(\boldsymbol{r},\omega)=0.$
In an homogeneous dielectric $\varepsilon(\boldsymbol{r},\omega)=\varepsilon(\omega)$
and (\ref{note11}) reduces to $\nabla\cdot\boldsymbol{E}(\boldsymbol{r},\omega)=0$
--- which implies $\ \nabla\cdot
\mbox{\boldmath\ensuremath{\mathcal{E}}}
(\boldsymbol{r},t)=0$. In a situation where there is free charge as well, the $0$ on the
RHS of (\ref{note11}) is merely replaced by $\rho_{f}(\boldsymbol{r},\omega)$
(denoted simply $\rho(\boldsymbol{r},\omega)$ in many textbooks).
Likewise, (\ref{J02max4}) ends up as

\begin{equation}
c^{2}\nabla\times\boldsymbol{B}(\boldsymbol{r},\omega)=i\omega\varepsilon^{\prime}(\boldsymbol{r},\omega)\boldsymbol{E}(\boldsymbol{r},\omega)\qquad\text{\textsf{{\small\{linear and isotropic\}}}\emph{{\small,}}}
\end{equation}
or, equivalently, $\nabla\times\boldsymbol{H}(\boldsymbol{r},\omega)=i\omega\boldsymbol{D}(\boldsymbol{r},\omega).$
The remaining Maxwell equations (\ref{J02max2}) and (\ref{J02max3})
are not modified as they do not involve any interaction with the material.}

Eq. (\ref{J02weq3}) displays the key effect of the dielectric almost
blatantly. With $\varepsilon^{\prime}=1,$ (\ref{J02weq3}) is obviously
the wave equation in vacuum; this can also be checked by taking the
FT of (\ref{J02vacio}). Therefore, the wave equation in the dielectric
is formally identical to that in vacuum, with the only difference
that $c/n(\omega)$ appears in place of $c$. The index $n(\omega)$
is in general a \emph{complex} quantity {[}see (\ref{J02epspri}),
(\ref{J02PEm}) and (\ref{J02x}){]}. Calling the real and imaginary
parts $n_{r}$ and $-n_{i},$ respectively (the minus sign is chosen
for later convenience), we write\footnote{The notation $n_{\text{complex}}=n-i\kappa$ is also frequently employed,
with $\kappa$ called the \emph{extinction coefficient}. We see that
it relates to our convention by $n\equiv n_{r},$ $\kappa\equiv n_{i}$
and $n_{\text{complex}}\equiv n.$ (Actual symbols used for $n_{\text{complex}}$
can be $n_{c},$ $\tilde{n},$ $\bar{n},$ $N...$)} 
\begin{equation}
n(\omega)=n_{r}(\omega)-in_{i}(\omega).\label{J02nrni}
\end{equation}
\emph{If }$n(\omega)$ \emph{were real}, we could readily conclude,
from the form of (\ref{J02weq3}), that plane waves are again solutions
of the wave equation in an homogeneous dielectric, only with their
(frequency-dependent) velocity of propagation being $c/n(\omega)$
rather than $c.$ But since a complex velocity is meaningless, we
will have to solve (\ref{J02weq3}) explicitly.

Assuming a linearly-polarized ($\boldsymbol{E=\hat{u}}E$) plane wave
propagating in the $z$ direction ($\nabla^{2}=\partial^{2}/\partial z^{2}$),
a simple scalar wave equation is obtained: 
\begin{equation}
\dfrac{d^{2}E(z,\omega)}{dz^{2}}+\dfrac{\omega^{2}}{c^{2}}n^{2}(\omega)E(z,\omega)=0.\label{J02weq4}
\end{equation}

\subsection{Attenuated monochromatic waves}

\label{J02sbsctAMW}

The general solution of (\ref{J02weq4}) is 
\begin{equation}
E(z,\omega)=A(\omega)e^{-i(\omega/c)n(\omega)z}+B(\omega)e^{i(\omega/c)n(\omega)z},\label{J02eqz1}
\end{equation}
where $A(\omega)$ and $B(\omega)$ are the arbitrary integration
constants (with respect to $z;$ remember $\omega$ is a mere parameter
here). The first and second summands obviously correspond to forward
and backward travelling waves along $z,$ respectively.\footnote{According to our sign convention, as explained in the Appendix.}
We will assume forward propagation only and set $B(\omega)=0.$ Then
$A(\omega)=E(0,\omega)$ is the FT of the electric field at $z=0.$
We now consider a purely\emph{\ monochromatic wave} of discrete frequency
$\omega_{0}.$ The corresponding phasor (Subsection \ref{JApAfourierseries})
is 
\begin{equation}
\tilde{E}(z)=Ae^{-i(\omega_{0}/c)n(\omega_{0})z}=Ae^{-(\omega_{0}/c)n_{i}(\omega_{0})z}e^{-i(\omega_{0}/c)n_{r}(\omega_{0})z},
\end{equation}
with $A\equiv A(\omega_{0}),$ which we assume real. The decomposition
(\ref{J02nrni}) has been employed in the second equality. The real
temporal field is, thus, 
\begin{equation}
\mathcal{E}(z,t)=\operatorname{Re}\left[\tilde{E}(z)e^{i\omega_{0}t}\right]=Ae^{-\,n_{i}(\omega_{0})(\omega_{0}/c)z}\cos\left(\omega_{0}t-n_{r}(\omega_{0})\frac{\omega_{0}}{c}z\right).\label{J02onda+ atenuada(t)}
\end{equation}
The argument of the cosine in (\ref{J02onda+ atenuada(t)}) can be
cast in the form $-n_{r}(\omega_{0})(\omega_{0}/c)[z-t\,c/n_{r}(\omega_{0})]$,
which shows explicitly a functional dependence of the type $f(z-v_{\text{ph}}t)$,
with 
\begin{equation}
v_{\text{ph}}=v_{\text{ph}}(\omega_{0})=\frac{c}{n_{r}(\omega_{0})}.\label{J02vfase}
\end{equation}
Therefore, $v_{\text{ph}},$ called \emph{phase velocity,} is the
velocity of propagation of \emph{the phase of} a sinusoidal (\emph{i.e.},
perfectly \emph{monochromatic}) wave of frequency $\omega_{0}.$ This
velocity does \emph{not} describe the ``velocity of the wave''\ as
one would commonly understand it. This point will be clarified in
Subsection \ref{J02sbsctGV}.

\begin{figure}[h]
\centering{}\includegraphics[width=2.0075in,height=1.2047in,keepaspectratio]{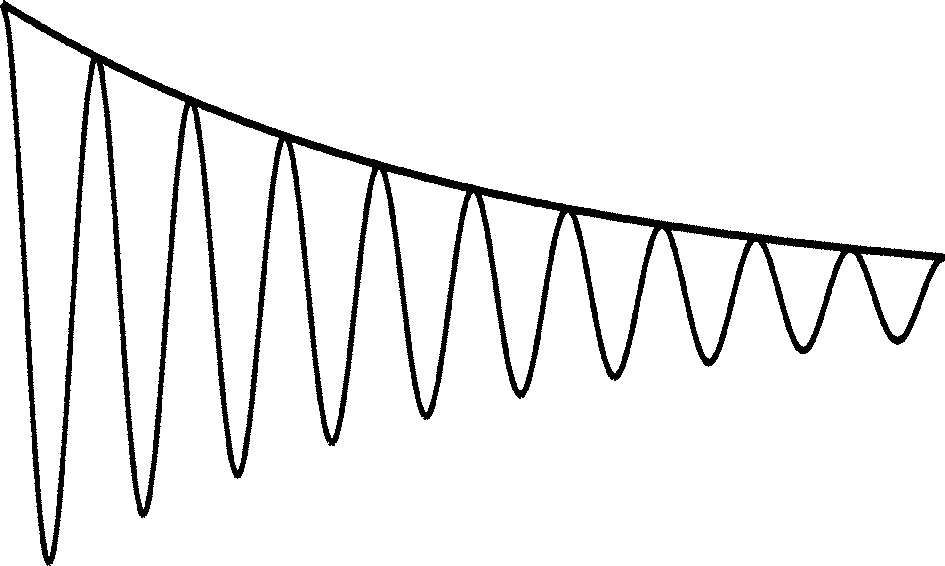}\caption{Form of the forward-propagating solution (\ref{J02onda+ atenuada(t)}).
(The exponential decay is greatly exagerated for illustration purposes.)
Regardless of the spatial damping of the envelope, at each coordinate
$z,$the wave is perfectly monochromatic.}
\label{J02fdamp} 
\end{figure}

We conclude that the phase velocity of a monochromatic wave in an
homogeneous dielectric is $c$ divided by the \emph{real part }of
the refractive index. The imaginary part $-n_{i},$ which is always
negative in normal, non-amplifying materials ($n_{i}>0$), ultimately
arises from the imaginary-valued ``friction''\ term $i\omega(\zeta/m)$
in (\ref{J02x}), and accounts for the material losses. It results
in the decreasing exponential factor in (\ref{J02onda+ atenuada(t)}).
The wave is thus progressively attenuated as it propagates, and $\mathcal{E}(z,t)$
has the damped form sketched in Fig. \ref{J02fdamp}.\footnote{When, contrary to these notes, the convention $kz-\omega t$ is employed,
the complex index is written as $n=n_{r}+in_{i}$ (or $n_{\text{complex}}=n+i\kappa$),
so that a \emph{positive }imaginary part of the refractive index ($n_{i}$)
will yield attenuated propagation. Note that, with our choice $\omega t-kz,$
the imaginary part of the refractive index, which is $-n_{i}$ \emph{including
the minus sign}, should be negative for such a lossy medium, which
results in \emph{our} $n_{i}$ being positive.} In many practical cases, when the dielectric is almost transparent
at the frequencies of interest, the approximations $n_{i}\simeq0$
and $n\simeq n_{r}$ can be made, so that $v_{\text{ph}}\simeq c/n.$

The fact that the speed of light in a dielectric\emph{\ depends on
its frequency} is the key result of this section. This phenomenon
of dispersive nature has abundant and important implications in Photonics.

\subsection{The\ ``true''\ frequency dependence of the refractive index}

\label{J02sbsctTFDRI}

Lorentz's classical model for the electron-field interaction displays
two features which are qualitatively correct: the linear relationship
between the polarization and the electric field (when the latter is
not too large), and the dispersive nature of the dielectric response,
which results in a frequency-dependent complex susceptibility $\chi(\omega)$.
However, quantum-mechanical calculations show, to start with, that
$\chi(\omega)$ is in reality made up of \emph{several} (not only
one) resonant contributions of the form (\ref{J02chiL}), which thus
appear in the actual forms of $\varepsilon^{\prime}(\omega)$ and
$n(\omega)$ as well: 
\begin{equation}
n(\omega)=\sqrt{\varepsilon^{\prime}(\omega)}=\sqrt{1+\chi(\omega)}=\sqrt{1+\sum\limits _{j=1}^{N}\frac{A_{j}\,\omega_{j}^{2}}{-\omega^{2}-i\omega\,b_{j}+\omega_{j}^{2}}},\label{J02ncomp}
\end{equation}
or, in terms of the vacuum wavelength,\footnote{Writing $n(\omega)$ and $n(\lambda)$ simultaneously is a ``sloppy''\ yet
not so rare habbit in the literature (just as writing $E(t)$ and
$E(\omega)$ for a function and its FT, for example). Obviously, it
is mathematically wrong to use the same function name, $n,$ for two
\emph{different} functions. If the refractive index is going to be
expressed regularly as a function of $\lambda,$ the symbols can always
be redefined as $n_{\omega}(\omega)$ and $n(\lambda)$ to shorten
the notation.} 
\begin{equation}
n_{\lambda}(\lambda)\equiv n(2\pi c/\lambda)=\sqrt{1+\sum\limits _{j=1}^{N}\dfrac{A_{j}\lambda^{2}}{\lambda^{2}-i\gamma_{j}\lambda-\lambda_{j}^{2}}},\label{J02nland}
\end{equation}
with $\lambda_{j}=2\pi c/\omega_{j}$ and $\gamma_{j}=b_{j}\lambda_{j}^{2}/(2\pi c).$

The parameters $A_{j},$ $b_{j}$ and $\omega_{j}$ must be calculated
through the quantum-mechanical formalism (or measured) for each material.
One possible origin of the coefficients $b_{j}$ (responsible for
the imaginary part of the refractive index) is the \emph{spontaneous
emission}. We briefly anticipate the quantum model of light-matter
interaction, which essentially is viewed as the interplay of three
processes. In the simplest picture, the monochromatic radiation of
frequency $\nu=\omega/(2\pi)$ is comprised of a flux of photons with
energy $E_{\text{p}}=h\nu$ each. If $E_{\text{p}}$ equals the energy
difference between two atomic levels, some photons may be absorbed
by atoms of the dielectric (\emph{absorption}), which thus become
temporarily excited. Thus, the photon disappears and its energy incorporates
to the absorbing atom through the excitation\ of one of its electrons,
which ``jumps''\ from its original orbital to the higher-energy
orbital; in the notation of (\ref{J02ncomp}), we have\ $\nu=$ $\omega_{j}/(2\pi),$
\emph{i.e.}, the radiation frequency has to coincide with one of the
resonant frequencies. Next, the atom re-emits the\footnote{We use the article \emph{``the''} rather loosely, because there
is no justification to think that the re-emitted photon is ``the
same photon''\ that was absorbed.} photon, hopefully in the same direction and with the appropriate
phase relationship --- somehow ``influenced''\ by the presence
of the electromagnetic field (\emph{stimulated emission}) --- so
as to keep contributing coherently to the global propagating field.
The atom returns to its unexcited energy level, but the new photon
can in turn be absorbed again by another atom of the material, which
then will re-emit it, and so on. However, not all absorbed photons
are re-emitted by stimulated emission; some excited atoms simply de-excite
emitting a photon in a random direction and with unpredictable phase
(\emph{spontaneous emission}). Although such photons do not really
disappear, they can be considered as ``lost''\ for all propagation
purposes. To be precise, some photons may be spontaneously emitted
having, by chance, the right direction and the right phase to add
coherently to the propagating field. However, statistically, the number
of ``lost''\ spontaneous photons is by far greater.

It should be noted that, although the photon ``flies''\ across
the interatomic vacuum at speed $c$, the accumulated processes of
absorption-emission amount to modifying the \emph{phase} of the resulting
electromagnetic field so has to make $v_{f}=c/n_{r}(\omega).$ This
modification of the phase velocity occurs at all frequencies, but
it is particularly abrupt around the resonances, at which, in the
quantum picture, actual absorptions and emissions of photons are much
more likely to take place.

Note that, contrary to what one might at first think and is sometimes
implied in the literature, the described absorption-emission cycles
do \emph{not} result in the \emph{slowing down} of the \emph{pure}
monochromatic wave. A true monochromatic wave never starts or ends,
so the absorption/emission events have been happening ``forever''\ and
it is meaningless to speak of ``propagation delay''. Things will
change in Section \ref{J02sctPS}.

\begin{figure}[h]
\centering{}\includegraphics[scale=0.35]{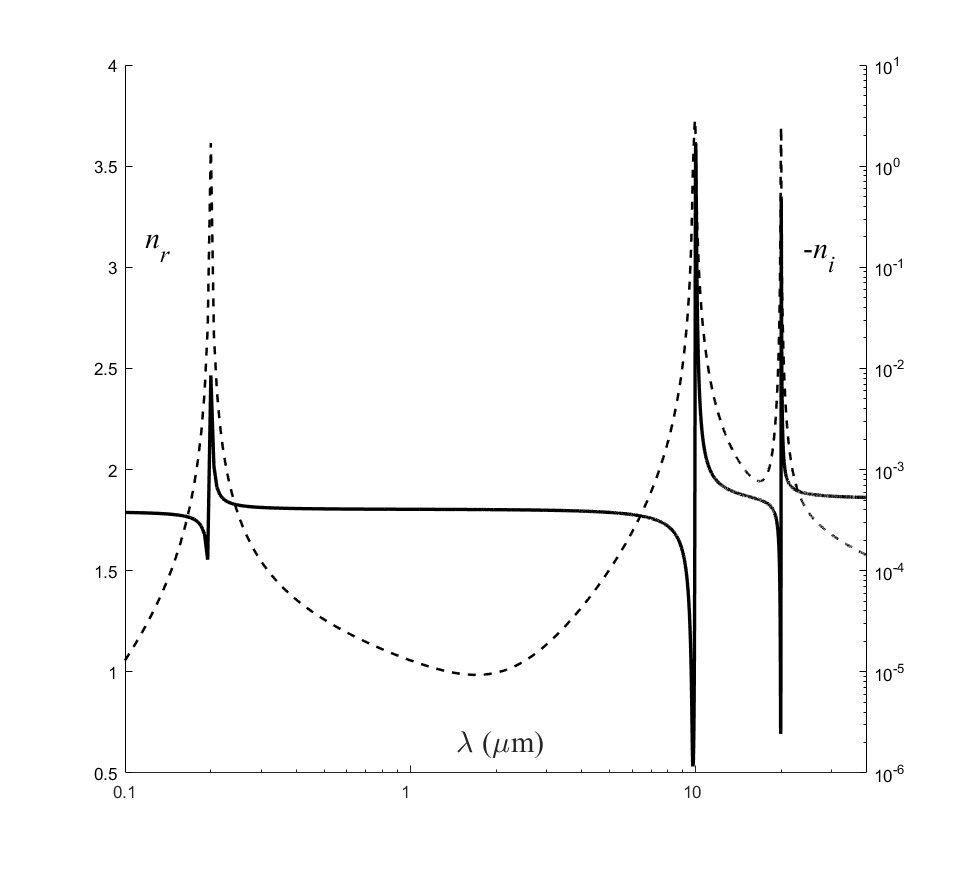}\caption{Real (solid line) and imaginary (dotted line) parts of the refractive
index of an hypothetical dielectric material highly transparent in
the near-IR, with resonances in the UV and mid-IR spectral ranges,
as predicted by formula (\ref{J02ncomp}).}
\label{J02FIGnrniMatlab} 
\end{figure}

It can be seen in Fig. \ref{J02FIGnrniMatlab} that the real refractive
index ``wiggles''\ at any resonance frequency and can reach a value
actually smaller than 1 on the right side. This is related to a phase
shift (most easily described in classical terms, considering the interference
between the incident wave and the wave radiated by the induced atomic
dipoles), and of course is without consequences by itself.

\begin{figure}[h]
\centering{}\includegraphics[width=4.4715in,height=3.2495in,keepaspectratio,viewport=0bp 0bp 16.9963in 12.3228in]{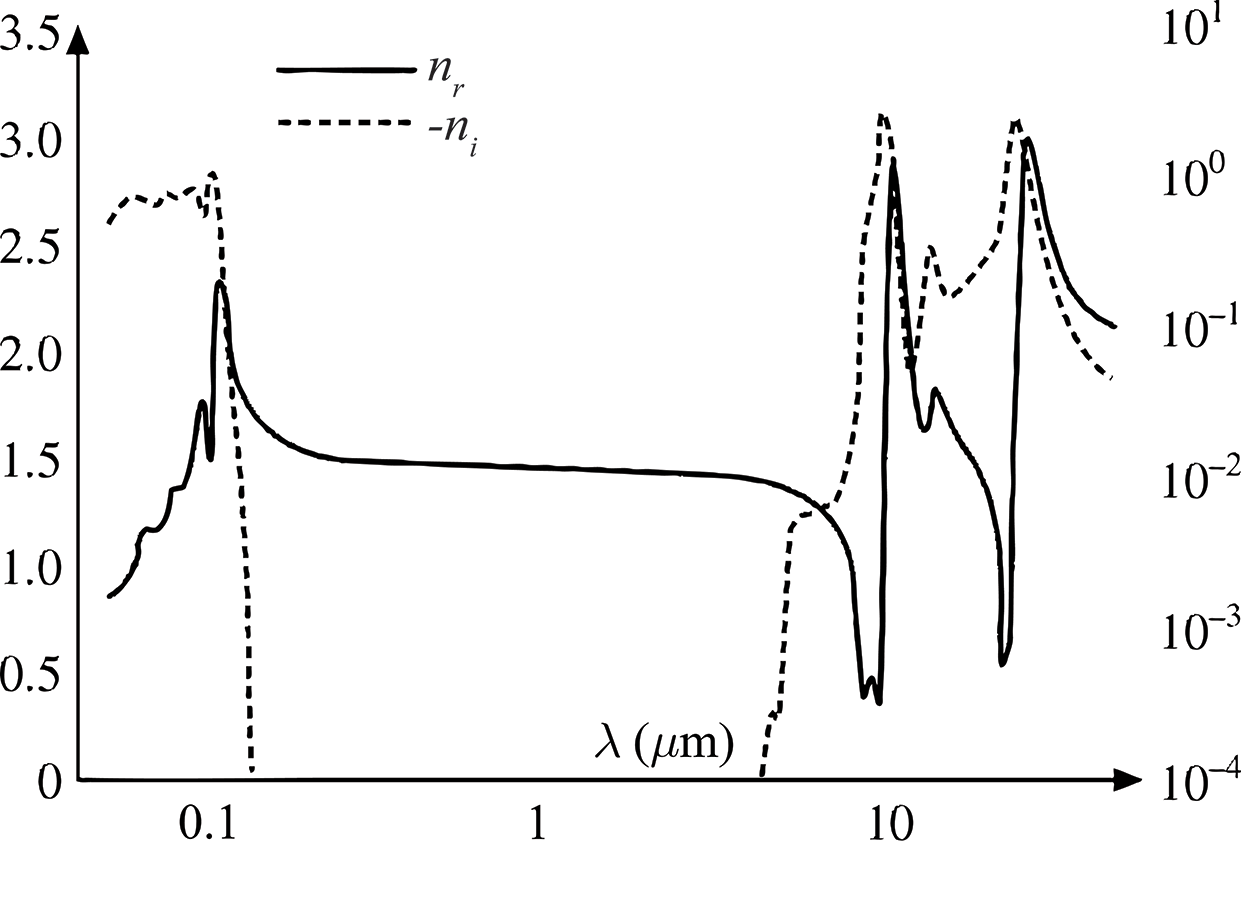}\caption{Actual shapes of the real and imaginary parts of the refractive index
of the silica in the spectral range considered in the Fig. \ref{J02FIGnrniMatlab}.}
\label{J02FIGsilica} 
\end{figure}

Concurring with the simple process of spontaneous emission described
above, in real solids there are other mechanisms whereby photons may
be effectively removed from the propagating field. Interaction with
\emph{phonons }is an important example. Phonons can be described as
quanta of the (unavoidable) thermally-excited vibration modes of the
molecules (these mechanical oscillations are energy-quantified in
phonons, much like the electromagnetic field is in photons). Under
certain circumstances, phonons can be excited at the expense of the
energy of photons, which is thus lost as heat.

Fig. \ref{J02FIGnrniMatlab} illustrates the form of $n(\omega)$
according to (\ref{J02ncomp}) for an hypothetical material with three
absorption peaks. However, by way of example, Fig. \ref{J02FIGsilica}
shows the actual real and imaginary parts of the refractive index
of the \emph{fused silica} glass, which is an amorphous (or vitreous)
form of silicon dioxide or silica, SiO$_{2},$ the most important
material in the current technology of optical fibers. Although there
is some resemblance between the curves, the measured response of SiO$_{2}$
also shows remarkable differences with the simple behavior predicted
by the Lorentz oscillators. It is obvious that a more complete, and
presumably complicated, model of the material is required to describe
its dielectric properties realistically. This is far beyond the scope
of this presentation, so we will only make some qualitative considerations.

In solids, atoms are packed very close to each other, with the result
that their outer orbitals overlap and interact strongly. As a result,
the original discrete energy levels of the isolated atoms broaden
and become (quasi) continuous \emph{bands}. In fused silica, the electronic
transitions (which we have attempted to describe by means of the Lorentz
oscillators) involve tightly bound valence electrons of the SiO$_{2}$
molecules which excitation requires high energy photons corresponding
to the ultraviolet (UV) spectral range. Other effects, such as interactions
with \emph{excitons,}\footnote{An exciton is a ``quasi-particle''\ formed by an electron and a
hole strongly coupled to each other. See for example \cite{chuang_physics_1995}.} make the modelling of the UV absorption even more complex. Thus,
the continuous character of the spectral absorption in the UV region,
makes it more difficult to model it by simple Lorentz's oscillators.

Examining Fig. \ref{J02FIGsilica}, we see that, moving to lower frequencies
from the UV zone, there is a wide, almost transparent spectral range
that reaches the near-infrared zone. As wavelength increases further
within the infrared (IR) zone, the extinction coefficient grows rapidly
again. The absorption in the IR, however, does not take place through
interactions with electrons as in the UV, but through \emph{vibrational}
transitions. Think, for example, of a two-atom molecule of polar character
(\emph{i.e.}, the spatial negative charge is located, say, closer
to one of the atoms than the other, resulting\ in a permanent dipole
moment). Thermal agitation may cause periodical stretching and shrinking
of the interatomic distance, thus modulating the dipole moment. The
situation resembles the modulation of the electronic dipoles depicted
in Fig. \ref{J02fmuelle}. In the electronic interaction, however,
the atoms or molecules were assumed to be nonpolar and it was the
external optical field which produced the electronic dipoles by displacing
the electronic charge from its equilibrium position. On the contrary,
in the present case the dipole moments are pre-existing and their
oscillations are thermally induced. As is inherent to the concept
of phonon, the mechanical oscillations are \emph{collective} (even
in noncrystalline solids\footnote{Vitreous SiO$_{2}$ is a covalent network based on tetrahedra with
SiO$_{4}$ units, but with variations in bond angles and distances,
and absence of perioding order beyond a few near neighbouring units
\cite{palik_handbook_1998}.}) and form wave patterns across the material. Therefore, the dipoles
``riding''\ on these phonon waves can interact coherently with
the IR electromagnetic waves. The phonons involved in these transitions
are of the type called \emph{``TO''\ }(transverse optical). As
a first approximation, the displacement of the ions can be modelled
by Lorentz oscillators, which yields an expression remarkably similar
to (\ref{J02chiL}) with the resonant frequency being the natural
vibrational frequency of the TO phonon mode \cite{fox_optical_2010}.

The model qualitatively described above only applies if the molecules
of the glass have a non-zero dipole moment available to interact directly
with the field. Silica is essentially non-polar, so the direct interaction
with TO phonons is not possible. Nevertheless, other higher-order
interactions involving several phonons can eventually give rise to
the appearance of dipolar charge distributions with which the electric
field can interact (multi-phonon absorption). In any event, even if
some separated peaks are distinguishable in the IR region in Fig.
\ref{J02FIGsilica}, the actual features of the IR absorption cannot
be accounted for by a simple Lorentz model.

In view of all the considerations made above, we may even wonder why
bring up the Lorentz model at all. The good news is that, in the \emph{almost
transparent} spectral regions, formulas (\ref{J02ncomp})--(\ref{J02nland})
work very satisfactorily for many materials. As far as optical fibers
are concerned, transparency is indeed the desired situation, which
in expression (\ref{J02nland}) occurs when $\lambda$ is between
two resonances and sufficiently far from both (hence, from all others
too), so that $|\lambda_{j}^{2}-\lambda^{2}|\,>>|\gamma_{j}\lambda|$
for all $\lambda_{j}.$ In this case the imaginary part of (\ref{J02nland})
is negligible and $n_{\lambda}(\lambda)$ becomes real: 
\begin{equation}
\fbox{\ensuremath{n_{\lambda}(\lambda)=\left(1+\sum\limits _{j=1}^{N}\dfrac{A_{j}\,\lambda^{2}}{\lambda^{2}-\lambda_{j}^{2}}\right)^{1/2}\qquad}{\small Sellmeier's formula}\ensuremath{\text{\emph{{\small\quad}}}}}\label{JO2sellmi}
\end{equation}

Many dielectric materials have their refractive index modelled by
a Sellmeier expression with three terms in the sum.\footnote{Sellemeier's expansion is the most popular approximation for representing
a real refractive index, but not the only one. Pikhtin-Yas'kov formula,
for example, adds one term to represent a broadband electronic contribution
to the index \cite{bass_handbook_1995}.} The parameters $A_{j}$ and $\lambda_{j},$ which are tabulated,
are typically obtained by fitting the experimental data to the model.

\subsection{Time and space frequencies, propagation constant, and wavelength}

\label{J02sbsctTSFPCW}

We will assume a dielectric with negligible losses hereafter and write
$n$ for the real index The field (\ref{J02onda+ atenuada(t)}) read:
\begin{equation}
\mathcal{E}(z,t)=A\cos\left(\omega t-n(\omega)\frac{\omega}{c}z\right)=A\cos\left(\omega t-k(\omega)z\right),\label{J02oplan}
\end{equation}
where we have replaced the symbol $\omega_{0}$ by $\omega$ to ease
the notation (but remember that $\omega$ is \emph{not} a variable:
it is a fixed \emph{parameter}!) In the last expression of (\ref{J02oplan})
we have introduced the \emph{propagation constant} or \emph{wavenumber
vector}, defined as 
\begin{equation}
\fbox{\ensuremath{k(\omega)\equiv n(\omega)\dfrac{\omega}{c}\equiv n(\omega)k_{0},}}\label{K02kw}
\end{equation}
where $k_{0}$ is, accordingly, the propagation constant \emph{in
vacuum}, in which $n(\omega)=1$ at all frequencies. Using (\ref{J02vfase}),
we see that 
\begin{equation}
\fbox{\ensuremath{v_{\text{ph}}(\omega)=\dfrac{c}{n(\omega)}=\dfrac{\omega}{k(\omega)}.}}\label{J02ddd11}
\end{equation}

\begin{figure}
\centering{}\includegraphics[scale=0.45]{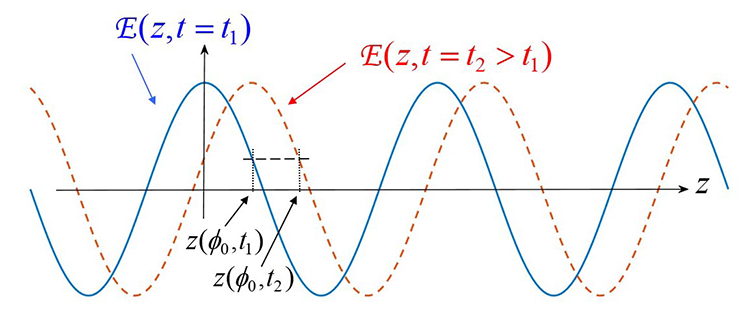}

\caption{The propagation constant or wavenumber $k$ is the spatial angular
frequency, just as $w$ is the time angular frequency. Its (spatial)
period in the medium is $\lambda$, just as $T$ is the period in
time. The figure shows a monochromatic wave across the spatial coordinate
at two different frozen times}

\label{J02FIGcongelado} 
\end{figure}

If in (\ref{J02oplan}) we fix the coordinate $z,$ we are left with
a periodic time function, the period ($T$) being given by the condition
$\omega(t+T)=\omega t+2\pi.$ This is, 
\begin{equation}
T=\frac{2\pi}{\omega}.
\end{equation}
If we fix time $t$ instead, we obtain a periodic spatial function,
the spatial period ($\lambda$) being given by the condition $kz=k(z+\lambda)+2\pi,$
as sketched in Fig. \ref{J02FIGcongelado}. The spatial period is
better known as \emph{wavelength}. We thus obtain 
\begin{equation}
\lambda=\frac{2\pi}{k}.\label{J02ddd2}
\end{equation}
So we see that the propagation constant $k$ represents, in space,
the ``spatial frequency''\ expressed in radians/meter, just as
$\omega$ represents, in time, the temporal frequency expressed in
en radians/second. It also follows that $v_{\text{ph}}=\lambda/T=\lambda\nu,$
with $\omega=2\pi\nu.$

\begin{figure}[h]
\centering{}\includegraphics[scale=0.2]{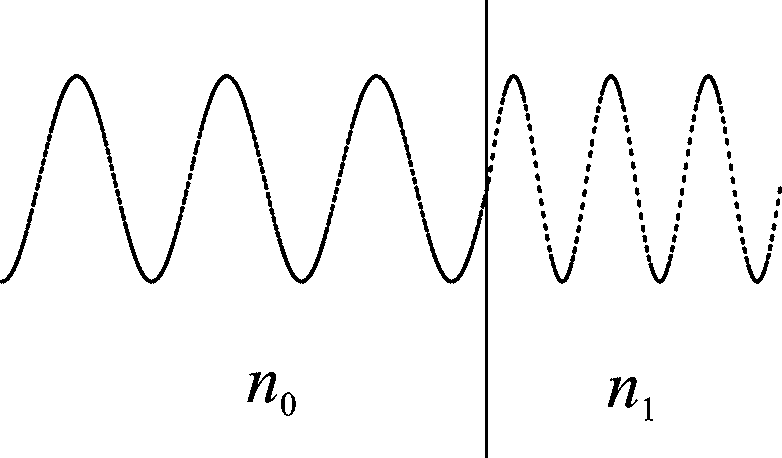}\caption{The propagation constant or wavenumber $k$ is the spatial angular
frequency, just as $\omega$ is the time angular frequency. Its (spatial)
period in the medium is $\lambda$, just as $T$ is the period in
time. The figure shows a monochromatic wave across the spatial coordinate
at two different frozen times.}
\label{J02FIGlanda} 
\end{figure}

An important point to notice is that $\lambda$ is a function of $n$:
\begin{equation}
\lambda=\frac{2\pi}{k(\omega)}=\frac{1}{n(\omega)}\frac{2\pi c}{\omega}.\label{J02landavaria}
\end{equation}
Consequently, since $c$ is a constant and the angular frequency $\omega$
is the same in all media\footnote{Except in a situation where there is relative \emph{movement} between
different media, in which case the\emph{\ Doppler effect} would indeed
yield different frequencies. Never in this book shall we encounter
such situation.}, it follows from (\ref{J02landavaria}) that \emph{the wavelength
is different in different media} (according to\emph{\ }$n$). When
a laser is said to emit ``at 1,55 $\mu$m,''\ it is understood
that the given value is that \emph{in vacuum} (or, to most practical
effects, in air). In a different medium with an index $n,$ the wavelength
would be different: $\lambda=\lambda_{0}/n,$ where $\lambda_{0}$
is the vacuum value. Fig. \ref{J02FIGlanda} illustrates this point.

\section[Propagation of signals]{Propagation of signals \sectionmark{Propagation of signals}}

\label{J02sctPS}

The results obtained in Subsection \ref{J02sbsctAMW} can only be
a starting point in our discussion. Although a purely monochromatic
wave (which, strictly speaking, does not exist) can be realized approximately
by keeping a highly-coherent light source switched-on for a very long
(``infinite'') time, it is utterly useless for the purpose of transmitting
information. Think of a digital signal, for example. In order to transmit
the sequence of ones and zeros, some kind of variation has to be added
to the wave shape so as to single out the symbols. For example, an
increased amplitude may mean the presence of a ''one''. Obviously,
the envelope-modulated wave is no longer monochromatic. We are thus
confronted with new situations that will require new concepts.

\subsection{Group velocity}

\label{J02sbsctGV}

\begin{figure}[h]
\centering{}\includegraphics[scale=0.55]{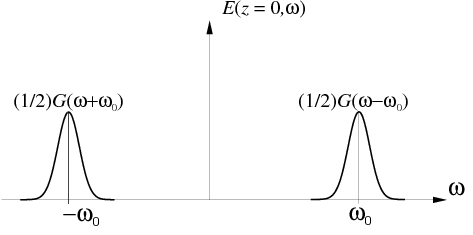}\caption{Optical spectrum. The bandwidth of the envelope-modulating pulse $g(t)$
is extremely small compared with the carrier frequency. It has been
greatly exaggerated in the figure for visualization purposes.}
\label{J02FIGespectro} 
\end{figure}

Suppose we modulate the amplitude of an optical monochromatic wave
(the \emph{``carrier''}), of angular frequency $\omega_{0},$ with
a signal $g(t),$ as shown in Fig. \ref{J02FIGespectro}. This modulated
(plane) wave enters an homogeneous lossless dielectric of semi-infinite
length, starting at the plane $z=0$. The initial electric field is
thus $\mathcal{E}(z=0,t)=g(t)\cos(\omega_{0}t),$ and the wave propagates
across the dielectric in the $+z$ direction. Inspired by (\ref{J02onda+ atenuada(t)}),
one might think that, at a distance $z=L,$ the propagated field would
be: $\mathcal{E}(z=L,t)=g(t-L/v_{\text{ph}})\cos(\omega_{0}t-k(\omega_{0})L)=g(t-L/v_{\text{ph}})\cos[\omega_{0}(t-L/v_{\text{ph}})];$
this is, a mere time-delayed version of $\mathcal{E}(z=0,t)$, the
delay being $L/v_{\text{ph}}.$ However, this is wrong. The concept
of phase velocity is inherent to \emph{monochromatic} waves, like
(\ref{J02onda+ atenuada(t)}), whereas $\mathcal{E}(z=0,t)$ is made
up of the aggregation of different spectral contributions, as actually
expressed by its FT.\footnote{$E(0,\omega),$ the spectrum of $\mathcal{E}(0,t),$ is a \emph{continuous}
function of $\omega$. Consequently, $E(0,\omega)$ evaluated at any
particular point $\omega^{\prime}$ in the frequency axis, \emph{i.e.},
$E(0,\omega=\omega^{\prime}),$ cannot represent the amplitude of
a true (measurable) sinusoidal electric field of frequency $\omega^{\prime}$
--- in the sense of $A$ in (\ref{J02oplan}). Actually, $E(0,\omega)$
is not a field amplitude (in V/m), but a spectral field \emph{density}
{[}in (V/m)/Hz{]} which can only make physical sense when combined
``continuously''\ (through the Fourier integral) with other frequencies.
A differential contribution $E(0,\omega^{\prime})d\omega$ is at least
needed to accomodate an infinitesimal amount of electromagnetic power
around $\omega^{\prime}$. Still, the concept of phase velocity, although
derived within the phasor formalism for a ``tangible''\ sinusoidal
wave, can also be used with a continuous spectral density. In this
case, regarding $v_{\text{ph}}(\omega^{\prime})$ as ``the phase
velocity that a \emph{monochromatic wave} of frequency $\omega^{\prime}$
propagating in the medium \emph{would have}''\ is perfectly correct
--- only, this should not mislead us into believing that the density
$E(0,\omega=\omega^{\prime})$ represents the amplitude of such a
monochromatic wave.}

At this point, the following argument is sometimes found in the literature:
``Because the phase velocity $v_{\text{ph}}(\omega)=c/n(\omega)$
varies with frequency, `each frequency' $\omega$ of the spectrum
of the initial field $\mathcal{E}(z=0,t)$ will propagate at a different
velocity $v_{\text{ph}}(\omega).$ Therefore, the different spectral
components will not `arrive' at the coordinate $z=L$ at the same
time, and the pulse shape will be distorted (broadened).''\ This
reasoning is very unfortunate for two reasons: (a) the concept of
\emph{arrival time }is meaningless for a spectral component, as discussed
above; (b) the field envelope does indeed get distorted \emph{when}
$v_{\text{ph}}$ is frequency-dependent, but not \emph{because }$v_{\text{ph}}$
is frequency-dependent; at least not with the naive (and erroneous)
interpretation just expressed. We will go back to this point later,
but we will first attempt to obtain the form $\mathcal{E}(z,t)$ under
certain simplifying conditions.

We have been able to find a general solution of the wave equation
\emph{in the frequency domain}, (\ref{J02weq4}), which we reproduce
here assuming forward-propagation only ($B(\omega)=0$): 
\begin{equation}
E(\omega,z)=A(\omega)e^{-i(\omega/c)n(\omega)z},\label{J02wz}
\end{equation}
where $A(\omega)=E(z=0,\omega)$ is the spectral distribution or FT
of the initial field $\mathcal{E}(z=0,t).$ Now, 
\begin{equation}
\mathcal{E}(t,z=0)=g(t)\cos(\omega_{0}t),\label{J02init}
\end{equation}
with $g(t)$ the time-varying carrier envelope, which in our example
of a digital signal has the shape of a pulse, but it could certainly
be any arbitrary function of time representing an analog signal too.
Taking the FT of (\ref{J02init}), we obtain 
\begin{equation}
E(\omega,z=0)=A(\omega)=\tfrac{1}{2}[G(\omega-\omega_{0})+G(\omega+\omega_{0})].\label{J02tf}
\end{equation}
Using (\ref{J02init}) in (\ref{J02wz}), we arrive at the expression
of the FT of the propagated field at a distance $z$: 
\begin{equation}
E(\omega,z)=\tfrac{1}{2}[G(\omega-\omega_{0})+G(\omega+\omega_{0})]e^{-i(\omega/c)n(\omega)z}.\label{J02tfz}
\end{equation}

Before calculating $\mathcal{E}(t,z),$ an important point must be
noted. According to (\ref{J02tfz}) and (\ref{J02tf}), the following
relationship holds: 
\begin{equation}
\dfrac{E(\omega,z)}{E(\omega,0)}=e^{-i\tfrac{\omega}{c}n(\omega)z}\equiv H_{d}(\omega).\label{J02ftt}
\end{equation}
So, the dielectric medium of length $z$ can be considered as a \emph{linear
system} characterized by a transfer function $H_{d}(\omega)$ relating
the output and input fields. Since we have assumed that $n(\omega)$
is real, $|H_{d}(\omega)|=1;$ hence the dielectric is an \emph{all-pass
filter}, with flat amplitude response. To be precise, even admitting
a small attenuation, the losses in optical fibres are essentially
frequency-independent \emph{within the spectral width of the transmitted
signals}. Thus $H_{d}(\omega)=K\exp\left(-i\,(\omega/c)\,n_{r}(\omega)z\right)$
with $K<1$ but constant, so the transfer function is still flat in
modulus. Then, any signal distortion will be entirely due the to phase
of the dielectric response of the dielectric, as we will soon see.

There only remains to obtain $\mathcal{E}(t,z)$ by simply taking
the FT$^{-1}$ of (\ref{J02tfz}): 
\begin{align}
\mathcal{E}(t,z) & =\frac{1}{2\pi}\int\nolimits _{-\infty}^{\infty}E(\omega,z)\,e^{i\omega t}d\omega\nonumber \\
 & =\frac{1}{2\pi}\int\nolimits _{-\infty}^{\infty}\tfrac{1}{2}[G(\omega-\omega_{0})+G(\omega+\omega_{0})]\,e^{i\,[\omega\,t-k(\omega)\,z]}d\omega.\label{J02sol}
\end{align}
In (\ref{J02sol}) we have introduced the propagation constant (\ref{K02kw}).
The physical interpretation of expression (\ref{J02sol}) is almost
straightforward, but contains a subtle point. The propagated field
is described as the sum of infinite \emph{quasi}-monochromatic differential
contributions $d\mathcal{E}(t,z)$ of spectral width $d\omega$, each
located around a frequency $\omega$. A phase velocity $v_{\text{ph}}(\omega)$
seems naturally associated to the contribution $d\mathcal{E}(z,t)$
centered at $\omega$, so one might think that each of such partial
fields propagates at the velocity $v_{\text{ph}}(\omega).$\footnote{Despite the fact that $E(\omega,z)\,e^{i\omega t}$ is a periodic
function of $t$ ($\omega$ and $z$ are parameters here), a \emph{continuous}
frequential accumulation of such densities, $\int E(\omega,z)\,e^{i\omega t}d\omega,$
results mathematically in a non-periodic, \emph{time-limited} signal,
$\mathcal{E}(t,z).$ However, if the accumulation is differential,
$E(\omega,z)\,e^{i\omega t}d\omega,$ the resulting $d\mathcal{E}(t,z)$
will be rather similar to a tangible monochromatic wave of frequency
$\omega$, but starting and ending in time; hence the tempting conjecture
that it should move at a velocity $v_{\text{ph}}(\omega).$} This is mathematically wrong, as we see below.

In principle, in order to integrate (\ref{J02sol}) we need to know
$G(\omega)$ --- hence the pulse shape $g(t)$ --- and, of course,
the specific form of $n(\omega).$ Besides, the integration would
have to be performed (most likely, numerically) for each and every
$g(t)$ and $n(\omega)$ we were to consider. Fortunately, we can
do better than that to some extent. Note that the optical carrier
frequency $\nu_{0}=(2\pi)^{-1}\omega_{0}$, corresponding to a vacuum
wavelength $\lambda_{0}\approx1$ $\mu$m, will always be in the range
$\nu_{0}\gtrsim10^{14}$ Hz in our applications. On the contrary,
the bandwidth $\Delta f$ of the transmitted signals --- in this
example, $g(t)$ --- can typically be up to some $10$ -- $20$
GHz (a bit rate of 10 Gb/s is standard). Then $\Delta f<<\nu_{0}$
always, and the spectrum of the optical field has indeed the form
sketched in Fig. \ref{J02FIGespectro}, which has the features a \emph{very
narrow} band-pass signal. In other words, $G(\omega-\omega_{0})$
is virtually zero as soon as $\omega$ moves a little bit away (relatively)
from $\omega_{0}.$ The same applies to $G(\omega+\omega_{0})$ with
respect to $-\omega_{0}.$ Consequently, we will assume as a start
that we can approximate $n(\omega)$ around $\omega_{0}$ through
a Taylor series with few terms, perhaps two or three, in the summand
of $G(\omega-\omega_{0})$ in (\ref{J02tfz}). Such approximation
breaks down when $\omega$ is away from $\omega_{0}$, but we do not
care since the distant frequencies do not contribute to the integral
anyway, as $G(\omega-\omega_{0})\simeq0$ there. An analogous approximation
applies to $-\omega_{0}.$

Writing $\omega=\omega_{0}+\Omega,$ we have, for $\omega$ near $\omega_{0},$
\begin{equation}
n(\omega)=n(\omega_{0}+\Omega)=n_{0}+n_{1}\Omega+\tfrac{1}{2}n_{2}\Omega^{2}+\tfrac{1}{6}n_{3}\Omega^{3}+\cdots,\label{J02ntay}
\end{equation}
with $n_{0}=n(\omega_{0})$ and $n_{j}=d^{(j)}n(\omega)/d\omega^{j}|_{\omega_{0}}.$
We can also expand $k(\omega)$ directly: 
\begin{equation}
k(\omega)=k(\omega_{0}+\Omega)=k_{0}+k_{1}\Omega+\tfrac{1}{2}k_{2}\Omega^{2}+\tfrac{1}{6}k_{3}\Omega^{3}+\cdots,\label{J02ktay}
\end{equation}
with\footnote{The symbol $k_{0}$ was used previously to denote the propagation
constant in vacuum ($n=1$), \emph{i.e.,} $k_{0}=\omega/c.$ Here
we reuse $k_{0}$ as $k_{0}=k(\omega_{0})=(\omega_{0}/c)n(\omega_{0}),$
the propagation constant of the media at $\omega=\omega_{0}.$ A similar
ambiguity occurs with $n_{0}.$ The context should avoid any confusion.} $k_{0}=k(\omega_{0})$ and $k_{j}=d^{j}k(\omega)/d\omega^{j}|_{\omega_{0}}.$
The relationship between the coefficients $k_{j}$ and $n_{j}$ is
straightforward: 
\begin{equation}
k(\omega_{0}+\Omega)=n(\omega_{0}+\Omega)\frac{\omega_{0}+\Omega}{c}=\underset{k_{0}}{\underbrace{\frac{\omega_{0}}{c}n_{0}}}+\underset{k_{1}}{\underbrace{\frac{1}{c}(n_{0}+n_{1}\omega_{0})}}\Omega+\underset{(1/2)k_{2}}{\underbrace{\frac{1}{c}\left(n_{1}+\tfrac{1}{2}\omega_{0}n_{2}\right)}}\Omega^{2}+\cdots\label{J02tay}
\end{equation}

For the moment we will assume that $n(\omega)$ can be approximated
around $\omega_{0}$ with only two terms, \emph{i.e.}, $n_{2}=n_{3}=\cdots\simeq0.$
Also, since $\Omega<<\omega_{0},$ we will neglect $\tfrac{1}{c}n_{1}\Omega^{2}$
as compared with $\frac{1}{c}n_{1}\omega_{0}\Omega.$ Then only $k_{0}$
and $k_{1}$ need be retained in (\ref{J02ktay}). Moreover, the unquestionable
band-pass character of $\mathcal{E}(z,t)$ allows us to make use of
all associated approximations, as explained in Subsection \ref{JApAanalytical}.
We can thus operate only with the positive frequency spectrum, $G(\omega-\omega_{0}).$
Denoting $\mathcal{\tilde{E}}(z,t)$ the analytical field, and setting
$k(\omega)=k(\omega_{0}+\Omega)\simeq$ $k_{0}+k_{1}\Omega,$ (\ref{J02sol})
yields 
\begin{align}
\mathcal{\tilde{E}}(z,t) & \simeq\frac{1}{2\pi}\int\nolimits _{-\infty}^{\infty}G(\Omega)\,e^{i[(\omega_{0}+\Omega)t-(k_{0}+k_{1}\Omega)z]\,}d\Omega\nonumber \\
 & =e^{i(\,\omega_{0}t-k_{0}z)\,}\frac{1}{2\pi}\int\nolimits _{-\infty}^{\infty}G(\Omega)\,e^{i[\Omega\,(t-k_{1}z)]\,}d\Omega=e^{i(\,\omega_{0}t-k_{0}z)\,}g(t-k_{1}z),\label{J02tan}
\end{align}
where the property (\ref{JApAprop3}) has been used in the last equality.
Thus, we finally obtain 
\begin{equation}
\mathcal{E}(z,t)=\operatorname{Re}[\mathcal{\tilde{E}}(z,t)]=g(t-z/k_{1}^{-1})\cos(\omega_{0}t-k_{0}z).\label{J02Ezt}
\end{equation}

\begin{figure}[h]
\centering{}\includegraphics[scale=0.18]{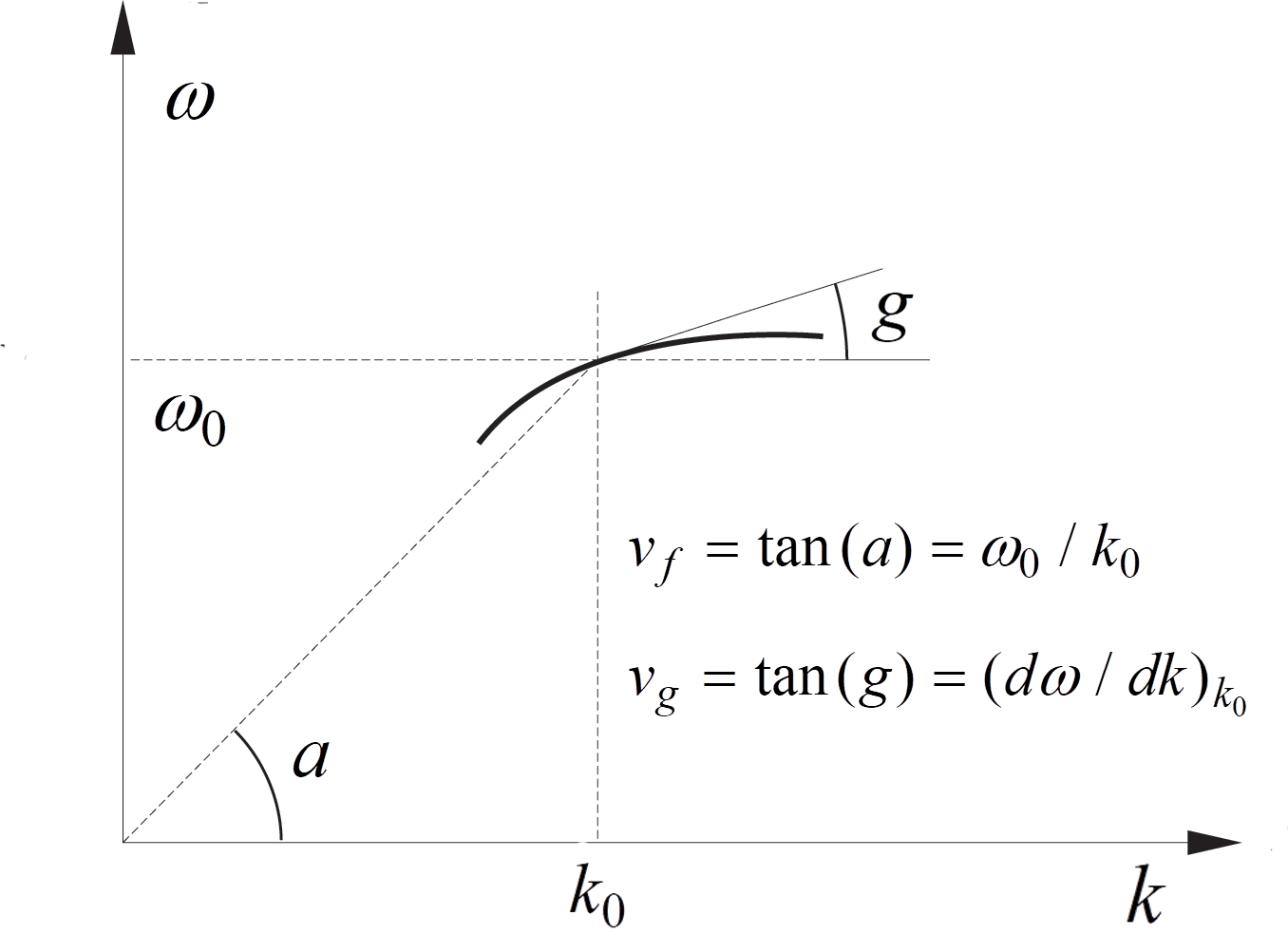}\caption{Geometrical interpretation of the phase and group velocities.}
\label{J02FIGgrupofase} 
\end{figure}

Expression (\ref{J02Ezt}) shows that the monochromatic carrier propagates
at its corresponding phase velocity $v_{\text{ph}}(\omega_{0})=c/n_{0}=\omega_{0}/k_{0},$
but the \emph{envelope} --- which contains the transmitted information
--- propagates at a \emph{different} velocity, given by $k_{1}^{-1}=(dk/d\omega)_{\omega=\omega_{0}}^{-1}=(d\omega/dk)_{\omega=\omega_{0}}.$
We will call this the \emph{group velocity} (``at $\omega_{0}$''):
\begin{equation}
\fbox{\ensuremath{v_{g}(\omega_{0})=\left.\dfrac{d\omega}{dk}\right\vert _{\omega=\omega_{0}}.}}\label{J02vg}
\end{equation}
Compare (\ref{J02vg}) with the phase velocity of the carrier, (\ref{J02ddd11}),
which we re-write here: 
\begin{equation}
\fbox{\ensuremath{v_{p}(\omega_{0})=\dfrac{\omega_{0}}{k_{0}}.}}\label{J02vp}
\end{equation}
Fig. \ref{J02FIGgrupofase} illustrates the difference between both
velocities geometrically.

The result (\ref{J02Ezt}) does not display any distortion of the
wave shape, but just a pure propagation delay of value $z/v_{g}.$
This is mathematically understandable because, in truncating the expansion
of $k(\omega)$ to just two terms, the only effect of $k(\omega)$
is the time shift induced by the linear summand $i\Omega k_{1}z$
in the exponent of the Fourier integral. We have then arrived at an
important conclusion:\ \emph{Distortionless propagation requires
that the propagation constant have a linear dependence with frequency,
}$k(\omega)=k(\omega_{0})+k_{1}(\omega-\omega_{0})$ (at least, within
the spectral range of the signal). Now, when is $k$ \emph{exactly}
(not approximately) linear with $\omega$? Equation (\ref{J02tay})
gives the answer: only when $n(\omega)$ is strictly constant; if
$n(\omega)$ does not vary with $\omega$ around $\omega_{0}$ ($n_{1}=n_{2}=n_{3}=\cdots=0$),
all $k_{j}$ with $j>1$ are exactly zero and $k$ is linear. Moreover,
in this case the phase and group velocity coincide, since $v_{g}^{-1}=k_{1}=n_{0}/c$.
Note that the linear dependence that we assumed to obtain (\ref{J02Ezt})
was an approximation in the first place, as we neglected the contribution
of $n_{1}\neq0$ to the coefficient of $\Omega^{2}.$

In a situation with a truly constant $n,$ the phase velocity is the
same at all frequencies (as in vacuum), which means that no spectral
component experiences any phase modification different from all other
components; as a result, there is no distortion. Unfortunately, Sellmeier's
formula (\ref{JO2sellmi}) warns us that the idea of a constant refractive
index $n$ in some spectral range is unreal, and even a still more
convincing argument will be outlined in Subsection \ref{J02sbscKK}.
Thus, one or more higher-order terms must usually be considered, according
to the degree of accuracy sought. As we will soon see, these terms
unavoidably bring about distortion.

If, however, the distortion is moderate (or, at least, no so dramatic
that the pulse becomes unrecognizable), there is no reason to discard
the concept of group velocity to describe the velocity at which the
envelope (whether being distorted or not) moves. Even more, it will
be useful to define a \emph{``group refractive index''} such that,
analogous to the relationship $v_{\text{ph}}=c/n,$ we can write 
\begin{equation}
\fbox{\ensuremath{v_{g}(\omega)=\dfrac{c}{n_{g}(\omega)}\qquad}{\small definition of group index}}\label{Jo2dng}
\end{equation}
Using (\ref{Jo2dng}), (\ref{J02vg}) and (\ref{K02kw}), the relationship
between both indices is obtained: 
\begin{equation}
\fbox{\ensuremath{n_{g}(\omega)=n(\omega)+\omega\dfrac{dn(\omega)}{d\omega}.}}\label{J02reln}
\end{equation}

Formula (\ref{J02reln}) is consistent with the fact that, if (hypothetically)
$n(\omega)$ is constant, then $n_{g}=n$ and $v_{\text{ph}}=v_{g},$
as mentioned above. Other than in this ideal case, Fig. \ref{J02FIGnrniMatlab}
shows that, in the lossless frequency ranges between resonances, the
(virtually real) index $n(\omega)$ is always an increasing function
of $\omega$. Then, $dn/d\omega>0$ and $n_{g}(\omega)>n_{\text{ph}}(\omega),$
implying that $v_{g}(\omega)<v_{\text{ph}}(\omega).$ Expression (\ref{J02reln})
can be written in terms of $\lambda$ through the definition(\ref{J02nland}):\footnote{Recall the (much-used) relations 
\[
\dfrac{d}{d\omega}=-\dfrac{\lambda^{2}}{2\pi c}\dfrac{d}{d\lambda}\quad\leftrightarrows\quad\dfrac{d}{d\lambda}=-\dfrac{\omega^{2}}{2\pi c}\dfrac{d}{d\omega}.
\]
} 
\begin{equation}
\fbox{\ensuremath{n_{g\lambda}(\lambda)=n_{\lambda}(\lambda)-\lambda\dfrac{dn_{\lambda}(\lambda)}{d\lambda},}}\label{J0enlanda}
\end{equation}
with $n_{g\lambda}(\lambda)\equiv n_{g}\left(2\pi c/\lambda\right).$

Obviously, the time it takes the pulse envelope to travel a specified
distance ($L$), when the wavelength of the optical carrier is $\omega_{0},$
denoted $\tau_{g}(\omega_{0})$ {[}$=\tau_{g}(2\pi c/\lambda_{0})\equiv\tau_{g\lambda}(\lambda_{0})${]}
and formally called \emph{group delay}, is given by 
\begin{equation}
\tau_{g}(\omega_{0})=\dfrac{L}{v_{g}(\omega_{0})}\qquad\text{{\small definition\ of\ group\ delay.}}\label{J02tau}
\end{equation}

The meaning of the group velocity under ``unusual''\ circumstances
(for example, near the resonances) is briefly discussed in Subsection
\ref{J02sctFSL}.

\subsection{Dispersion and envelope distortion}

\label{J02sbsctDED}

A true frequency-independent refractive index is physically unreal.
A non-dis\-pers\-ive index can only be taken, at most, as a first
approximation which accuracy will suffice or not depending on the
specific problem at hand. We will first consider that, at least, one
additional summand is required in the Taylor expansion of $k(\omega);$
this is, we will the keep the $\tfrac{1}{2}k_{2}\Omega^{2}$ term
in (\ref{J02ktay}). In general, the presence of this new quadratic
term in the exponent of (\ref{J02tan}) yields the integral analytically
unsolvable. Therefore, numerical integration would be required for
each specific form of $g(t).$ Fortunately, we can still obtain some
\emph{estimative}, general results which will provide us with insight
into the effects of the dispersion. Even if we cannot calculate the
shape of the propagated pulse without actually solving the Fourier
integral, we anticipate that the pulse envelope will most likely \emph{broaden,}
and we will attempt to estimate such broadening.

\begin{figure}[h]
\centering{}\includegraphics[width=5in,height=4in,keepaspectratio]{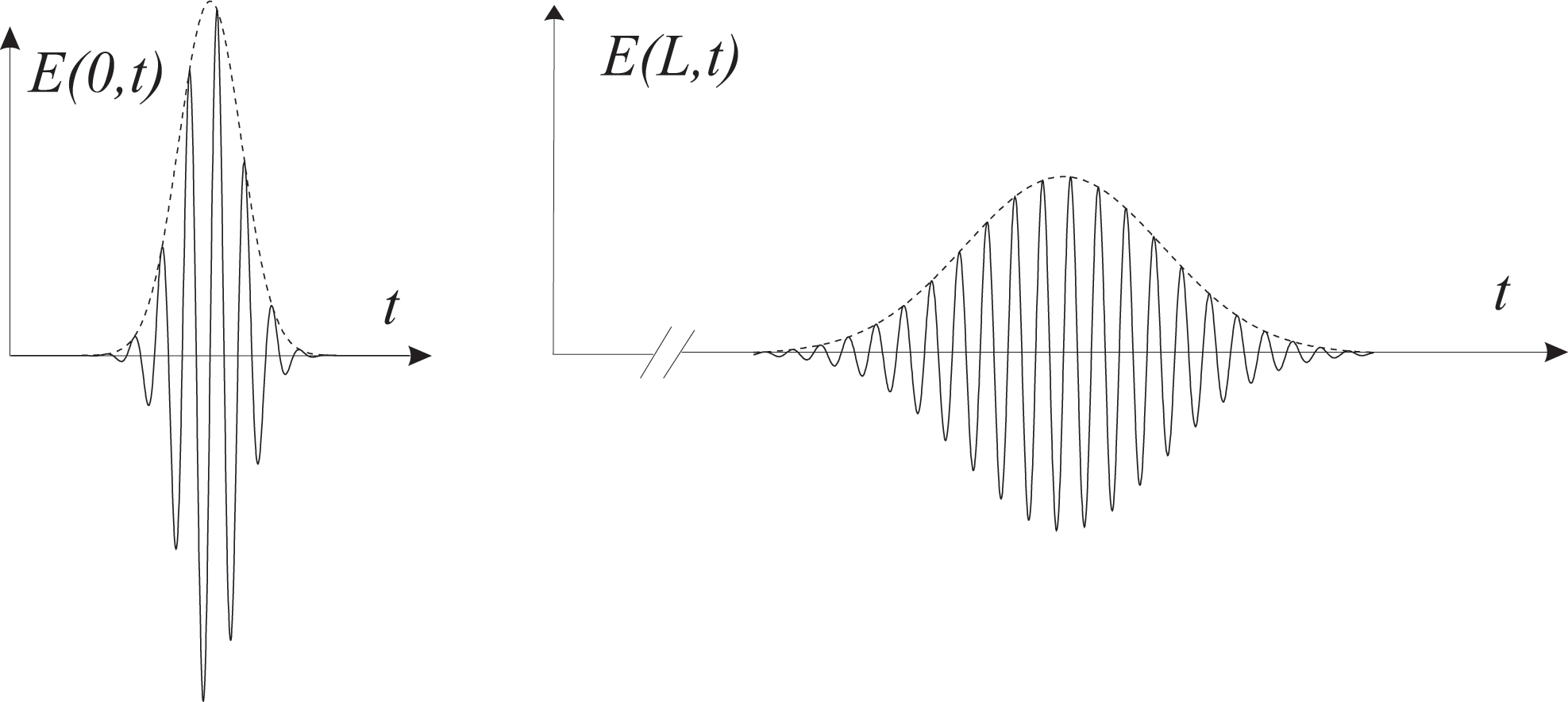}\caption{Sketch of the expected pulse broadening.}
\label{J02FIGensancha} 
\end{figure}

\begin{figure}[h]
\centering{}\includegraphics[width=5in,height=3.5in,keepaspectratio]{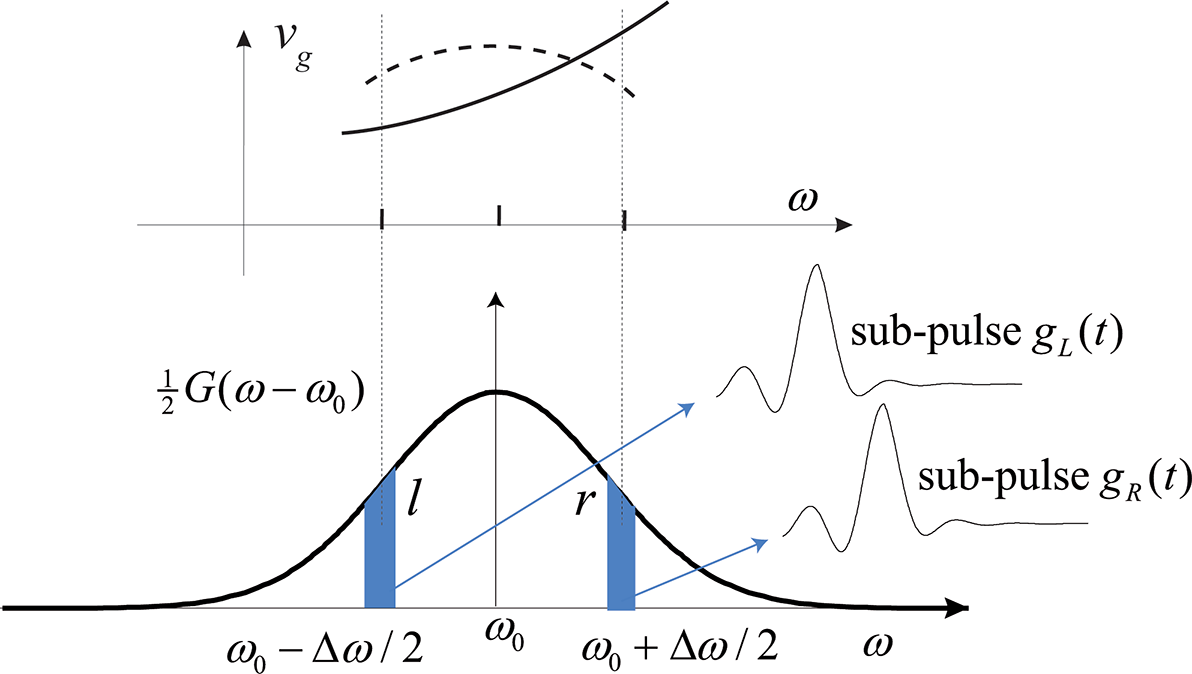}\caption{Rough estimation of the pulse broadening by considering the spectral
dependence of $v_{g}(\omega).$ See text.}
\label{J02FIGsubespectros} 
\end{figure}

The left part of Fig. (\ref{J02FIGensancha}) shows the form of the
spectrum of the optical signal at $z=0.$ As mentioned, we only need
to consider the positive frequencies. Assume that the frequency dependence
of the group velocity around $\omega_{0}$ is as sketched in Fig.
\ref{J02FIGsubespectros} in solid line. The result that the envelope
propagates, as a whole, at a velocity $v_{g}(\omega_{0})$ was obtained
within the approximation $k(\omega_{0}+\Omega)\simeq k_{0}+k_{1}\Omega.$
To obtain some information about the distortion of the pulse shape,
we need to be more subtle. Let us pick two narrow slices from the
spectrum, as shown in Fig. \ref{J02FIGsubespectros}. We will call
them ``sub-spectra''\ $R$ and $L$, respectively; they also include
the corresponding negative-frequency parts, not shown in the figure.
Their separation is chosen so that it roughly corresponds to the relevant
spectral width of $G(\Omega).$To the narrow sub-spectra $R$, if
considered isolated from the rest of the spectrum, there corresponds
a ``sub-pulse''\ envelope $g_{R}(t)$ in the time domain. Owing
to a basic property of the Fourier-transformed pairs, $g_{R}(t)$
is expected to be less abrupt and longer in duration than the whole
pulse envelope $g(t)$. Similarly, another sub-pulse $g_{L}(t)$ will
correspond to the sub-spectra $L$. But, in view that the energy corresponding
to $g_{R}(t)$ is closely located around $\omega_{0}+\Delta\omega/2,$
we see that the group velocity at which $g_{R}(t)$ propagates will
be given more accurately by $v_{R}=v_{g}(\omega_{0}+\Delta\omega/2),$
rather than $v_{g}(\omega_{0}).$ Likewise, $g_{L}(t)$ will propagate
at the velocity $v_{L}=v_{g}(\omega_{0}-\Delta\omega/2).$ Formally
stated, we expand $\beta(\omega)$ to first order again, but this
time around $\omega=\omega_{0}-\Delta\omega/2$ for $g_{L}(t),$ and
around $\omega=\omega_{0}-\Delta\omega/2$ for $g_{R}(t)$, and we
can write both ``partial''\ optical fields as\footnote{Actually, in our example both sub-pulses happen to have the same shape,
\emph{i.e.}, $g_{L}=g_{R},$ which can be proved easily by noting
the symmetric positions of the spectral ``slices''\ around $\omega_{0}$
in this example (Fig. \ref{J02FIGsubespectros}), and the fact that
the baseband spectrum $G(\Omega)$ is even in modulus and odd in phase
(as $g(t)$ is a real signal).\ But, even if this were not case,
it would be irrelevant since only a rough estimate of the broadening
of the whole pulse $g(t)$ is sought.} 
\begin{align}
\mathcal{E}_{\text{sub 1}}(t,z) & =g_{L}(t-z/v_{L})\cos(\omega_{0}t-\beta_{L}z)\\
\text{ }\mathcal{E}_{\text{sub 2}}(t,z) & =g_{R}(t-z/v_{R})\cos(\omega_{0}t-\beta_{R}z).
\end{align}

According to the graphic of $v_{g}(\omega)$ in Fig. \ref{J02FIGsubespectros},
$g_{R}(t)$ propagates faster than $g_{L}(t);$ therefore, the resulting
total pulse envelope will necessarily broaden, as sketched in Fig.
\ref{J02FIGensancha}. We will assume that the broadening of the whole
original pulse $g(t)$ after propagation can be reasonably estimated
as the accumulated temporal delay between the two isolated sub-pulses.
Consequently, we focus on these subpulses and their associated spectral
``slices''. Then, in travelling from $z=0$ to $z=L,$ the envelope
$g(t)$ will broaden by an amount $\Delta\tau,$ given by 
\begin{equation}
\Delta\tau\simeq\left\vert \dfrac{L}{v_{R}}-\dfrac{L}{v_{L}}\right\vert =L\left\vert v_{g}^{-1}(\omega_{0}+\Delta\omega/2)-v_{g}^{-1}(\omega_{0}-\Delta\omega/2)\right\vert .\label{J02deltaa}
\end{equation}
But 
\begin{align}
v_{g}^{-1}(\omega_{0}+\Delta\omega/2) & \simeq v_{g}^{-1}(\omega_{0})+\left(\dfrac{\Delta\omega}{2}\right)\left.\dfrac{dv_{g}^{-1}}{d\omega}\right\vert _{\omega_{0}}+\frac{1}{2}\left(\dfrac{\Delta\omega}{2}\right)^{2}\left.\dfrac{d^{2}v_{g}^{-1}}{d\omega^{2}}\right\vert _{\omega_{0}}\nonumber \\
 & =v_{g}^{-1}(\omega_{0})+\left(\dfrac{\Delta\omega}{2}\right)\frac{1}{2}k_{2}+\left(\dfrac{\Delta\omega}{2}\right)^{2}k_{3}\label{J02expa}\\
v_{g}^{-1}(\omega_{0}-\Delta\omega/2) & \simeq v_{g}^{-1}(\omega_{0})-\left(\dfrac{\Delta\omega}{2}\right)\frac{1}{2}k_{2}+\left(\dfrac{\Delta\omega}{2}\right)^{2}k_{3},\label{J02expa2}
\end{align}
where (\ref{J02vg}) has been used, and $k_{2}$ and $k_{3}$ are
defined in (\ref{J02ktay}), and now named as follows: 
\begin{align}
k_{2} & \equiv\,\left.\frac{d^{2}k}{d\omega^{2}}\right\vert _{\omega_{0}}\qquad\text{{\small Dispersion Parameter or GVD, \textquotedblleft Group Velocity Dispersion\textquotedblright} }\label{J02k2}\\
k_{3} & \equiv\,\left.\frac{d^{3}k}{d\omega^{3}}\right\vert _{\omega_{0}}\qquad\text{{\small Dispersion Parameter of 2nd order.}}\label{def beta3}
\end{align}
Then, replacing (\ref{J02expa}) and (\ref{J02expa2}) in (\ref{J02deltaa}),
we obtain 
\begin{equation}
\fbox{\ensuremath{\Delta\tau=k_{2}L\Delta\omega\qquad\text{\textsf{{\small\{1st-order dispersion only\}}}}}}\label{J02delta}
\end{equation}

Quite reasonably, the pulse broadening turns out to be proportional
to the propagation distance and increases with the spectral width
of the propagated signal. The coefficient $k_{2}$ obviously depends
on the dielectric material through $n(\omega).$ Note that if $k_{2}=0$
no pulse broadening is predicted, but only because the higher-order
coefficients have not been taken into account.

Suppose now that $v_{g}(\omega)$ is not a monotonously increasing
(or decreasing) function around $\omega_{0},$ but has an extreme
(maximum or minimum) precisely at $\omega_{0},$ as shown in Fig.
\ref{J02FIGsubespectros} in dashed line. Then $k_{2}=0,$ but the
dispersion does not disappear as $v_{g}$ still varies with $\omega.$
As noted above, this should be accounted for by $k_{3}.$ However,
for symmetry reasons, the terms with $k_{3}$ happen to cancel out
when subtracting (\ref{J02expa}) and (\ref{J02expa2}), so we need
to change the locations of the sub-spectra to allow for $k_{3}$ to
manifest itself. We will compute the delay between a sub-pulse corresponding
to $\omega_{0}$ {[}which is $L/v_{g}(\omega_{0}$ {]} and that of
the sub-pulse at $\omega_{R}=\omega_{0}+\Delta\omega/2$ (choosing
$\omega_{L}$ instead is indistinct).

From (\ref{J02expa}) with $k_{2}=0,$ we obtain 
\begin{equation}
|v_{g}^{-1}(\omega_{0}+\Delta\omega/2)-v_{g}^{-1}(\omega_{0})|=\frac{1}{8}\left(\Delta\omega\right)^{2}k_{3}\,,
\end{equation}
Thus, 
\begin{equation}
\fbox{\ensuremath{\Delta\tau=\frac{1}{8}k_{3}L(\Delta\omega)^{2}\qquad\text{\textsf{{\small\{2nd-order dispersion only\}}}}}}\label{J02delta3}
\end{equation}

Results (\ref{J02delta}) and (\ref{J02delta3}) display the linear
dependence of $\Delta\tau$ with the propagation distance $L.$ Naturally,
the frequency-dependent nature of the refractive index, often referred
to as \emph{chromatic dispersion,} is the ultimate cause of these
distortive effects.

Expression (\ref{J02delta3}) also shows that, when 2nd order dispersion
dominates, the pulse broadening is proportional to the signal bandwidth
\emph{squared.} The wavelength $\lambda_{\text{ZD}}$ at which $k_{2}(\omega)$
is zero ($k_{2}(2\pi c/\lambda_{\text{ZD}})=0$) is usually called
\emph{``zero-dispersion wavelength,''} but it must be understood
that it refers to the \emph{1st-order} dispersion only; in general,
the dispersion will be minimum, but not zero, at $\lambda=\lambda_{\text{ZD}}.$

\subsection{Dispersion and residual frequency modulation or ``chirp''}

\label{J02sbsctRFMC}

In addition to the distortion of the time envelope of the pulse, which
we have roughly estimated above, the 1st-order dispersion ($k_{2}$)
necessarily gives rise to another effect: the frequency modulation
(FM) of the optical carrier. It is easy to understand why. Working
with the analytical part of (\ref{J02sol}) and expanding $k(\omega_{0}+\Omega)$
to 2nd order in $\Omega$ around $\omega_{0},$ we have, in the customary
narrow-band situation, 
\begin{align}
\tilde{\mathcal{E}}(z,t) & =\frac{1}{2\pi}\int\nolimits _{-\infty}^{\infty}G(\Omega)\,e^{i\,[(\omega_{0}+\Omega)\,t-k(\omega_{0}+\Omega\,)\,z]}d\Omega\nonumber \\
 & \simeq e^{i\,(\omega_{0}\,t-k_{0}\,z)}\frac{1}{2\pi}\int\nolimits _{-\infty}^{\infty}G(\Omega)\,e^{i\,\Omega(t-z/v_{g})}e^{-i\frac{1}{2}\,k_{2}\Omega^{2}z}d\Omega.\nonumber \\
 & \equiv e^{i\,(\omega_{0}\,t-\beta_{0}\,z)}\tilde{g}(t,z)\equiv e^{i\,(\omega_{0}\,t-\beta_{0}\,z)}|\tilde{g}(t,z)|e^{i\gamma(t,z)},\label{J02vay}
\end{align}
with $\tilde{g}(t,z)=\,$FT$^{-1}[G(\Omega)\exp(-i\tfrac{1}{2}\,k_{2}\Omega^{2}z)].$
Note that $\tilde{g}(t,z)$ is a ``baseband''\ signal as its spectrum
contains no optical frequencies. The real field is 
\begin{equation}
\mathcal{E}(z,t)=\operatorname{Re}[\tilde{\mathcal{E}}(z,t)]=|\tilde{g}(t,z)|\cos[\omega_{0}\,t-\beta_{0}\,z+\gamma(t,z)].
\end{equation}
with 
\begin{equation}
\tilde{g}(t,z)=|\tilde{g}(t,z)|e^{i\gamma(t,z)}.\label{mofas}
\end{equation}

\begin{figure}[h]
\centering{}\includegraphics[width=2.668in,height=1.7789in]{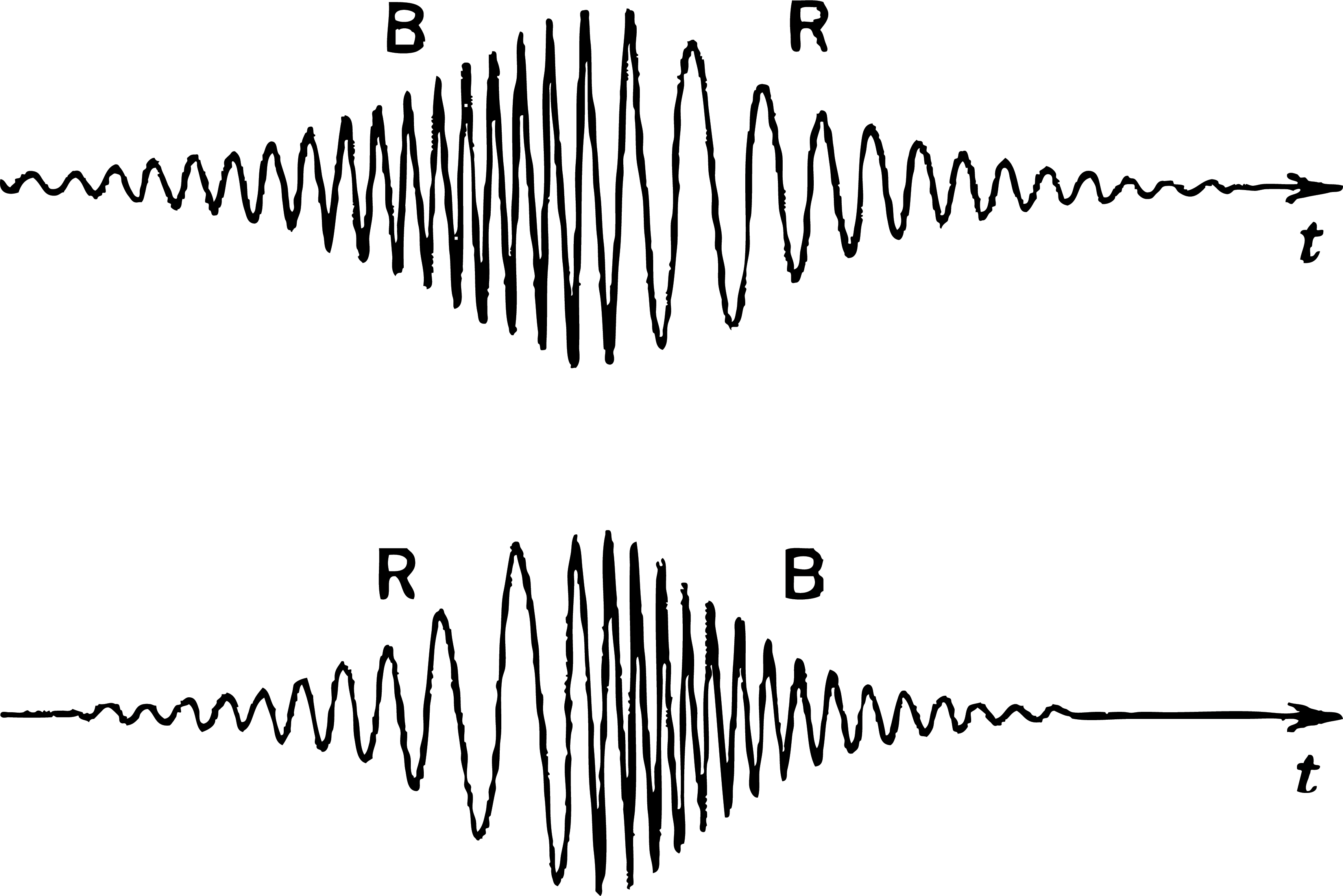}\caption{Example of a chirped pulse. Linear chirp has been assumed: $\nu(t)=\nu_{0}+at$,
with $a$ being a real constant. \emph{Above:} $a<0$ In this case,
$\nu(t)$ is higher at the beginning of the pulse, which thus contains
the (instantaneous) ``blue''\ frequencies, while the tail contains
the ``red'', lower frequencies. Below: $a>0,$ which results in
the opposite situation. For the sake of visualization, the frequency
modulation has been greatly exaggerated.}
\label{J02FIGbr} 
\end{figure}

Now the \emph{instantaneous frequency} is defined as the time derivative
of the phase (divided by $2\pi$): 
\begin{equation}
\nu_{\text{ins}}(t)=\frac{1}{2\pi}\frac{d}{dt}[\omega_{0}\,t-\beta_{0}\,z+\gamma(t,z)]=\nu_{0}+\frac{1}{2\pi}\frac{d\gamma(t,z)}{dt},
\end{equation}
with $\nu_{0}=\omega_{0}/(2\pi).$ Thus in general, \emph{only if
}$\gamma(t,z)=0$ there will be no frequency modulation\footnote{Or if $\gamma(t,z)$ is constant with time, but this situation is
hardly realistic.}. In view of (\ref{mofas}), this condition is equivalent to $\tilde{g}(t,z)$
being a real valued-function. However, $\tilde{g}(t,z)$ is generally
not real, a fact that can be readily verified by checking that $\tilde{g}(t,z)\neq\tilde{g}^{\ast}(t,z)$:
\begin{align}
\tilde{g}^{\ast}(t,z) & =\left[\frac{1}{2\pi}\int\nolimits _{-\infty}^{\infty}G(\Omega)\,e^{i\,\Omega(t-z/v_{g})}e^{-i\frac{1}{2}\,k_{2}\Omega^{2}z}d\Omega\right]^{\ast}\nonumber \\
 & =\frac{1}{2\pi}\int\nolimits _{-\infty}^{\infty}G(\Omega)\,e^{i\,\Omega(t-z/v_{g})}e^{i\frac{1}{2}\,k_{2}\Omega^{2}z}d\Omega\neq\tilde{g}(t,z),\label{J02igual}
\end{align}
unless $k_{2}=0.$ Consequently, except in the non-dispersive case,
there will be an instantaneous variation of the frequency around its
``nominal''\ value $\omega_{0},$ which will depend on $k_{2}$
and the pulse shape at each $z$. This residual frequency modulation
is called \emph{chirp}. Fig. \ref{J02FIGbr} illustrates this concept.

Note that if there were \emph{only} 2nd-order dispersion, a factor
$\exp(-i\frac{1}{6}k_{3}\Omega^{3}z),$ odd in $\Omega,$ would replace
$\exp(-i\frac{1}{2}k_{2}\Omega^{2}z)$ in the first integral of (\ref{J02igual}),
with the final result that $\tilde{g}(t,z)=\tilde{g}^{\ast}(t,z).$
Therefore, 2nd-order dispersion alone would cause no chirp.

\subsection{Dispersion in terms of the wavelength}

\label{J02sbsctDTW}

Relations (\ref{J02delta}) and (\ref{J02delta3}) are frequently
written in an equivalent form, but with all parameters expressed in
terms of the wavelength rather than the frequency. With this purpose,
we define a parameter called simply \emph{``dispersion''} and denoted
by $D$,\ which, as we will see, plays the same role as $k_{2}$.
By definition, 
\begin{equation}
D(\lambda)\equiv\frac{1}{L}\frac{d\tau_{g}(\lambda)}{d\lambda}\qquad\text{{\small Definition of dispersion.}}\label{J02defD}
\end{equation}
Consequently, $D(\lambda)$ expresses how the propagation time varies
with the wavelength because of the group velocity dispersion, which
is the ultimate origin of the time envelope distortion. Namely, we
see that $D(\lambda)$ represents the temporal broadening of the propagated
pulse envelope per unit propagation length, per unit wavelength width
of the spectrum, when the latter is located around $\lambda$ (the
optical carrier). The common units of $D$ are ps/(km$\times$nm).
Now, using (\ref{J02tau}) and (\ref{J02vg}) in (\ref{J02defD}),
it follows that 
\begin{equation}
D=-\frac{2\pi\,c}{\lambda^{2}}k_{2},\label{J02D}
\end{equation}
and, using (\ref{J02D}) and {[}$\Delta\omega=-(2\pi c/\lambda^{2})\Delta\lambda${]},
expression (\ref{J02delta}) can be written in the following alternative
form: 
\begin{equation}
\fbox{\ensuremath{\Delta\tau=DL\Delta\lambda\qquad\text{\textsf{{\small\{1st-order dispersion only\}}}}}}\label{J02qsw}
\end{equation}

In (\ref{J02qsw}) and (\ref{J02delta}), it is understood that only
the absolute value of $\Delta\tau$ matters. For example, $D<0$ ($k_{2}>0$)
simply means that $v_{g}$ increases with $\lambda$ (because $n_{g}$
decreases with $\lambda$) --- a situation labelled as \emph{normal
dispersion}. In this case, the sign of $\Delta\tau$ comes out negative
in (\ref{J02qsw}), which only means that the subpulse $g_{R}(t)$
arrives before than $g_{L}(t)$. When $D>0$ ($k_{2}<0$), $n_{g}$
increases with $\lambda$ ($v_{g}$ decreases with $\lambda$), which
is called \emph{anomalous dispersion}.

The alternative for $k_{3}$ is the \emph{dispersion slope} or \emph{differential
dispersion,} $S,$ defined as: 
\begin{equation}
S\equiv\frac{dD}{d\lambda}=\frac{4\pi\,c}{\lambda^{3}}k_{2}+\left(\frac{2\pi\,c}{\lambda^{2}}\right)^{2}k_{3}\qquad\text{{\small Definition of differential dispersion.}}\label{J02S}
\end{equation}
The relation (\ref{J02D}) has been used in the second equality of
(\ref{J02S}). Thus, unlike $D$ and $k_{2}$ in (\ref{J02D}), $S$
cannot generally be written as a function of $k_{3}$ only. However,
on most occasions $k_{3}$ is employed precisely because $k_{2}=0$
at a particular wavelength. In this case, 
\[
S=\left(\frac{2\pi\,c}{\lambda^{2}}\right)^{2}k_{3}\qquad\text{{\small at }}{\small\lambda=\lambda}_{\text{{\small ZD}}}{\small,}
\]
where $\lambda_{\text{ZD}}$ is the so-called \emph{zero-dispersion
wavelength} --- even if it only refers to the \emph{first-order}
dispersion ---; \emph{i.e.}, $k_{2}(\lambda_{\text{ZD}})=0$.

\subsection{Dispersion, losses and causality. The Kramers-Kronig relations}

\label{J02sbscKK}

The results in Section \ref{J02sbsctGV} gave us a hint on the physical
impossibility of finding materials without dispersion, that is, having
a constant, frequency-independent refractive index $n$. Now, such
results were obtained starting from \emph{ad hoc} radiation-matter
interaction models, so one might still wonder if the key conclusions
that followed are really universal. We will now address this issue.

\begin{figure}[h]
\centering{}\includegraphics[scale=0.4]{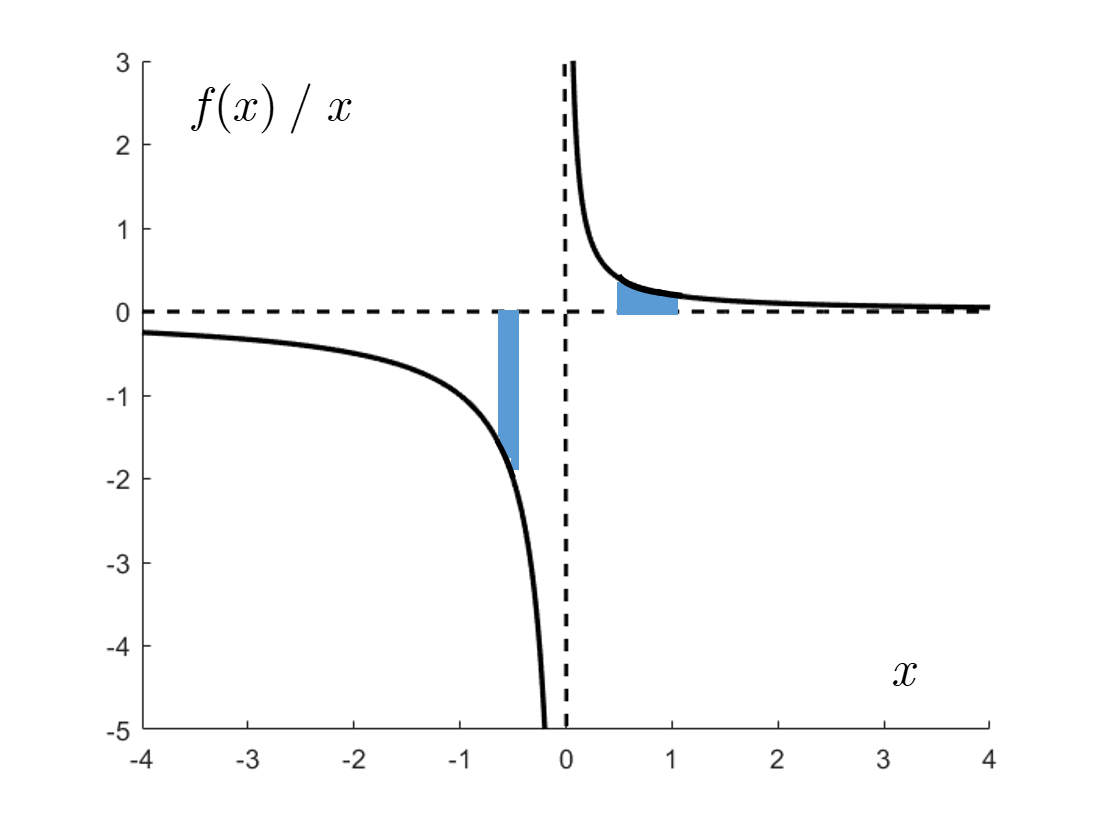}\caption{The principal value of the integral yields the true finite value of
the net area under the curve $f(x)/x.$}
\label{J02fPV} 
\end{figure}

We need first to introduce, or refresh, the concept of \emph{principal
value}. Consider the following integral: 
\begin{equation}
I=\wp\int_{x_{1}}^{x_{2}}\dfrac{f(x)}{x-a}dx\equiv\lim_{\delta\rightarrow0^{+}}\left[\int_{x_{1}}^{a-\delta}\dfrac{f(x)}{x-a}dx+\int_{a+\delta}^{x_{2}}\dfrac{f(x)}{x-a}dx\right].\label{J02pint}
\end{equation}
The symbol $\wp$ denotes the so-called Cauchy principal value (PV)
of the integration along the real axis from $x_{1}$ to $x_{2},$
where $x_{1}<a<x_{2}.$ The PV is a device to deal with the singularity
at $x=a,$ which would otherwise lead to ill-defined results. The
key point is that, in (\ref{J02pint}), the singularity is approached
symmetrically from both sides. For example, consider $a=0$ and the
function $f(x)$ defined as $\{f(x)=1$ if $x<0;$ $f(x)=1/5$ if
$x>0\}.$ The integrand $f(x)/x$ is shown in Fig. \ref{J02fPV}.

The direct integration yields an indeterminacy of the form $\infty-\infty$:
\begin{equation}
\int_{x_{1}}^{x_{2}}\dfrac{f(x)}{x}dx=\int_{x_{1}}^{0}\dfrac{dx}{x}+\int_{0}^{x_{2}}\dfrac{dx}{5x}=\dfrac{1}{5}\ln x_{2}-\ln|x_{1}|\,+\infty-\infty.
\end{equation}
However, the graphic in Fig. \ref{J02fPV} suggests that the total
area under the curve $f(x)/x$ might be finite, as the infinite negative
area on the left seems to compensate\ for the infinite positive area
on the right.\footnote{In fact, had we defined $f(x)=1$ for all $x,\ $and set $x_{1}=-x_{2},$
the positive and negative branches in Fig. \ref{J02fPV} would be
perfectly symmetrical and the total area would clearly be zero ---
which can indeed be verified in (\ref{J02check}) if the $1/5$ is
replaced by $1.$} Using the PV technique, we obtain 
\begin{equation}
\wp\int_{x_{1}}^{x_{2}}\dfrac{f(x)}{x}dx=\lim_{\delta\rightarrow0^{+}}\left[\int_{x_{1}}^{-\delta}\dfrac{dx}{x}+\int_{\delta}^{x_{2}}\dfrac{dx}{5x}\right]=\dfrac{1}{5}\ln x_{2}-\ln|x_{1}|,\label{J02check}
\end{equation}
as presumed.

We will now make use of the following relation: 
\begin{equation}
\wp\int_{-\infty}^{\infty}\dfrac{\chi(\omega^{\prime})}{\omega^{\prime}-\omega}d\omega^{\prime}=i\pi\chi(\omega).\label{J02prop}
\end{equation}
We have used the symbol $\chi$ because we are going to apply this
result to the (linear) susceptibility, but (\ref{J02prop}) is a general
mathematical property, valid for any function provided that it tends
to zero\footnote{\label{J02notaCau}Expressed more accurately, the requirement is that
the function should be analytical in the upper-half complex plane,
when $\omega$ is considered the real part of a complex frequency
$z$, $z=\omega+i\omega_{i}$ ($\omega_{i}>0$). Relation (\ref{J02prop})
can then be proved by contour integration in such semiplane.} when $|\omega|\,\rightarrow\infty.$ It is not by chance that this
condition is indeed satisfied by the particular form (\ref{J02chiL})
of $\chi(\omega)$; this regular behavior is actually based on physical
grounds. We know that 
\begin{equation}
\chi(\omega)=\int_{-\infty}^{\infty}\breve{\chi}(t)e^{-i\omega t}dt=\int_{0}^{\infty}\breve{\chi}(t)e^{-i\omega t}dt,\label{J02intp}
\end{equation}
where the second equality follows from the causality of the dielectric
response demanding $\breve{\chi}(t)=0$ for $t<0,$ as it was remarked
with regard to (\ref{J02conv}). As it happens, the restriction $t>0$
in the integration of (\ref{J02intp}) is essential in the derivation
of (\ref{J02prop}). If this were not the case, the contour integration
mentioned in footnote \ref{J02notaCau} might diverge and the result
(\ref{J02prop}) would not be generally ensured. In short, because
causality guarantees the validity of (\ref{J02prop}), \emph{any}
physically acceptable (\emph{i.e.}, causal) susceptibility\emph{\ }$\chi(\omega),$
regardless of its particular mathematical form, must satisfy the condition
(\ref{J02prop}).

Separating the real and imaginary parts of $\chi$, 
\begin{equation}
\chi(\omega)=\chi_{r}(\omega)+i\chi_{i}(\omega),\label{J02chiri}
\end{equation}
the expression (\ref{J02conv}) can be split in two equations: 
\begin{align}
\chi_{r}(\omega) & =\dfrac{\wp}{\pi}\int_{-\infty}^{\infty}\dfrac{\chi_{i}(\omega^{\prime})}{\omega^{\prime}-\omega}d\omega^{\prime}\label{J02kk1}\\
\chi_{i}(\omega) & =-\dfrac{\wp}{\pi}\int_{-\infty}^{\infty}\dfrac{\chi_{r}(\omega^{\prime})}{\omega^{\prime}-\omega}d\omega^{\prime}.\label{J02kk2}
\end{align}

Expressions (\ref{J02kk1})--(\ref{J02kk2}) reveal a fundamental,
unbreakable link between the dispersive behavior of a causal dielectric,
described by $\chi_{r}(\omega)$, and its losses, described by $\chi_{i}(\omega).$
We note in passing that $\chi_{r}(\omega)$ and $\chi_{i}(\omega)$
are the \emph{Hilbert transform} of each other \cite{coulon_signal_1986}.

The \emph{Kramers-Kronig }(KK)\emph{\ relations} \cite{landau_electrodynamics_1995}
are just the equations (\ref{J02kk1})--(\ref{J02kk2}) rewritten
in an alternative form which involves the positive frequencies only.
This is possible thanks to the real character of $\breve{\chi}(t),$
which implies that $\chi(\omega)$ satisfies the property (\ref{JApAwreal})
or, equivalently, 
\begin{equation}
\chi_{r}(-\omega)=\chi_{r}(\omega)\text{\qquad and\qquad}\chi_{i}(-\omega)=\chi_{i}^{\ast}(\omega).\label{J02simchi}
\end{equation}
Splitting the integrals in (\ref{J02kk1})--(\ref{J02kk2}) as $\int_{-\infty}^{\infty}=\int_{-\infty}^{0}+\int_{0}^{\infty}$
and using the symmetry properties (\ref{J02simchi}), it is immediate
to arrive at the following relations: 
\begin{align}
\chi_{r}(\omega) & =-\dfrac{2}{\pi}\wp\int_{0}^{\infty}\dfrac{\omega^{\prime}\chi_{i}(\omega^{\prime})}{\omega^{2}-\omega^{\prime2}}d\omega^{\prime}\label{J02kkk1}\\
\chi_{i}(\omega) & =\dfrac{2\omega}{\pi}\wp\int_{0}^{\infty}\dfrac{\chi_{r}(\omega^{\prime})}{\omega^{2}-\omega^{\prime2}}d\omega^{\prime}.\label{J02kkk2}
\end{align}

The KK relations (\ref{J02kkk1})--(\ref{J02kkk2}) provide valuable
insight into the form of the material response. For example, because
the integral in (\ref{J02kkk2}) will generally be nonzero except
at $\omega=0$, we cannot envision an hypothetical lossless dielectric
if there is dispersion. In fact, if there existed a \emph{non-dispersive}
dielectric, \emph{i.e.}, one with $\chi_{r}(\omega^{\prime})=$ constant,
then $\chi_{i}(\omega)=0$ (since $\wp\int_{0}^{\infty}(\omega^{2}-\omega^{\prime2})^{-1}d\omega^{\prime}=0$)
and the material would indeed be lossless in this case. Therefore,
dispersion is necessarily associated to losses; since no medium other
than vacuum is without dispersion, we find that a perfect transparent
dielectric cannot exist.

As another example, it can be shown that if an isolated absorption
peak exists at frequency $\omega_{0},$ then the real susceptibility
in the quasi-transparent region near $\omega_{0}$ is of the form
$\chi_{r}(\omega)\simeq B(\omega_{0}^{2}-\omega^{2})^{-1}$ \cite{mills_nonlinear_1998}.
This feature can be verified for $\chi_{r}(\omega)$ in the form (\ref{J02ncomp}),
but it is a general result that follows from the KK relations without
any mention to the particular physical origin or model of the absorption
peak.

The KK relations also indicate that it is unnecessary to know both
the real and imaginary parts of the susceptibility; the knowledge
of one of them --- albeit over the whole $0$ to $\infty$ frequency
range --- suffices to compute the other one. In fact, it is common-place
to derive data of the real refractive index of a material from its
absorption spectrum, which is usually easier to measure.

\subsection[``Fast''\ and ``slow''\ light]{``Fast''\ and ``slow''\ light}

\label{J02sctFSL}

Early in the development and consolidation of the electromagnetic
theory, it was realized that the formalism describing the propagation
of waves might lead, apparently, to some bizarre results. This is
readily seen when the expression of the group index (\ref{J02reln})
is replaced in (\ref{Jo2dng}), yielding 
\begin{equation}
v_{g}(\omega)=\dfrac{c}{n(\omega)+\omega\dfrac{dn(\omega)}{d\omega}}.\label{J02deno1}
\end{equation}
Assuming that $n(\omega)$ $>0,$ one notices that if there is a frequency
range with anomalous dispersion, this is, $dn(\omega)/d\omega<0$
($dn_{\lambda}(\lambda)/d\lambda>0$), then the denominator of (\ref{J02deno1})
might be smaller than 1, in which case we would have $v_{g}>c.$ The
situation $dn_{\lambda}/d\lambda>0$ occurs indeed within the absorption
peaks, as appreciated in Fig. \ref{J02FIGnrniMatlab}. On the other
hand, it should not be forgotten that expression (\ref{J02deno1})
was derived assuming an almost transparent dielectric, which is obviously
not the case in spectral regions with significant losses. Therefore,
we must take a step back and check the validity of (\ref{J02deno1})
for a complex $n(\omega).$

Writing $n(\omega)=n_{r}(\omega)-in_{i}(\omega),$ the propagation
constant becomes complex: 
\begin{equation}
k(\omega)=k_{R}(\omega)-ik_{I}(\omega)=(\omega/c)n_{r}(\omega)-i(\omega/c)n_{i}(\omega).
\end{equation}
Following the same steps that led to (\ref{J02tan}), we arrive at
the expression 
\begin{equation}
\mathcal{\tilde{E}}(z,t)\simeq\frac{1}{2\pi}\int\nolimits _{-\infty}^{\infty}G(\Omega)\,e^{i[(\omega_{0}+\Omega)t-(k_{0}+k_{1}\Omega+\tfrac{1}{2}k_{2}\Omega^{2}+\cdots)z]\,}d\Omega.
\end{equation}
But now all the coefficients $k_{j}$ of the Taylor expansion of $k(\omega)$
are complex. Writing $k_{j}=k_{jR}-ik_{jI},$ we have 
\begin{align}
\mathcal{\tilde{E}}(z,t)\simeq e^{i(\omega_{0}t-k_{0R}z)\,}e^{-k_{0I}z\,}\frac{1}{2\pi}\int\nolimits _{-\infty}^{\infty} & G(\Omega)\,e^{i[\Omega(t-k_{1R}\Omega z)]\,}e^{-i\left(\tfrac{1}{2}k_{2R}\Omega^{2}z+\cdots\right)}\nonumber \\
 & \times e^{-\left(k_{1I}\Omega+\tfrac{1}{2}k_{2I}\Omega^{2}+\cdots\right)}d\Omega.\label{J02inthe}
\end{align}

\begin{figure}[h]
\centering{}\includegraphics[scale=0.09]{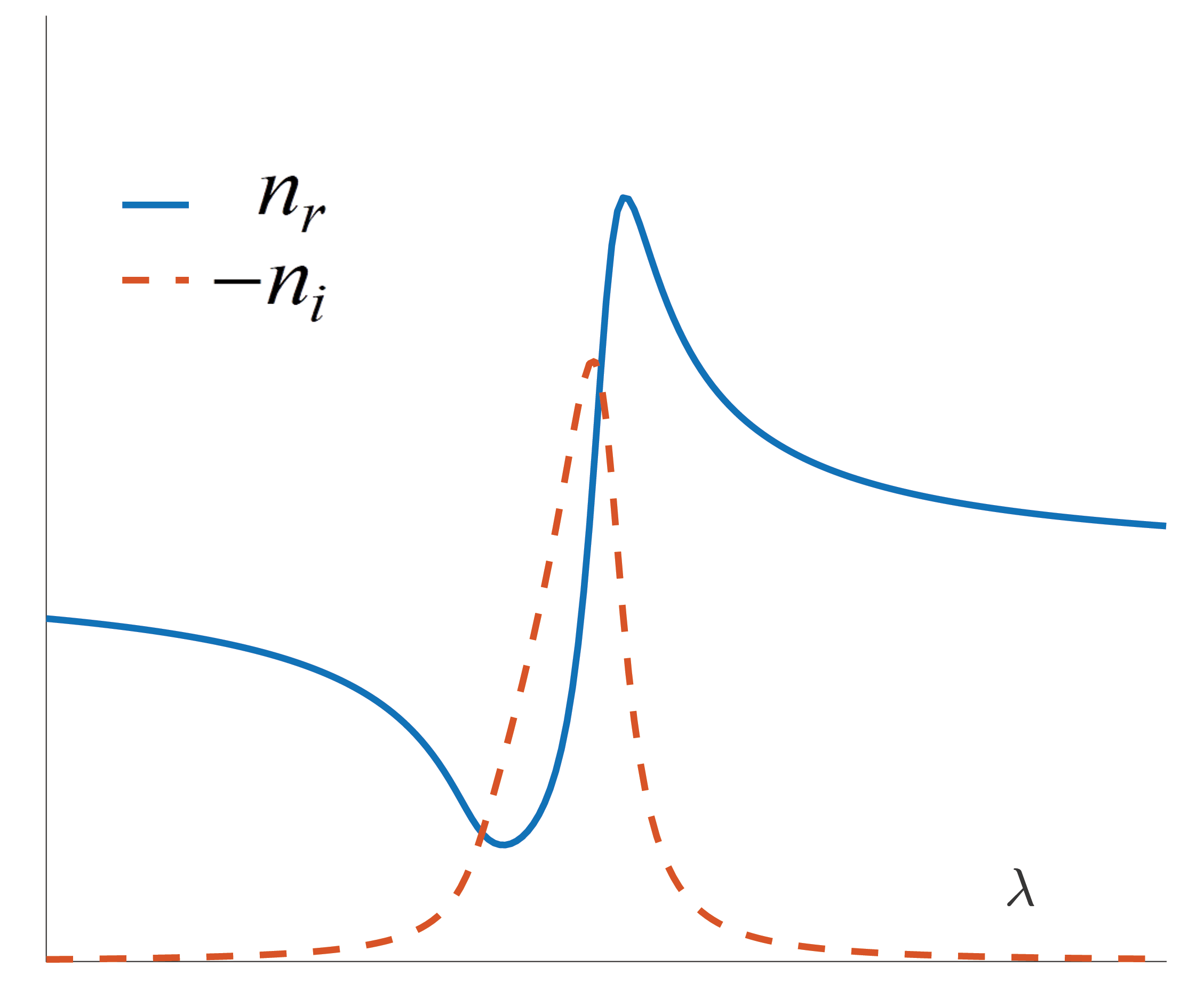}

\caption{Real (blue) and imaginary (dashed red) parts of the refractive index
around an isolated resonance. In a narrow spectral range there is
anomalous dispersion.}

\label{J02FIGrayita} 
\end{figure}

It comes as no surprise that the presence of $k_{0I}=\operatorname{Im}k(\omega_{0})$
results in an exponential decay of the wave amplitude. In absence
of the third exponential factor in the integrand, we would conclude
that, apart from the global, non-distortive attenuation due to $k_{0I},$
the pulse would propagate as in the lossless case but with a group
velocity $v_{g}=(dk_{1R}/d\omega)_{\omega=\omega_{0}}^{-1};$ \emph{i.e.},
expression (\ref{J02deno1}) would apply with $n_{r}=\operatorname{Re}(n)$
in place of $n.$ Of course, some amount of distortion would be present,
according to the values of $k_{2R},$ $k_{3R},$ etc. The key point
is that the term $\exp[k_{1I}\Omega+(1/2)k_{2I}\Omega^{2}+\cdots]$
in (\ref{J02inthe}) \emph{cannot} be neglected. In our model, anomalous
dispersion occurs precisely inside the absorption peaks, where losses
are not only large, but also significantly dispersive, as shown in
Fig. \ref{J02FIGrayita}. Thus $k_{1I},$ $k_{2I}$, etc. cannot be
ignored, which prevents drawing an analytical result of the form $\sim g(t-z/v_{g})$
from (\ref{J02inthe}).

One can insist on calling $v_{g}(\omega),$ as defined in (\ref{J02deno1})
(with $n_{r}$ instead of $n$), the ``group velocity''\ --- and
it is common practice ---, but its meaning is not quite the same
anymore. The fact that $v_{g}>c$ does not imply now that the wave
envelope (the information) travels faster than light in vacuum, and
there is of course not any affront to the special theory of relativity.
Typically, the pulse becomes so attenuated and distorted (the latter
being contributed not only by $n_{i}(\omega),$ but also by the strong
dispersion of $n_{r}(\omega)$ around the resonances), that it can
hardly be identified as an entity with a single propagation velocity.
In any case, no portion of the arriving waveform travels faster than
$c.$ In a better situation, when the pulse shape remains reasonably
intact --- even if severely attenuated ---, the condition $v_{g}>c$
reflects the fact that the \emph{peak} of the \emph{propagated pulse}
appears at the output when the \emph{peak} of the \emph{input pulse}
has not yet entered the propagation section (see for example \cite{chamorro-posada_superluminal_2009}
and references therein). This phenomenon is useless to all practical
purposes, however, as the pulse front never arrives before a time
lapse $L/c$ has passed, $L$ being the propagation length.

There is no reason why $n_{g}$ could not even be negative. In one
such situation, the pulse peak has been predicted to be moving \emph{backwards}
within the medium \cite{gehring_observation_2006}, a funny consequence
of the peculiar interferences taking place during the propagation,
but once again without any implication of the faster-than-light type.

From the corpuscular perspective, it seems quite obvious that the
presence of any atomic media, dielectrics included, can never result
in an effective velocity greater than $c.$ When photons are not interacting
with matter, they can only be ``flying''\ across the interatomic
vacuum, naturally at speed $c.$ Absorptions and emissions, in turn,
are processes that take time, so they can only but slow down the light
travel further. In spite of these sensible objections, the misleading
and somewhat pompous term ``\emph{superluminal}''\ has been coined
and accepted to refer to any situation in which $v_{g}>c$ (or $n_{g}<1$).
``\emph{Fast light}''\ is another generic name sometimes used to
refer to these phenomena.

A common type of superluminality is that arising in one-resonance
absorbing atomic media as well as in artificial structures such as
optical coupled resonators. This kind of superluminal propagation
can be explained in a unified manner as an interference process \cite{chamorro-posada_superluminal_2009}.
Superluminality may occur in active (amplifying) media as well, in
a manner rather similar to the absorption case. Other proposals make
use of diverse nonlinear optical processes.

``\emph{Slow light}''\ is another popular term. It refers to the
ability of achieving very low group velocities. In this case, the
condition $v_{g}<<c$ does represent a signal (typically a pulse)
which propagates very slowly or, at least, at a velocity remarkable
smaller than $c/n_{r}$, the phase velocity in the background material.\footnote{Coupled microring \emph{structures} can be taylored to achieve slow
light propagation (see for example \cite{chamorro-posada_fast_2009}
and references therein). In this case $n_{r}$ would be the (modal)
refractive index of the constituent waveguide.} Needless to say, the electromagnetic waves always propagate at speed
$c,$ as discussed above; therefore, behind the expression ``slow
light''\ there is just a light-matter interaction process in an
atomic medium which causes a particularly significant delay. In artificial
structures, the delay is obtained basically by keeping the light going
around a semi-closed path for as long as possible. Likewise, the also
employed expression ``stopped light''\ does not imply that a photon
flux can been ``stopped''\ at all, but simply means that the light
is absorbed by a suitable medium which can remain excited indefinitely,
and then re-emitted coherently at any desired time. (In the ``meantime'',\ the
energy and the information are stored in a different physical form
inside the atomic medium; it is senseless to pretend that the absorbed
and the re-emitted photons are ``the same ones''.)

To a great extent, the research in slow light is aimed at developing
optical retarders with\emph{\ high propagation delays} over \emph{large
spectral widths}. Both requirements are hard to meet simultaneously.
In fact, a theoretical limit to the maximum delay$\times$bandwidth
\emph{product} is often found in the basic types of the retarding
structures (including bulk atomic media).\ In order to overcome these
deeply-rooted limitations, nonlinear processes and other strategies
are being explored (see for example \cite{cestier_resonance_2010}).
Slow light --- or, more generally, the control of the speed of light
--- has enormous applications in optical communications and optical
(classical and quantum) processing in general \cite{boyd_applications_2006}.

\section[Electromagnetism of amplifying media]{Electromagnetism of linear amplifying media \sectionmark{Electromagnetism
of amplifying media}}

\label{J03sctEMAM}

\begin{figure}[bh]
\centering{}\includegraphics[scale=0.55]{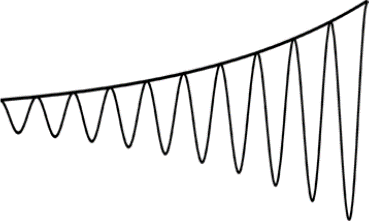}\caption{The graphic shows, for any fixed $t=t_{0},$ the form of the optical
field $\mathcal{E}(z,t_{0})$ as it propagates along an amplifying
medium having a refractive index with a negative imaginary part.}
\label{examp} 
\end{figure}

By a simple extension of the results of Subsection \ref{J02sbsctAMW},
it is easy to model electromagnetically an optically \emph{amplifying}
material. For the sake of simplicity we will, again, consider plane
waves propagating across an homogeneous, isotropic medium in the $z$
direction. Assuming a linearly polarized, purely monochromatic wave,
the electric field has the form (\ref{J02onda+ atenuada(t)}), which
we reproduce here: 
\begin{equation}
\mathcal{E}(z,t)=Ae^{-\,n_{i}(\omega_{0})(\omega_{0}/c)z}\cos\left(\omega_{0}t-n_{r}(\omega_{0})\frac{\omega_{0}}{c}z\right).\label{J03field1}
\end{equation}
All that must be done for (\ref{J03field1}) to represent an amplified,
rather than attenuated, wave propagating along the $+z$ direction,
is to switch the sign of $n_{i}(\omega_{0})$ from positive to negative.
In this way, (\ref{J03field1}) may describe an \emph{exponentially
amplified} wave, as illustrated in Fig. \ref{examp}.

One can therefore say that, in order for a medium to amplify the light
of frequency $\omega_{0},$ the imaginary part of its refractive index
need be positive at that frequency, $-n_{i}(\omega_{0})>0.$ Apart
from its usefulness by itself, optical amplification is essential
to achieve laser oscillation. In Subsection \ref{pflux} below, we
will study in more detail the operation of the amplifier, while the
fundamentals of the laser oscillation will be briefly presented in
Subsection \ref{oscilla}.

\subsection{Dielectric susceptibility and population inversion}

\label{pflux}

We assume that the susceptibility of a dielectric, whether passive
or active (amplifying), can be written as in (\ref{J02ncomp}), 
\begin{equation}
\chi(\omega)=\sum\limits _{j=1}^{N}\frac{A_{j}\,\omega_{j}^{2}}{-\omega^{2}-i\omega\,b_{j}+\omega_{j}^{2}}.\label{J03sus1}
\end{equation}
We further assume that the radiation frequency $\omega$ is far from
all the resonance frequencies of the material but one, $\omega_{r},$
so that 
\begin{equation}
\chi(\omega)\simeq\sum\limits _{j\neq r}^{N}\frac{A_{j}\,\omega_{j}^{2}}{\omega_{j}^{2}-\omega^{2}}+\frac{A_{r}\,\omega_{r}^{2}}{-\omega^{2}-i\omega\,b_{r}+\omega_{r}^{2}}\equiv\bar{\chi}(\omega)+\chi_{\text{res}}(\omega)\qquad\text{(near }\omega_{r}\text{),}\label{par}
\end{equation}
where $\bar{\chi}(\omega),$ including all the terms in the parenthesis
in (\ref{par}), is the (real-valued) ``non-resonant''\ susceptibility,
and 
\begin{equation}
\chi_{\text{res}}(\omega)=\frac{A_{r}\,\omega_{r}^{2}}{-\omega^{2}-i\omega\,b_{r}+\omega_{r}^{2}}\equiv\chi^{\prime}(\omega)-i\chi^{\prime\prime}(\omega),\label{J03imim}
\end{equation}
which describes the effect of the resonant transition at $\omega_{r}$
on the propagating wave. We next define the ``background index'',
\begin{equation}
\bar{n}(\omega)\equiv\sqrt{1+\bar{\chi}(\omega)},\label{J03nbar}
\end{equation}
which can be thought of as the (real-valued) refractive index, at
frequencies near $\omega_{r},$ contributed by all other resonances
(Fig. \ref{J02FIGnrniMatlab}). The actual refractive index \emph{near
the resonance} can then be expressed as follows: 
\begin{align}
n(\omega) & =\sqrt{1+\chi(\omega)}\simeq\sqrt{1+\bar{\chi}(\omega)+\chi^{\prime}(\omega)-i\chi^{\prime\prime}(\omega)}=\bar{n}(\omega)\sqrt{1+\dfrac{\chi^{\prime}(\omega)-i\chi^{\prime\prime}(\omega)}{\bar{n}^{2}(\omega)}}\nonumber \\
 & \simeq\,\bar{n}(\omega)\left(1+\dfrac{\chi^{\prime}(\omega)}{2\bar{n}^{2}(\omega)}\right)-i\dfrac{\chi^{\prime\prime}(\omega)}{2\bar{n}(\omega)}\equiv n_{r}(\omega)-in_{i}(\omega)\qquad\text{(near }\omega_{r}\text{)}\label{J03nn}
\end{align}
if $|\chi_{\text{res}}(\omega)|/\bar{n}^{2}(\omega)<<1$.

The effect of the resonance is twofold. On one hand, it slightly modifies
$\bar{n}(\omega)$ through $\chi^{\prime}(\omega)$; on the other,
it brings about an imaginary part $-n_{i}(\omega)$ which will cause
attenuation or amplification, depending on its sign. Actually, the
classical intensity (W/m$^{2}$) of this monochromatic plane wave
propagating through a dielectric medium of index $n=n_{r}-in_{i}$
is given by by the well-known expression 
\begin{equation}
I_{\text{Cl}}=\sqrt{\frac{\epsilon_{0}}{\mu_{0}}}n_{r}|\tilde{E}(z)|^{2},\label{J03ca}
\end{equation}
with $\tilde{E}(z)$ the complex amplitude or phasor of the field
(\ref{J03field1}), 
\begin{equation}
\widetilde{E}(z)=Ae^{-\,n_{i}(\omega)\frac{\omega}{c}z}e^{-\,in_{r}(\omega)\frac{\omega}{c}z}.\label{J03cb}
\end{equation}
Replacing (\ref{J03cb}) in (\ref{J03ca}), we obtain 
\begin{equation}
I_{\text{Cl}}=\sqrt{\frac{\epsilon_{0}}{\mu_{0}}}n_{r}A^{2}\exp\left[-2\dfrac{\omega}{c}n_{i}(\omega)z\right],\label{J03icl}
\end{equation}
which will grow along the propagation direction $z$ as long as $-n_{i}(\omega)>0.$

\begin{figure}[h]
\centering{}\includegraphics[scale=0.25]{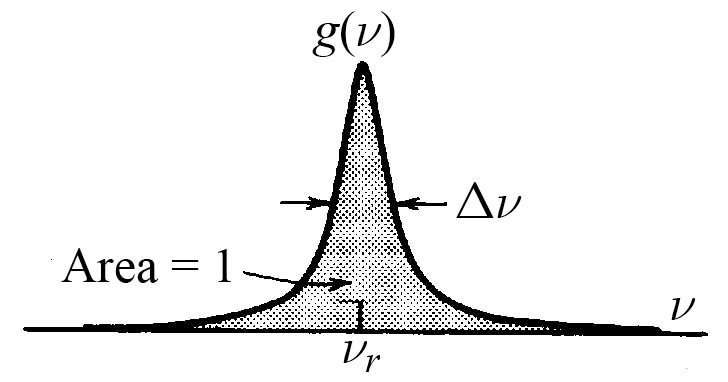}\caption{Lorentzian lineshape around the central frequency $\nu_{r}.$}
\label{loren} 
\end{figure}

Now, what is the physical origin of a \emph{positive} refractive index
(negative $n_{i}$)? In principle, a quantum-mechanical model of the
electromagnetic radiation would be necessary to answer this question.
Without going that far, Einstein's model for the interaction between
light and two-level atoms provides sufficient insight for our purposes.
We will basically follow a rather standard derivation such as that
presented in \cite{yariv_photonics_2007}. We call $N_{1}$ and $N_{2}$
the average atomic density (atoms/m$^{3}$) of atoms (or molecules)
in the ground state (energy $E_{1}$) and excited state (energy $E_{2}>E_{1}$),
respectively. The energy of the photons must match the atomic energy
difference: $E_{2}-E_{1}\approx h\nu_{r}=\hbar\omega_{r}.$ $N_{1}$
and $N_{2}$ will generally depend of the specific site within the
material (i.e. they are functions of $z$), but for low optical intensity
(``small signal'') they can be considered approximately constant.
We will not elaborate further on this model, which can be found in
countless textbooks of photonics and related fields. Assuming our
(dielectric) material is made up of such two-level atoms or molecules,
the optical intensity is found to evolve according to the expression
\begin{equation}
I=I(0)\exp[\sigma_{\nu}(N_{2}-N_{1})z]\,\qquad\text{(valid for small signal).}\label{J03iq}
\end{equation}
In (\ref{J03icl}), $\sigma_{\nu}$ is the so-called \emph{cross-section},
which depends on the properties of the material and on the frequency
$\nu.$ Namely, 
\begin{equation}
\sigma_{\nu}=\frac{c^{2}}{8\pi\bar{n}^{2}\nu^{2}\tau_{\text{sp}}}\bar{g}(\nu)\label{J03sigma1}
\end{equation}
with $\tau_{\text{sp}}$ the\emph{ spontaneous emission time} (the
average time it takes an excited atom at level $E_{2}$ to spontaneously
deexcite to level $E_{1}$ by emitting a photon of energy $E_{2}-E_{1},$
with random direction and phase). In (\ref{J03sigma1}), $\bar{g}(\nu)$
is the normalized ($\int_{0}^{\infty}$ $\bar{g}(\nu)d\nu=1$) \emph{spectral
lineshape} or \emph{lineshape function}, which has a probabilistic
interpretation: roughly speaking, $\bar{g}(\nu)d\nu$ is the probability
that a photon having an energy between $\nu$ and $\nu+d\nu$ will
be emitted/absorbed by the atom, if an emission/absorption is to occur.
The simplest form of $\bar{g}(\nu)$ is the so-called Lorentzian lineshape
(Fig. \ref{loren}): 
\begin{equation}
\bar{g}(\nu)=\frac{1}{2\pi}\frac{\Delta\nu}{(\nu-\nu_{r})^{2}+(\Delta\nu/2)^{2}},\label{J03g}
\end{equation}
peaked at $\nu=\nu_{r}.$

Note that (\ref{J03icl}) is also a ``small signal''\ result, like
(\ref{J03iq}), since it was derived within the framework of linear
optics, thus excluding nonlinear phenomena such as gain saturation
or others.

Expressions (\ref{J03icl}) and (\ref{J03iq}) should then be equivalent,
which demands that the exponents be equal. Recalling (\ref{J03sigma1}),
this leads to 
\begin{equation}
n_{i}=-\frac{1}{\tau_{\text{sp}}}\frac{c^{3}}{32\pi^{2}\bar{n}^{\,2}\nu^{3}}\bar{g}(\nu)(N_{2}-N_{1}).\label{J03eq}
\end{equation}
Formula (\ref{J03eq}) shows the connection between the imaginary
part of the refractive index and the ``quantum state''\ of the
atoms of the material. An amplifying medium requires $n_{i}<0,$ hence
$N_{2}>N_{1}$ --- there must exist \emph{population inversion} in
the medium, i.e. more atoms in the excited state than in the ground
state.\footnote{In order not to overload the notation with subscripts, functional
symbols like $n_{i}$ will loosely refer to either $n_{i}(\omega)$
or $n_{i}(\nu)$ hereafter, depending on the context, in the understanding
that really denote different mathematical functions.}

\emph{Near the resonance} at $\nu_{r}=$ $\omega_{r}/(2\pi),$ some
frequency-dependent parameters in (\ref{J03eq}) can be considered
practically constant within the whole narrow linewidth span $\Delta\nu$;\ namely,
$\nu^{3}\simeq\nu_{r}^{3}$ and $\bar{n}^{\,2}(\nu)\simeq\bar{n}^{\,2}(\nu_{r})$
(recall that $\bar{n}$ excludes the resonance peak at $\nu_{r}$).
Thus the frequency dependence of $n_{i}$ is essentially given by
the Lorentzian line $\bar{g}(\nu)$ (\ref{J03g}), which does display
an abrupt variation. The linewidth $\Delta\nu$ can typically be of
the order of hundreds of GHz. Of course, $n_{i}$ given in (\ref{J03nn})
should have the same Lorentzian form. Let us check this. Computing
$\chi^{\prime\prime}=-\operatorname{Im}(\chi_{\text{res}}),$ we obtain
\begin{align}
n_{i} & \simeq\dfrac{\omega_{r}}{8\bar{n}(\omega_{r})}\frac{A_{r}b_{r}}{(\omega-\omega_{r})^{2}+b_{r}^{2}\omega_{r}^{2}/4}\text{\qquad\ [}\omega\text{ (}\nu\text{) near }\omega_{r}\text{ (}\nu_{r}\text{)]}\label{J03opp}\\
 & =\text{sign}[A_{r}]\dfrac{\nu_{r}}{8\bar{n}(\nu_{r})}\frac{1}{2\pi}\frac{\Delta\nu}{(\nu-\nu_{r})^{2}+(\Delta\nu/2)^{2}}=\text{sign}[A_{r}]\dfrac{\nu_{r}}{8\bar{n}(\nu_{r})}\bar{g}(\nu).\label{J03opp1}
\end{align}
The approximations $\omega_{r}^{2}-\omega^{2}=(\omega_{r}+\omega)(\omega_{r}-\omega)\simeq2\omega_{r}(\omega_{r}-\omega),$
$\omega^{2}\simeq\omega_{r}^{2}$ and $\bar{n}^{\,2}(\omega)\simeq\bar{n}^{\,2}(\omega_{r})$
have been used in (\ref{J03opp}). All are justified by the assumption
that $\omega$ is close to $\omega_{r}$ within a range of the order
$\Delta\omega=2\pi\Delta\nu<<\omega_{r},$ as stated above. In this
way, (\ref{J03eq}) and (\ref{J03opp})--(\ref{J03opp1}) are consistent
as they have the same Lorentzian frequency dependence. We also see
that the relation $\Delta\nu$ $=|A_{r}|b_{r}$ must hold, in the
understanding that $A_{r}$ will be negative in an amplifying material.

A useful expression showing explicitly the frequency dependence of
$\chi^{\prime\prime}$ can be obtained from the first equality in
(\ref{J03opp}) and (\ref{J03eq}), again with $\nu^{3}\simeq\nu_{r}^{3}$
and calling $\lambda_{r}=c/\nu_{r},$ the resonance (vacuum) wavelength:
\begin{align}
\chi^{\prime\prime}(\nu) & =\frac{1}{\tau_{\text{sp}}}\frac{c^{3}}{16\pi^{2}\bar{n}(\nu_{r})^{\,}\nu_{r}^{3}}\bar{g}(\nu)(N_{1}-N_{2})\nonumber \\
 & =\frac{\lambda_{r}^{3}}{8\pi^{3}\bar{n}(\lambda_{r})\tau_{\text{sp}}^{\,}\Delta\nu}\frac{N_{1}-N_{2}}{1+\left(\dfrac{\nu-\nu_{r}}{\Delta\nu/2}\right)^{2}}\text{\qquad(near }\nu_{r}\text{).}\label{J03inv}
\end{align}
It is easy to check that $\chi^{\prime}(\nu)$ is related to $\chi^{\prime\prime}(\nu)$
through the simple expression 
\begin{equation}
\chi^{\prime}(\nu)=2\frac{\nu_{r}-\nu}{\Delta\nu}\chi^{\prime\prime}(\nu).\label{J03rel}
\end{equation}

\subsection{Laser oscillation}

\label{oscilla}

\begin{figure}[h]
\centering{}\includegraphics[scale=0.4]{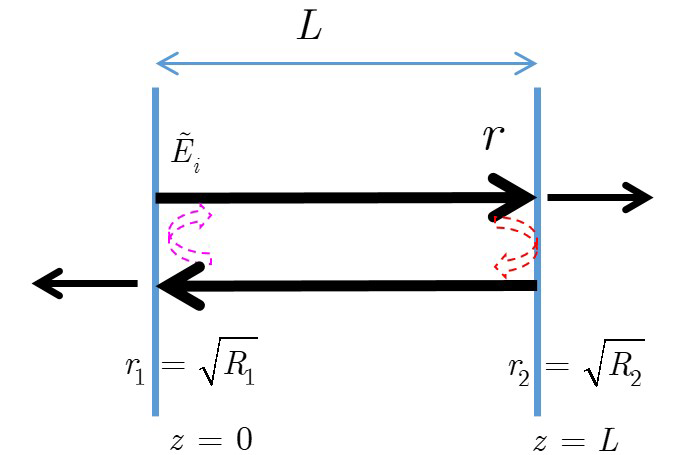}\caption{A Fabry-Pérot laser. The cavity between the mirrors is filled with
a dielectric material that can be made to amplify the light.}
\label{cavity} 
\end{figure}

In the preceding subsection we have explored how a suitable dielectric
material can be turned into an optical amplifier. Amplification is
indeed the first requirement to make a laser, but some mechanism providing
output-to-input feedback must be added in order to obtain optical
oscillation. Not surprisingly, the most straightforward way to achieve
signal feedback at optical frequencies is to use mirrors. The basic
structure is shown in Fig. \ref{cavity}. The active dielectric medium
is confined between two partially reflecting plane mirrors separated
by a distance $L$. Calling $R_{1}$ and $R_{2}$ the power reflectivities
of the mirrors ($R_{1}\leq1,$ $R_{2}<1$), the fraction of the field
amplitude reflected at each end will be$\sqrt{R_{1}}$ and $\sqrt{R_{2}},$
respectively, for a plane wave propagating in the $\pm z$ direction,
as assumed. Such a structure is called a \emph{Fabry-Pérot (FP)} cavity.

Roughly speaking, a self-maintained electromagnetic oscillation will
occur if the amplification of the light within the cavity compensates
for all the propagation losses. Assume that a monochromatic wave exists
inside the FP cavity with an amplitude $E_{1}$ at, for example, $z=0,$
immediately to the right of mirror 1 (any other $z$ location could
be chosen). The existence of such a wave across the structure is only
possible if the following consistency condition is fulfilled: 
\begin{equation}
E_{1}e^{-i\tilde{k}L}\sqrt{R_{2}}e^{-i\tilde{k}L}\sqrt{R_{1}}=E_{1},\label{J03round}
\end{equation}
with 
\begin{equation}
\tilde{k}=\frac{\omega_{L}}{c}n(\omega_{L})=\frac{\omega_{L}}{c}[n_{r}(\omega_{L})-in_{i}(\omega_{L})]\equiv k(\omega_{L})-ik_{i}(\omega_{L}),\label{J03defk}
\end{equation}
the complex propagation constant, where $\omega_{L}$ is the angular
wave frequency, still \emph{unknown} but assumed to be close to $\omega_{r}.$
On the LHS of (\ref{J03round}), $E_{1}e^{-i\tilde{k}L}=(E_{1}e^{-k_{i}L})e^{-ikL}\equiv\tilde{E}_{2}$
simply describes the amplitude and phase of a right-propagating wave
at $z=L$ (just before mirror 2) if, as assumed, it has a real amplitude
$E_{1}$ at $z=0.$ A reflected wave must then exist with an amplitude
$\tilde{E}_{2}\sqrt{R_{2}}\equiv\tilde{E}_{3}$ at $z=L$ (it is assumed
that the mirrors do not introduce any phase shift). The field of this
backward-propagating wave will necessarily be\footnote{The reader may be wondering why we have written a factor $\exp(-i\tilde{k}z)=\exp(-ikz)\exp(-k_{i}z)$
for the backward-propagating wave, when it should apparently be $\exp(+ikz)\exp(-k_{i}z).$
For one thing, if a propagation factor $\exp(+ikz)$ were attributed
to the wave returning to the left mirror, the accumulated phase at
the starting point $z=0$ would always turn out to be zero regardless
of the value of $L,$ which obvioulsy makes no sense. But that of
course is not an explanation. The formal justification starts by noting
that (\ref{J03round}) is \emph{not} really a phasorial equation,
since time is not ``frozen'': using phasors implies ignoring the
time altogether by fixing an arbitrary instant at which all amplitudes
and phases are compared. Expression (\ref{J03round}), however, involves
several different times. A rigorous derivation --- yielding the same
result anyway --- requires referring all complex amplitudes \emph{to
the same instant}. We will take the time at which the wavefront returns
to $z=0$ after a round trip, which we will call $t_{\text{round}}.$
Now, if at $t=0$ the starting complex field at $z=0$ was $E_{1}$
(assuming an initial zero phase) its phase at $t=$ $t_{\text{round}}$
will be different; namely, it will have been increased by an amount
$\omega_{L}t_{\text{round}}$. Therefore, \emph{at }$t=t_{\text{round}},$
the reference forward-propagating complex field at $z=0$ is $E_{1}e^{i\omega_{L}t_{\text{round}}}\equiv E_{1}^{f},$
with

\[
\omega_{L}t_{\text{round}}=\omega_{L}\frac{2L}{v_{\text{ph}}}=\frac{2L\omega_{L}n_{r}(\omega_{L})}{c}=2Lk.
\]
Thus,

\[
E_{1}^{f}=E_{1}e^{i2kL}.
\]
On the other hand, the 1-round trip field at $z=0$ \emph{and }$t=t_{\text{round}},$
denoted $E_{1}^{b},$ will be (note that we use the \emph{correct}
signs for the imaginary exponents now)

\[
E_{1}^{b}=E_{1}e^{-ikL}e^{-k_{i}L}\sqrt{R_{2}}e^{+ikL}e^{-k_{i}L}\sqrt{R_{1}}=E_{1}\sqrt{R_{1}R_{2}}e^{-2k_{i}L}.
\]
Making $E_{1}^{b}=E_{1}^{f}$ yields

\[
e^{i2kL}=\sqrt{R_{1}R_{2}}e^{-2k_{i}L},
\]
which is equivalent to (\ref{J03round}).} $\tilde{E}_{3}e^{-i\tilde{k}L}\equiv\tilde{E}_{4}$ at $z=0,$ and
$\tilde{E}_{4}\sqrt{R_{1}}\equiv\tilde{E}_{5}$ immediately after
reflection at $z=0,$ where it is identified with the assumed right-propagating
field we started with, i.e. $\tilde{E}_{5}=E_{1}.$ This gives the
result (\ref{J03round}). Equating modulus and phase separately, (\ref{J03round})
yields 
\begin{align}
\sqrt{R_{1}R_{2}}e^{-2k_{i}L} & =1\qquad\qquad\qquad\qquad\text{(modulus)}\label{J03cond1}\\
kL & =m\pi,\quad m=1,2,\ldots\quad\text{(phase).}\label{J03cond2}
\end{align}

We will consider the modulus condition first. Using (\ref{J03nn})
in (\ref{J03defk}), we get 
\begin{equation}
\tilde{k}(\omega_{L})=k(\omega_{L})-ik_{i}(\omega_{L})=\frac{\omega_{L}}{c}\bar{n}(\omega_{L})\left(1+\dfrac{\chi^{\prime}(\omega_{L})}{2\bar{n}^{2}(\omega_{L})}\right)-i\left(\frac{\omega_{L}}{c}\dfrac{\chi^{\prime\prime}(\omega_{L})}{2\bar{n}(\omega_{L})}+\dfrac{\alpha_{\text{int}}}{2}\right).\label{J03erin}
\end{equation}
The extra term $-i\alpha_{\text{int}}/2$ has been introduced \emph{ad
hoc} to account for additional propagation losses which are difficult
to model electromagnetically in a rigorous fashion. For example, diffraction
losses cannot possibly arise in our idealized one-dimensional model
supporting plane waves of infinite extent, but they do exist in real
devices. Other contributions to $\alpha_{\text{int}}$ may include
polarization losses, scattering losses, etc.\footnote{In writing out the wave equation for a laser medium, some authors
keep the conduction current term, $\boldsymbol{J}_{c}=\sigma\boldsymbol{E}$
$,$ which ends up as an additional summand $\sigma/(\epsilon_{0}\omega_{L})$
in the expression of $k_{i}(\omega_{L}).$ We think this is not very
fortunate conceptually as it suggests that the optical field can generate
an electrical current of optical frequency, which is by no means possible
($\sigma(\omega_{L})=0$). Actually, the dubius physical meaning of
$\sigma/(\epsilon_{0}\omega_{L})$ is without consequences as it is
common practice to simply assimilate this term to unspecified ``additional
losses''\ and rename it as $\alpha_{\text{int}}/2$ --- as we have
done directly.}

Using (\ref{J03erin}) in (\ref{J03cond1}) leads to 
\begin{equation}
g(\omega_{L})=\alpha_{m}+\alpha_{\text{int}},\label{J03gan}
\end{equation}
with 
\begin{equation}
g(\omega_{L})\doteq-\frac{\omega_{L}}{c}\dfrac{\chi^{\prime\prime}(\omega_{L})}{\bar{n}(\omega_{L})},
\end{equation}
the \emph{material gain}, and 
\begin{equation}
\alpha_{m}\doteq\frac{1}{2L}\ln\frac{1}{\sqrt{R_{1}R_{2}}},
\end{equation}
the \emph{mirror losses}. The latter accounts for the radiation escaping
through the mirrors. Obviously, if both mirrors were perfectly reflecting
($R_{1}=R_{2}=1$), then $\alpha_{m}=0,$ but in this case the laser
would be useless; at least one of the mirrors must be partially reflecting
to let some light out.

The material gain $g$ (not to be confused with the lineshape function
$\bar{g}$) will be positive if $\chi^{\prime\prime}<0,$ which according
to (\ref{J03inv}) demands the material to be in population inversion,
this is, $N_{2}>N_{1}.$ In this case, the interpretation of condition
(\ref{J03gan}) is clear:\ maintaining a sustained optical wave in
the resonator requires that the amplification of the active medium
compensate exactly for all the propagation losses in the cavity, including
of course the energy lost through the mirrors. Generally, one of the
mirrors is almost perfectly reflecting ($R_{1}\simeq1$) while the
other provides the output. Laser diodes are a remarkable exception
as both mirrors radiate equally.

No situation other than (\ref{J03gan}) permits self-sustained oscillation.
If an initial oscillation is assumed to exist in the cavity but the
gain is smaller than the losses, the radiation will extinguish after
a few round-trips. On the other hand, if the gain were larger than
the losses, the optical field would increase in each round-trip, approaching
an infinite value. Such a situation cannot be physically envisioned
and, more precisely, gain saturation is the mechanism which prevents
such phenomenon from taking place, as we will discuss later.

Gain is one half of the story. Even if the field ``returns''\ to
the same point of the cavity with the same intensity, it also needs
to have the same phase; otherwise, destructive interference would
prevent the oscillation. Using (\ref{J03erin}) in the phase condition
(\ref{J03cond1}) leads to an implicit expression for the set of potential
lasing frequencies $\omega_{L}\in\{\omega_{m}\}$: 
\begin{equation}
\nu_{m}=\frac{\omega_{m}}{2\pi}=\frac{mc}{2\bar{n}(\omega_{m})\left(1+\dfrac{\chi^{\prime}(\nu_{m})}{2\bar{n}^{2}(\nu_{m})}\right)L},\qquad m=1,2,\ldots\label{J03frqs}
\end{equation}
Note that, if the active resonance could be removed from the cavity
material, then $\chi^{\prime}=0$ and (\ref{J03frqs}) would reduce
to 
\begin{equation}
\tilde{\nu}_{m}=\frac{mc}{2\bar{n}(\tilde{\nu}_{m})L},\qquad m=1,2,\ldots\qquad\text{(cavity of real index }\bar{n}\text{, no }\nu_{r}\text{ resonance),}\label{J03cav}
\end{equation}
where we have renamed the resulting oscillation frequencies in this
case as $\tilde{\nu}_{m}$ ($\tilde{\omega}_{m}=2\pi\tilde{\nu}_{m}$).
From (\ref{J03rel}) and (\ref{J03inv}), $\chi^{\prime}(\nu)=0$
when $N_{2}=N_{1}.$ The frequencies $\tilde{\nu}_{m}$ are sometimes
called the ``cold cavity''\ resonant frequencies or modes\footnote{In the literature, it is often stated that $\{\tilde{\nu}_{m}\}$
are the modes of the ``passive''\ or unpumped cavity. This is misleading
as in the absence of pumping, $N_{1}<N_{2}$ ($N_{2}/N_{1}=\exp[-(E_{2}-E_{1})/(k_{B}T)]$
in thermal equilibrium), thus $\chi^{\prime}\neq0.$ The condition
$N_{2}=N_{1},$ which leaves the material gainless and lossless (asumming,
as we have done, that the background index $\bar{n}$ is real), does
require some pumping.}. Expression (\ref{J03frqs}) can then be rewritten as 
\begin{equation}
\nu_{m}=\frac{\tilde{\nu}_{m}}{1+\dfrac{\chi^{\prime}(\nu_{m})}{2\bar{n}^{2}(\nu_{m})}},\qquad m=1,2,\ldots
\end{equation}

Formula (\ref{J03rel}) shows that, if an actual oscillation frequency
$\nu_{m}$ turns out to be equal to the actual material resonance
frequency, $\nu_{r}$, then $\chi^{\prime}(\nu_{m})=0$ for any value
of the population inversion $N_{2}-N_{1},$ in which case $\nu_{m}=\tilde{\nu}_{m}=\nu_{r}.$
In other words, when the center frequency of the atomic line $\bar{g}(\nu)$
coincides with one of the ``cold cavity''\ modes, the frequency
of this cavity mode will be the actual lasing frequency. In any other
case, $\chi^{\prime}(\nu_{m}),$ even if not zero, is expected to
be small if $\nu_{m}$ is very close to $\nu_{r}$ {[}c.f. (\ref{J03rel}){]},
in which case the correction will be small. A detailed analysis shows
that the presence of $\chi^{\prime}$ tends to shift $\nu_{m}$ slightly
from $\tilde{\nu}_{m}$ toward $\nu_{r},$ leading to the so-called
\emph{frequency-pulling} \cite{yariv_photonics_2007}, \cite{siegman_lasers_1990}.

\begin{figure}[h]
\centering{}\includegraphics[scale=0.2]{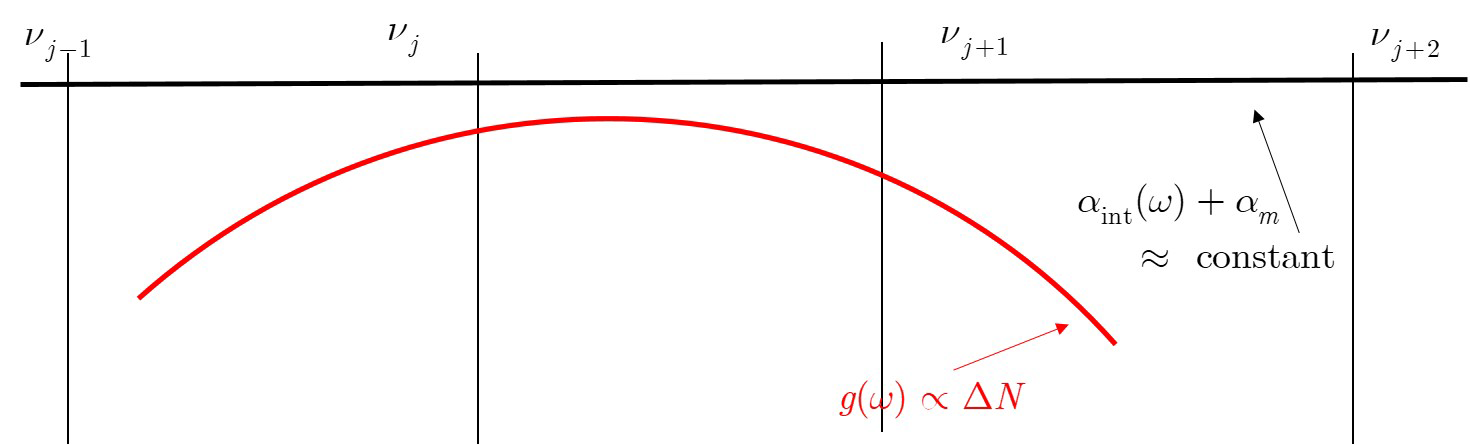}\caption{Spectral loss (black) and gain (red) curves of the cavity. Some possible
oscillation frequencies of the laser (assumed to be in the vicinity
of the atomic transition $\nu_{r}$) are also indicated. The red arc
could represent, for example, the peak of the Lorentzian curve shown
in Fig. \ref{loren}. In the case depicted, there is no (or not enough)
population inversion $\Delta N=N_{2}-N_{1}.$ Thus, the net loss exceeds
the net loss at all frequencies and the cavity will not lase.}
\label{laser5} 
\end{figure}

We will now explain the operation of the laser graphically. First,
ignoring the effect of $\chi^{\prime}$ for simplicity, expression
(\ref{J03cav}) seems to suggest that the possible oscillation frequencies,
represented as vertical lines in Fig. \ref{laser5}, are equispaced
with a separation $c/(2\bar{n}L)$ between two adjacent ``lines''.
This is not exactly the case as $\bar{n}$ is not constant but frequency-dependent
itself. A more accurate result is obtained as explained next.

Expression (\ref{J03cav}) can be rewritten as $\tilde{\nu}_{m}\,\bar{n}(\tilde{\nu}_{m})=mc/(2L).$
In particular, calling $\delta\tilde{\nu}$ the frequency step, i.e.
$\tilde{\nu}_{m+1}=\tilde{\nu}_{m}+\delta\tilde{\nu},$ and writing
$\bar{n}(\tilde{\nu}_{m+1})=\bar{n}(\tilde{\nu}_{m}+\delta\nu)=\bar{n}(\tilde{\nu}_{m})+\delta\bar{n},$
the equality reads, for $m+1,$ 
\begin{equation}
(\tilde{\nu}_{m}+\delta\nu)(\bar{n}(\tilde{\nu}_{m})+\delta\bar{n})=(m+1)\frac{c}{2L}=\tilde{\nu}_{m}\,\bar{n}(\tilde{\nu}_{m})+\frac{c}{2L},\label{J03sim}
\end{equation}
Neglecting the 2nd-order infinitesimal $\delta\tilde{\nu}\delta\bar{n},$
(\ref{J03sim}) simplifies to 
\begin{equation}
\tilde{\nu}_{m}\delta\bar{n}+\bar{n}(\tilde{\nu}_{m})\delta\nu=\frac{c}{2L}\Longrightarrow\left[\tilde{\nu}_{m}\left.\frac{\partial\bar{n}}{\partial\nu}\right\vert _{\tilde{\nu}_{m}}+\bar{n}(\tilde{\nu}_{m})\right]\delta\nu=\frac{c}{2L}.\label{J03gr}
\end{equation}
Recognizing the expression of $\bar{n}_{g}(\tilde{\nu}_{m})$, the
group index, in the square brackets, we finally obtain 
\begin{equation}
\delta\nu=\frac{c}{2\bar{n}_{g}L}\qquad\text{(ignoring }\chi^{\prime}\text{ frequency pulling).}
\end{equation}
If the background \emph{group} index $\bar{n}_{g}$ remains reasonably
constant in the spectral range of interest, near $\nu_{r},$ the oscillation
frequencies will be equispaced.

\begin{figure}[h]
\centering{}\includegraphics[scale=0.3]{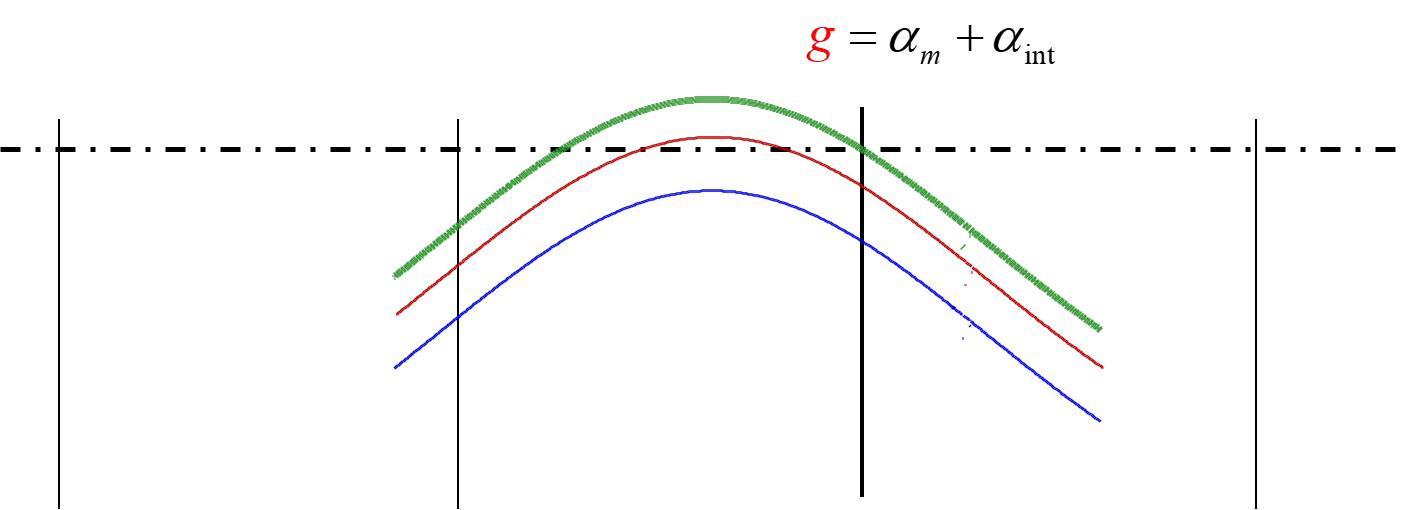}\caption{Increased pumping ends up creating sufficient population inversion
to build up a laser oscillation at one of the possible frequencies.
See text.}
\label{laser55} 
\end{figure}

Let us examine the compliance of equality (\ref{J03gan}). Regardless
of the population inversion $N_{2}-N_{1}$, which being a constant
simply determines the peak amplitude of the gain curve, the shape
of $g(v)$ will be Lorentzian, as stated. However, the right-hand
term of (\ref{J03gan}) is essentially flat. Even if both $\alpha_{\text{int}}$
and $\alpha_{m}$ are really frequency-dependent (for example, the
reflectivities $R_{1}$ and $R_{2}$ obviously depend on the wavelength),
they can be considered very approximately constant within the narrow
spectral width of interest. The situation is illustrated schematically
in Fig. \ref{laser55}. Consider that the material is at first unpumped,
so that $N_{1}-N_{2}>0$ ($g(\nu)$ is represented by the solid blue
curve) and the light could only experiment attenuation. In this case,
$g(\nu)<\alpha_{m}+\alpha_{\text{int}}$ at all frequencies and of
course there is no oscillation. Next, we enable the pumping, thus
increasing $N_{2}-N_{1}$ to the point that $g(\nu)$ (brown solid
line) surpasses the losses within some frequency range. However, none
of the permitted, phase-adapted oscillation frequencies lies in this
range, so still no oscillation can arise. Increasing $N_{2}-N_{1}$
further, the situation is finally reached where $g$ equals $\alpha_{m}+\alpha_{\text{int}}$
(the gain \emph{threshold value}) at one of the eligible frequencies
(green thick solid line). Lasing will then start at that specific
frequency only (at least in our simple description).

One may wonder how the oscillation actually builds up if there was
no light in the cavity in the first place. Just as an electronic oscillator
is started by the electronic noise existing the circuit, optical oscillation
in a laser is triggered by some photon generated by spontaneous emission
stemming from the fraction of atoms in the cavity which have been
thermically excited to the upper energy level. It is only necessary,
and statistically feasible, for one single so-emitted photon to have\ the
suitable direction and phase to match the modal characteristics of
the waves propagating in the cavity (an ideal, unphysical plane wave
pattern in our discussion, but a real specifically shaped optical
beam or waveguide mode in actual lasers).

What happens if the pumping is increased \emph{beyond} the threshold
value? In principle, $N_{2}-N_{1}$ should increase accordingly but,
as remarked above, such a situation cannot endure since the field
intensity inside the cavity grows to saturate the gain almost immediately.\footnote{\label{J03notapie1}Even if $g$ can be approximated as constant at
low intensity levels (``small signal''), strictly speaking the gain
is always intensity-dependent.} This will necessarily end up occurring even if $g$ is only ``infinitesimally''\ greater
$\alpha_{m}+\alpha_{\text{int}}$ (or, equivalently, $N_{2}-N_{1}$
``infinitesimally''\ greater than the threshold value of $N_{2}-N_{1}$).
The mechanisms involved can be qualitatively described as follows.

(a) At threshold, the photons generated by stimulated emission from
the net inverted atom population ($N_{2}-N_{1}$) are ``used up''\ to
balance the cavity round-trip losses.

(b) If pumping is increased, more atoms are lifted to the excited
state ($N_{2}-N_{1}$ increases), which will boost the stimulated
emission rate within the cavity. The new photons will quickly induce
still more stimulated emissions, producing more photons which will
in turn trigger further emissions, and so on. Now, for each emitted
photon, $N_{2}-N_{1}$ decreases by two units. The photon emission
rate, which grows dramatically, will very quickly surpass the fixed
pumping rate, so the photons in the cavity will soon significantly
outnumber the existing excited atoms. At this point, most of these
photons cannot stimulate emissions any more simply because there are
no more excited atoms available. This is, of course, nothing but a
gain saturation process ($N_{2}-N_{1}$ decreases, so the gain does
too). From an electromagnetic point of view, this process can be formally
accounted for by using a saturable, intensity-dependent dielectric
susceptibility in the model, rather than (\ref{J03sus1}); we will
not pursue this topic further.

(c) Now, all these newly-created photons will necessarily be evacuated
through the mirrors (not through the losses $\alpha_{\text{int}},$
which are already ``fed''\ by the photons generated at the threshold
value pumping).

(d) If the pumping could then be swiftly switched back to the threshold
value, an hypothetical observer would have simply seen a ``light
burst''\ corresponding to the transient exiting photons. However,
if the pumping above threshold is maintained, photons in excess will
be continuously created at a steady rate. These will be constantly
escaping through the mirrors. Meanwhile, as explained, the photon
density inside the cavity in steady state can only remain ``clamped''\ to
its threshold value (\ref{J03gan}) --- the only stable value ---
regardless of the level of the ``excess''\ pumping rate. As a side
note, it must be kept in mind that, even at the threshold level, the
gain may be significantly saturated (see footnote \ref{J03notapie1}).
Hence $g$ in equality (\ref{J03gan}) might not be well represented
by the ``small signal''\ gain and a full intensity-dependent expression
for $g$ might have to be employed, as mentioned above.

(e) Although the level of (above-threshold) pumping does not have
any effect on the equilibrium condition inside the cavity, we see
that it has an important effect: it generates a surplus of photons
that are radiated out. The higher the pumping rate is, the more photons
will have to leave the cavity. In other words, once the threshold
gain has been reached, all additional pumping will be used to generate
more \emph{optical power}.

The quantitative analysis of the operation and optimization of the
emitted laser power (where the reflectivity of the output mirror plays
a key role) can be found in innumerable textbooks.

\appendix

\section{APPENDIX\ on the Fourier transform and a few closely related topics}

Throughout these notes we consistently use the following definition
for the time-frequency Fourier transform pair:\footnote{Ocassionally, due to typographic shortage, we will also employ the
convention of asigning lowercase letters to the time functions and
uppercase to their Fourier transforms: $x(t)\longleftrightarrow X(\omega).$} 
\begin{align}
\mathcal{G}(t) & =\frac{1}{2\pi}\int_{-\infty}^{\infty}G(\omega)\,e^{i\omega t}d\omega\label{par1}\\
G(\omega) & =\int_{-\infty}^{\infty}\mathcal{G}(t)\,e^{-i\omega t}dt.\label{par2}
\end{align}
Here $\omega$ is the angular frequency in radians, related to the
ordinary frequency $\nu$ in hertzs by $\omega=2\pi\nu.$\footnote{The symbol $f$ is frequently used instead of $\nu$ in the literature.
Throughout this text, we follow the convention of using $f$ only
when the frequency range of the considered signal is ``electrical'',
rather than optical. In this case, we generally favour the symbol
$\Omega\equiv2\pi f$ for the angular frequency.} Unfortunately, there exist quite a few different (albeit equivalent)
conventions to define the transform pair in the literature. For example,
the factor $(2\pi)^{-1}$ can appear as a prefactor in (\ref{par2})
instead of (\ref{par1}), or as $(2\pi)^{-1/2}$ in both. The signs
$\pm$ in front of the exponent $i\omega$ in the integrals are also
a matter of taste. The choice (\ref{par1})--(\ref{par2}) is frequently
used in electrical engineering texts, while in the physics literature
the signs often appear exchanged. As a consequence, the propagation
factor of a monochromatic electromagnetic wave travelling in the $+z$
direction will read $\exp[i(\omega t-\beta z)]$ with our definition,
instead of the alternative form $\exp[i(\beta z-\omega t)]$. In order
to adapt one convention to the other, it will generally suffice to
substitute $i\rightarrow-i$ throughout.

\subsection{Some basic properties}

Here we only quote those properties of the Fourier transform (FT)
which are occasionally used in these notes. Apart from the linearity,
it is very easy to prove the following: 
\begin{align}
\text{FT}\left\{ \frac{d^{n}}{dt^{n}}\mathcal{G}(t)\right\}  & =(i\omega)^{n}G(\omega)\label{wder}\\
\text{FT}\{(-it)^{n}\mathcal{G}(t)\} & =\frac{d^{n}}{d\omega^{n}}G(\omega)\\
\text{FT}\left\{ \mathcal{G}(t-t_{0})\right\}  & =e^{-i\omega t_{0}}G(\omega)\label{wdos}\\
\text{FT}\left\{ e^{i\omega_{0}t}\mathcal{G}(t)\right\}  & =G(\omega-\omega_{0})\label{JApAprop3}\\
\text{FT}\left\{ \mathcal{G}(at)\right\}  & =\frac{1}{|a|}G\left(\frac{\omega}{a}\right)\qquad\text{(scaling).}\label{wprop}
\end{align}

If $\mathcal{G}(t)$ is real, then 
\begin{equation}
G(\omega)=G^{\ast}(-\omega)\qquad\text{(}\mathcal{G}(t)\text{ real),}\label{JApAwreal}
\end{equation}
which follows readily using (\ref{par1}) and forcing $\mathcal{G}(t)=\mathcal{G}^{\ast}(t).$

Writing $G(\omega)$ in the Euler form, $G(\omega)=|G(\omega)|\exp[i\gamma_{G}(\omega)],$
(\ref{JApAwreal}) implies that the modulus $|G(\omega)|$ is an even
function of $\omega$ while the phase $\gamma(\omega)$ is odd. In
the case of a linear, single-mode optical fiber of length $L$ with
negligible losses, a fiber transfer function $H(\omega)=|H(\omega)|\exp[i\gamma_{H}(\omega)]$
can be defined with $|H(\omega)|\simeq1$ and $\gamma_{H}(\omega)=-\beta(\omega)L$,
where $\beta(\omega)$ is the propagation constant of the field, which
can be written as $\beta(\omega)=(\omega/c)\bar{n}(\omega),$ $\bar{n}(\omega)$
being the real (in this lossless case) modal (or \emph{effective})
index. It then follows that $\bar{n}(\omega)$ is an even function
of $\omega.$ Naturally, the reality condition also has consequences
for the Taylor series expansion of $\beta(\omega)$ around the optical
carrier frequencies $\pm\omega_{0}$: 
\begin{align}
\beta(\omega) & \simeq\beta(\omega_{0})+(\omega-\omega_{0})\frac{1}{v_{g}(\omega_{0})}+\frac{1}{2}(\omega-\omega_{0})^{2}\beta_{2}(\omega_{0})+\frac{1}{6}(\omega-\omega_{0})^{3}\beta_{3}(\omega_{0})\qquad\text{near }\omega_{0}\\
\beta(\omega) & \simeq\beta(-\omega_{0})+(\omega+\omega_{0})\frac{1}{v_{g}(-\omega_{0})}+\frac{1}{2}(\omega+\omega_{0})^{2}\beta_{2}(-\omega_{0})+\frac{1}{6}(\omega+\omega_{0})^{3}\beta_{3}(-\omega_{0})\nonumber \\
 & =\beta(\omega_{0})-(\omega+\omega_{0})\frac{1}{v_{g}(\omega_{0})}+\frac{1}{2}(\omega+\omega_{0})^{2}\beta_{2}(\omega_{0})-\frac{1}{6}(\omega+\omega_{0})^{3}\beta_{3}(\omega_{0}).\qquad\text{near }-\omega_{0}.\label{wtrans}
\end{align}
The last equality in (\ref{wtrans}) follows because, $\beta(\omega)$
being odd, $\beta_{n}(\omega)=d^{n}\beta(\omega)/d\omega^{n}$ will
be even (odd) for $n$ odd (even). (Note: Remember that $1/v_{g}(\omega)=d\beta(\omega)/d\omega.$)

Although, in principle, the FT applies to sufficiently regular functions,
the Dirac delta (with which the reader is assumed to be familiar)
can be handled as well: 
\begin{align}
\text{FT}\left\{ \delta(t-t_{0})\right\}  & =e^{-i\omega t_{0}}\label{prop1}\\
\text{FT}\left\{ e^{i\omega_{0}t}\right\}  & =2\pi\delta(\omega-\omega_{0}).\label{prop2}
\end{align}
Equation (\ref{prop2}) is easily checked by putting $\mathcal{G}(\omega)=2\pi\delta(\omega-\omega_{0})$
in (\ref{par1}) and using the operational definition of the Dirac
delta: 
\begin{align}
 & \delta(y-y_{0})=0\qquad\text{if }y\neq y_{0}\label{wnulo}\\
\int_{a}^{b} & f(y)\delta(y-y_{0})dy=\left\{ \begin{tabular}{ll}
 \ensuremath{f(y_{0})}  &  \quad if \ensuremath{y_{0}\in(a,b)}\\
 \ensuremath{0}  &  \quad otherwise. 
\end{tabular}\right.\label{wpdelta}
\end{align}

Note that (\ref{prop2}) is consistent with (\ref{JApAprop3}) setting
$g(t)=1,$ since FT$\{1\}=2\pi\delta(\omega)$ {[}eq. (\ref{prop2}){]}.

\subsection{Relation with the Fourier series. Periodic signals and phasors}

\label{JApAfourierseries}

In textbooks, the FT is usually introduced as a limiting case of the
discrete Fourier \emph{series}. As is well known, any periodic function
of period $T$ can be written as a \emph{discrete}, usually infinite,
sum of harmonics, that is, sinusoids of angular frequencies $\omega_{n}=n(2\pi/T),$
each having a specific amplitude and phase. Formally, 
\begin{equation}
f(t)=\sum_{n=-\infty}^{\infty}c_{n}e^{i\omega_{n}t},\label{wseries}
\end{equation}
with 
\begin{equation}
c_{n}=\frac{1}{T}\int_{0}^{T}f(t)e^{-i\omega_{n}t}dt.\label{wserie}
\end{equation}
Analogous to (\ref{par1})--(\ref{par2}), if $f(t)$ is real, then
$c_{-n}=c_{n}^{\ast}$ and (\ref{wserie}) can be rewritten as a sum
of cosines, since $c_{n}\exp(i\omega_{n}t)+c_{-n}\exp(-i\omega_{n}t)=2|c_{n}|\cos(\omega_{n}t+\arg[c_{n}]).$

An aperiodic function can be thought of as a ``periodic''\ one
with $T\rightarrow\infty.$ Performing this limit properly, the sum
over $n$ becomes an integral over $\omega$ and the numerable $\{c_{n}\}$
evolve into a continuous function of $\omega$ --- the FT{}. However,
apart from the periodicity, an important quality distinguishes the
two representations, as we explain in the next paragraphs.

First, as can be learned from any elementary mathematics or communication
theory book, the FT can only be defined for functions that have \emph{finite}
energy, or, in a more mathematical language, are \emph{``square integrable''.}
For example, a real-valued, time-limited rectangular pulse of finite
amplitude $A$, $x(t)=A\pi(t/T),$ complies with this condition, its
energy --- in the mathematical sense --- being 
\begin{equation}
W_{x}=\int_{-\infty}^{\infty}|x(t)|^{2}dt=A^{2}\int_{-T/2}^{T/2}\pi^{2}(t/T)]dt=A^{2}/T^{2}<\infty.
\end{equation}
Of course, many functions with infinite temporal extent also have
finite energy. A typical example is a gaussian pulse $p(t)=A\exp(-t^{2}/2T^{2}),$
for which $W_{p}=A^{2}\sqrt{\pi}/T.$ It is quite evident that a necessary
condition for a function, say $x(t),$ to have finite energy is that
$\lim_{t\rightarrow\pm\infty}x(t)=0.$

The finite energy is connected to an important fact: the spectral
(energy) content of any non-periodic signal can not be defined \emph{at}
a \emph{specific} frequency. In effect, while we can accurately say,
for a signal $\mathcal{G}(t),$ that $|G(\omega_{0})|^{2}d\omega$
represents the differential amount of energy contained \emph{around}
$\omega_{0}$, the expression ``energy \emph{at} the frequency $\omega_{0}$''\ is
meaningless. This is so because $|G(\omega)|^{2}$ is a spectral energy
\emph{density} (in joules / Hz, for example), so that the product
$|G(\omega_{0})|^{2}d\omega$ is the area of a small rectangle of
height $|G(\omega_{0})|^{2}$ (J/Hz) and width $d\omega$ (Hz), yielding
a non-zero quantity of joules. However, $|G(\omega_{0})|^{2}$ by
itself has no area (its width is null) and thus contains no energy.
As a continuous density, it needs at least a differential frequency
interval to make any physical sense.

The situation is the opposite with the Fourier series, which ``lives
in another world'': in (\ref{wseries}), the frequencies are discrete;
there exist no frequencies between, say, $\omega_{n}$ and $\omega_{n+1}.$
In this case, $|c_{n}|^{2}$ does represent unambiguously the spectral
power of $f(t)$ \emph{at} the frequency $\omega_{n}.$ Note that
we use the power, not the energy, which, far from being zero as before,
is actually \emph{infinite} for any periodic function.\footnote{For example, the energy of $f(t)$ \emph{at} $\pm\omega_{n}$ is:

\begin{equation}
W_{f,\text{ }\omega_{n}}=\int_{-\infty}^{\infty}\{2|c_{n}|\cos(\omega_{n}t+\arg[c_{n}])\}^{2}dt=\infty.\label{winf}
\end{equation}
The result (\ref{winf}) is no wonder as the area of a squared cosine
is always positive, so when accumulated through the infinite periods
from $-\infty$ to $\infty,$ the integral diverges.The (average)
power at $\omega_{n}$, on the other hand, is:

\begin{equation}
P_{f,\text{ }\omega_{n}}=\frac{1}{T}\int_{0}^{T}[2|c_{n}|\cos(n\omega_{0}t+\arg[c_{n}])]^{2}dt=2|c_{n}|^{2},
\end{equation}
the $2$ factor accounting for both $c_{n}$ and $c_{-n}$.}

To sum up, the FT works with non-periodic functions but, at least
in principle, cannot be used for periodic functions. The latter, having
a discrete spectrum, can only be spectrally characterized through
the Fourier series.

All the same, it would be very convenient, for mathematical manipulations,
to have a unified tool for both aperiodic and periodic signals. Is
this possible? Yes, provided we find a way to incorporate the Dirac
delta ï¿œnto the otherwise continuous FT formalism.

In view that a continuous spectral density $|G(\omega)|^{2}$ cannot
single out a solitary point on the $\omega$ axis --- i.e., a discrete
frequency --- because its energy is zero, a possible solution would
be to relax the mathematical requirements of a ``well-behaved''\ Fourier
transform $G(\omega)$ to allow for frequency Dirac deltas as, in
fact, (\ref{prop2}) exemplifies. We will give a simple example to
illustrate this.

Consider an elementary series RLC electrical circuit. If $v(t)$ is
the voltage at the output of the generator and $i(t)$ is the current
through the circuit, the differential equation of the circuit reads
\begin{equation}
\frac{d}{dt}v(t)=R\frac{d}{dt}i(t)+L\frac{d^{2}}{dt^{2}}i(t)+\frac{1}{C}i(t).\label{wecdif}
\end{equation}
Taking the FT of equation (\ref{wecdif}) and using (\ref{wder}),
we obtain the FT of the current as a function of the FT of the input
voltage: 
\begin{equation}
I(\omega)=V(\omega)\frac{i\omega}{1/C-L\omega^{2}+i\omega R}\qquad\text{[A}/\text{Hz].}\label{wtra}
\end{equation}
Expression (\ref{wtra}) is valid for any $V(\omega)$ or, equivalently,
any arbitrary time-dependence of $v(t)$. But what if $v(t)=V_{0}\cos(\omega_{0}t)$?
In this case, according to (\ref{prop2}), 
\begin{equation}
V(\omega)=\frac{V_{0}}{2}[\pi\delta(\omega-\omega_{0})+\pi\delta(\omega+\omega_{0})]\qquad\text{[V}/\text{Hz].}\label{w2del}
\end{equation}
Substituting (\ref{w2del}) in (\ref{wtra}), we get 
\begin{equation}
I(\omega)=\frac{V_{0}}{2}\pi\lbrack\delta(\omega-\omega_{0})+\delta(\omega+\omega_{0})]\frac{i\omega}{1/C-L\omega^{2}+i\omega R}\qquad\text{[A}/\text{Hz].}\label{wIw}
\end{equation}
Although (\ref{wIw}) is still a density, we will now see that the
presence of the Dirac deltas allows to represent a finite power at
specific, discrete frequencies, so that the expected sinusoidal current
will certainly arise --- expression (\ref{wres}) below. In a way,
this can be viewed as a consequence of the delta having an infinite
amplitude, thus allowing that the area ``at''\ $\omega_{0}$, given
by $0\times\infty,$ be finite. Let us then compute the inverse FT
of (\ref{wIw}) using (\ref{wpdelta}): 
\begin{align}
i(t) & =\frac{1}{2\pi}\int_{-\infty}^{\infty}\frac{V_{0}}{2}\pi i\omega\frac{\delta(\omega-\omega_{0})+\delta(\omega+\omega_{0})}{1/C-L\omega^{2}+i\omega R}e^{i\omega t}d\omega\nonumber \\
 & =\frac{1}{2}\frac{V_{0}}{2}i\omega_{0}\frac{e^{i\omega_{0}t}}{1/C-L\omega_{0}^{2}+i\omega_{0}R}+\frac{1}{2}\frac{V_{0}}{2}(-i\omega_{0})\frac{e^{-i\omega_{0}t}}{1/C-L\omega_{0}^{2}-i\omega_{0}R}\nonumber \\
 & \equiv\frac{1}{2}F(\omega_{0})e^{i\omega_{0}t}+\frac{1}{2}F^{\ast}(\omega_{0})e^{-i\omega_{0}t}=|F(\omega_{0})|\cos(\omega_{0}t+\gamma_{F(\omega_{0})}).\label{wres}
\end{align}
Expression (\ref{wres}) is, naturally, the same result one obtains
by using the \emph{phasorial formalism}, routinely employed to analyze
linear systems when all signals are sinusoids of a unique, discrete
frequency. In the example at hand, one would write $\tilde{v}(t)=V_{0}e^{i\omega_{0}t}\equiv\tilde{V}e^{i\omega_{0}t}$
(we use the tilde to denote that only the positive frequency is used)
and expect a current of the form $\tilde{\imath}(t)=\tilde{I}e^{i\omega_{0}t}.$
Inserting these expressions in place of $v(t)$ and $i(t)$ in (\ref{wecdif}),
we obtain 
\begin{equation}
\tilde{I}=\tilde{V}\frac{i\omega_{0}}{1/C-L\omega_{0}^{2}+i\omega_{0}R}\qquad\text{[A]},\label{wfasor}
\end{equation}
which gives the phasor $\tilde{I}$ as a function of the phasor $\tilde{V}$
at the frequency $\omega_{0}.$ Multiplying (\ref{wfasor}) through
by $\exp(i\omega_{0}t)$ and taking the real part, (\ref{wres}) follows.
$\tilde{I}$ is not a current spectral density (A/Hz), as $I(\omega),$
but a full current in amperes.\footnote{This can be checked in expression (\ref{wcurr}) below noting that\ the
Dirac deltas have dimension of s$\,=\,$Hz$^{-1},$ since $\delta(\omega)=(2\pi)^{-1}\int_{-\infty}^{\infty}\exp(-i\omega t)dt.$}

Compare expression (\ref{wfasor}) with (\ref{wtra}). They appear
formally analogous, but in (\ref{wfasor}) the frequency $\omega_{0}$
is a fixed, discrete \emph{parameter} rather than a continuous variable,
as is $\omega$ in (\ref{wtra}). This is the reason why, in the context
of nonlinear optics, several discrete frequencies appear as a consequence
of the nonlinear processes, the notation $\tilde{E}_{\omega_{n}}$
(or $\tilde{E}_{_{n}}$) instead of $\tilde{E}(\omega_{n})$ is preferred
to refer to the complex amplitudes (phasors) of the different sinusoids.\footnote{Alas, this careless notation ($\omega_{n}$ as an \emph{argument})
is not rare in the literature...}

Note finally that (\ref{wfasor}) can be derived directly from (\ref{wtra}),
and this establishes the connection between the two approaches, which
lies on (\ref{wIw}). Again, calling the voltage phasor $\tilde{V}\equiv V_{0}$
(in this example $\tilde{V}$ is real, which is equivalent to saying
that the input voltage is chosen as the reference phase) and writing
\begin{equation}
I(\omega)=\tilde{I}[\pi\delta(\omega-\omega_{0})+\pi\delta(\omega+\omega_{0})],\label{wcurr}
\end{equation}
(\ref{wIw}) yields 
\begin{equation}
\tilde{I}[\pi\delta(\omega-\omega_{0})+\pi\delta(\omega+\omega_{0})]=\tilde{V}\pi\lbrack\delta(\omega-\omega_{0})+\delta(\omega+\omega_{0})]\frac{i\omega}{1/C-L\omega^{2}+i\omega R}.\label{wpaso}
\end{equation}
The equality (\ref{wpaso}) holds trivially ($0=0$) for all frequencies
other than $\omega_{0}$ or $-\omega_{0},$ due to (\ref{wnulo}).
A nontrivial equality can only occur at $\omega=\pm\omega_{0}.$ At
$\omega_{0},$ we obtain 
\begin{equation}
\tilde{I}\delta(\omega-\omega_{0})=\tilde{V}\delta(\omega-\omega_{0})\frac{i\omega_{0}}{1/C-L\omega_{0}^{2}+i\omega_{0}R}.\label{wfinal}
\end{equation}
Obviously, (\ref{wfinal}) implies that the coefficients multiplying
the deltas on both sides of the equation should be identical. This
leads to the result (\ref{wfasor}). The equation corresponding to
$-\omega_{0}$ is simply the complex conjugate of (\ref{wfasor}).

\subsection{Convolution}

\label{JApAconvolution}

An outstanding property of the FT is that, if 
\begin{equation}
y(t)=h(t)\ast x(t)\equiv\int_{-\infty}^{\infty}h(t-\tau)x(\tau)d\tau,\label{wconv}
\end{equation}
then the corresponding transforms are related by a product: 
\begin{equation}
Y(\omega)=H(\omega)X(\omega).\label{wfonv}
\end{equation}
This is, the FT of the \emph{convolution} of two functions is the
product of the Fourier transforms of the functions. The property (\ref{wfonv})
can be proved by writing $f(t-\tau)$ and $g(\tau)$ in terms of their
Fourier integrals in (\ref{wconv}), exchanging the order of the time
and frequency integrations, and using (\ref{prop2}). It is easy to
verify that the convolution is linear and commutative.

As is well known, the temporal response $y(t)$ of a linear invariant
system to an input $x(t)$ is given precisely by (\ref{wconv}) if
$h(t)$ is the impulse response of the system.\footnote{That is, if $x(t)=\delta(t),$ then $y(t)=\int_{-\infty}^{\infty}h(t-\tau)\delta(\tau)d\tau=h(t-0)=h(t).$
This proves, in passing, that

\begin{equation}
x(t)\ast\delta(t)=x(t)
\end{equation}
for any function $x(t).$ {[}It can also be derived through the transformed
relation (\ref{wfonv}){]}.} Therefore, the relation (\ref{wfonv}) provides a very simple way
to predict the output--input behavior through the Fourier transforms
of the signals.

The notation of (\ref{wconv})\footnote{$(x\ast y)(t)$ would perhaps be a more suitable notation.}
should not mislead into believing that, if $z(t)=$ $x(t)\ast y(t),$
then $x(at)\ast y(at)=^{{\small(!)}}z(at).$ {[}Wrong!{]} The correct
scaling relation is 
\begin{equation}
x(at)\ast y(at)=\frac{1}{|a|}z(at).\label{wsca}
\end{equation}

Finally, the dual relation of (\ref{wfonv}) is 
\begin{equation}
\text{FT}\{h(t)x(t)\}=\frac{1}{2\pi}H(\omega)\ast X(\omega).
\end{equation}

\subsection{Narrow band signals. Analytical representation}

\label{JApAanalytical}

The phasorial formalism reviewed in the previous section is the simplest
example of the analytical representation of a real signal. One has
\begin{align}
x(t) & =A\cos(\omega_{0}t+\varphi)=\tfrac{1}{2}Ae^{i(\omega_{0}t+\varphi)}+\tfrac{1}{2}Ae^{-i(\omega_{0}t+\varphi)}\nonumber \\
 & \equiv\tilde{X}e^{i\omega_{0}t}+\tilde{X}^{\ast}e^{-i\omega_{0}t}=\operatorname{Re}\left[\tilde{X}e^{i\omega_{0}t}\right],\label{wfas}
\end{align}
with $\tilde{X}=(A/2)\exp(i\varphi)$ being the phasor. Since the
``negative frequency''\ part of (\ref{wfas}), i.e. $\tilde{X}^{\ast}\exp(-i\omega_{0}t)$,
is redundant, operations can be performed with the positive part only,
knowing that the real time signal will be recovered at the end by
simply taking the real part of the result. Furthermore, the time dependence
$\exp(i\omega_{0})$ can be omitted because it affects all phasors
equally --- they all revolve at the same angular velocity $\omega_{0}$
--- and thus becomes superfluous. (The phasor diagram can be viewed
as a \emph{snapshot}, a ``time-frozen''\ image of the permanently
rotating vectors which permits to appreciate their amplitudes and
relative phases, the only relevant information.)

One would wish to extend the simplicity of using only the positive
frequencies to arbitrary signals with \emph{any} time dependence.
This can be done, but it is feasible only under certain conditions
--- fortunately, rather realistic --- to be discussed next.

If $\mathcal{F}(t)$ is an arbitrary real function, its \emph{analytical
signal} $\widetilde{\mathcal{F}}(t)$ is defined as the function which
contains only the \emph{positive} frequencies of $\mathcal{F}(t);$
more specifically, FT$[\widetilde{\mathcal{F}}(t)]$ is, by definition,
\begin{equation}
\widetilde{F}(\omega)=\text{FT}\mathcal{[}\widetilde{\mathcal{F}}(t)\mathcal{]}=\left\{ \begin{array}{cc}
2F(\omega), & \omega>0\\
F(0), & \omega=0\\
0, & \omega<0
\end{array}\right..\label{dsa}
\end{equation}
$\widetilde{\mathcal{F}}(t)$ is not real, as $\widetilde{F}(\omega)\neq\widetilde{F}^{\ast}(-\omega).$
Actually, it can be shown that the real and imaginary parts of $\widetilde{\mathcal{F}}(t)$
are given by 
\begin{equation}
\widetilde{\mathcal{F}}(t)=\mathcal{F}(t)+i\mathcal{\check{F}}(t),\label{hil}
\end{equation}
where $\mathcal{\check{F}}(t)$ is the \emph{Hilbert transform} of
$\mathcal{F}(t).$ The simplest example is the complex time exponential
at $\omega_{0}$: 
\begin{equation}
\underset{\widetilde{\mathcal{F}}(t)}{\underbrace{\exp(i\omega_{0}t)}}\,=\bigskip\underset{\mathcal{F}(t)}{\underbrace{\cos(\omega_{0}t)}}+\,i\,\underset{\mathcal{\check{F}}(t)}{\underbrace{\sin(\omega_{0}t)}};\label{ex}
\end{equation}
It is immediate to see that $\exp(i\omega_{0}t)$ and $\cos(\omega_{0}t)$
satisfy (\ref{dsa}).

In electromagnetism, it is frequent to write a field $\mathcal{F}(t)$
as the sum of its ``positive and negative frequency''\ parts, 
\begin{equation}
\mathcal{F}(t)=\mathcal{F}^{-}(t)+\mathcal{F}^{+}(t),\label{r1}
\end{equation}
defined as 
\begin{equation}
\mathcal{F}^{-}(t)\equiv\frac{1}{2\pi}\int\nolimits _{-\infty}^{0}F(\omega)\,e^{i\omega t}d\omega\qquad\text{and\qquad}\mathcal{F}^{+}(t)\equiv\frac{1}{2\pi}\int\nolimits _{0}^{\infty}F(\omega)\,e^{i\omega t}d\omega.\label{pn}
\end{equation}
It is then obvious that\footnote{And also that $\mathcal{\check{F}}(t)=i[\mathcal{F}^{-}(t)-\mathcal{F}^{+}(t)]$,
but we will not need to deal with the Hilbert transform here.} $\widetilde{\mathcal{F}}(t)$ is just $\mathcal{F}^{+}(t)$ save
for a 2 factor. 
\begin{equation}
\widetilde{\mathcal{F}}(t)=2\mathcal{F}^{+}(t).\label{g2}
\end{equation}

Applying this decomposition to the electric field propagated along
a single-mode fiber, we have 
\begin{equation}
\mathcal{E}(t,z)=\mathcal{E}^{+}(t,z)+\mathcal{E}^{-}(t,z)=\dfrac{1}{2\pi}\int_{0}^{\infty}E(\omega,z)\,e^{i\omega t}d\omega+\dfrac{1}{2\pi}\int_{-\infty}^{0}E(\omega,z)\,e^{i\omega t}d\omega.\label{pe}
\end{equation}
But $\mathcal{E}(t,z)$ is a \emph{real function}, this implying that\footnote{Note the similitude with the relation $c_{-n}=c_{n}^{\ast}$ in (\ref{wserie}).}
$E(-\omega,z)=E^{\ast}(\omega,z);$ therefore, by making the variable
change $\omega\rightarrow-u$ in the second integral of (\ref{pe}),
it easily follows that 
\begin{equation}
\mathcal{E}^{-}(t,z)=[\mathcal{E}^{+}(t,z)]^{\ast}.\label{wean}
\end{equation}
Consequently, it is only necessary to calculate one of the two components,
for example $\mathcal{E}^{+}(t,z).$

\begin{figure}[h]
\centering{}\includegraphics[scale=0.2]{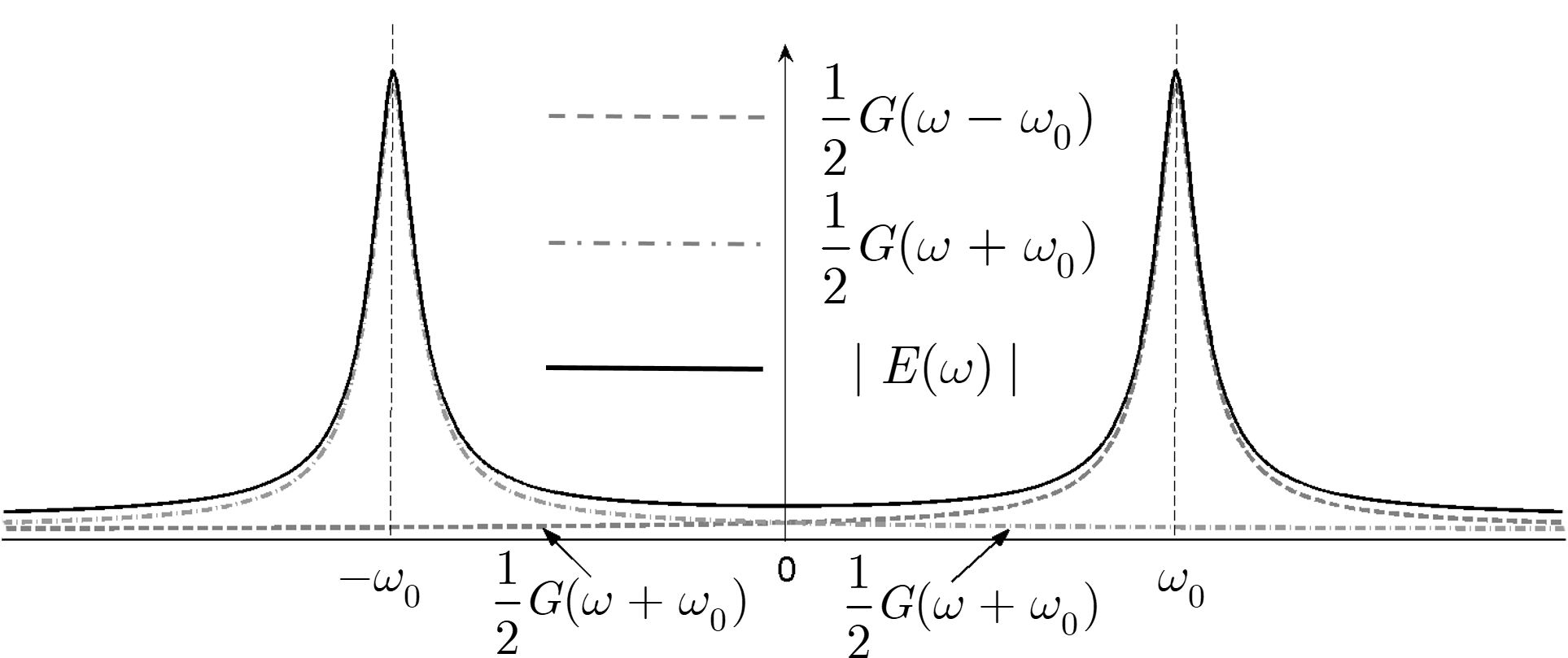}\caption{The positive-frequency part of $\mathcal{E}(t,z)$ is given by $\mathcal{E}^{+}(t,z)=\tfrac{1}{2\pi}\int_{0}^{\infty}E(\omega,0)e^{-i\beta(\omega)z}\,e^{i\omega t}d\omega\protect\neq\tfrac{1}{2\pi}\int_{-\infty}^{\infty}\tfrac{1}{2}G(\omega-\omega_{0})e^{-i\beta(\omega)z}\,e^{i\omega t}d\omega\ ,$
because neither $G(\omega-\omega_{0})$ is strictly null on the negative
semiaxis, nor $G(\omega+\omega_{0})$ is in the positive semiaxis,
as the ``tails''\ marked with arrows in the figure illustrate.\ In
practice, these contributions (greately exagerated in the figure for
visualization purposes) are negligible and the narrow-band approximations
can be made.}
\label{wfig1} 
\end{figure}

While all the properties stated so far are exact, a technical difficulty
might affect their usefulness in practice: An integral from $0$ to
$\infty$ is less likely to have a known closed solution than a Fourier
integral from $-\infty$ to $\infty.$\footnote{Note that we cannot make use of very convenient properties such as
(\ref{wprop}) and others if the integration limits are $0$ and $\infty,$
or $-\infty$ and $0.$ The narrow band approximation does allow to
use the advantages of the FT formalism.} If, at least, we could accurately make the approximations $\int_{0}^{\infty}(\cdot)d\omega\simeq\int_{-\infty}^{\infty}(\cdot)d\omega$
and $\int_{-\infty}^{0}(\cdot)d\omega\simeq\int_{-\infty}^{\infty}(\cdot)d\omega,$
things might be easier. Such approximations, illustrated in figure
\ref{wfig1}, do apply to the so-called \emph{narrow bandpass signals},
and fortunately this is our case. In all situations encountered in
practical optical transmission, the spectrum $E(\omega,z)$ has a
\emph{very} narrow relative bandwidth. We typically have 
\begin{equation}
E(\omega,z)=E(\omega,z=0)\,e^{-i\beta(\omega)z}=\dfrac{1}{2}[G(\omega-\omega_{0})+G(\omega+\omega_{0})]e^{-i\beta(\omega)z},
\end{equation}
which, as sketched in figure \ref{wfig1} has the form of two very
narrow peaks centered at $\omega_{0}$ and $-\omega_{0}.$ The width
of the peaks is extremely small compared with $\omega_{0}$ (for example,
$\sim20$ GHz against $\nu_{0}=\omega_{0}/(2\pi)\approx5\times10^{14}$
Hz), so we can safely say that $G(\omega-\omega_{0})\simeq0$ except
for $\omega$ very close to $\omega_{0}.$ Thus\footnote{Consider for example a gaussian-shaped spectrum, $G(\omega-\omega_{0})\propto\exp[-(\omega-\omega_{0})^{2}/(\Delta\omega)^{2}].$
Even with a bandwidth as large as $\Delta f=\Delta\omega/(2\pi)\approx100$
GHz, and $\nu_{0}=\omega_{0}/(2\pi)\approx10^{14}$ Hz, the relative
amplitude of $G(\omega-\omega_{0})$ at $\omega=0$ will be: $\exp[-\omega_{0}{}^{2}/(\Delta\omega)^{2}]=\exp(-(10^{14}/10^{11})^{2})\approx10^{-434295}$
(!), which more than justifies neglecting the contribution of $G(\omega-\omega_{0})$
on the negative semi-axis.} $G(\omega-\omega_{0})\simeq0$ for $\omega<0$ and, likewise, $G(\omega+\omega_{0})\simeq0$
for $\omega>0$. This will allows us to employ Fourier integrals:
\begin{align}
\mathcal{E}^{+}(t,z) & =\dfrac{1}{2\pi}\int_{0}^{\infty}E(\omega,0)e^{-i\beta(\omega)z}\,e^{i\omega t}d\omega\simeq\dfrac{1}{2\pi}\int_{0}^{\infty}\dfrac{1}{2}G(\omega-\omega_{0})e^{-i\beta(\omega)z}\,e^{i\omega t}d\omega\nonumber \\
 & \simeq\dfrac{1}{2\pi}\int_{-\infty}^{\infty}\dfrac{1}{2}G(\omega-\omega_{0})e^{-i\beta(\omega)z}\,e^{i\omega t}d\omega\qquad\text{(narrow band approximation).}\label{bes}
\end{align}

Likewise,

\begin{equation}
\mathcal{E}^{-}(t,z)\simeq\dfrac{1}{2\pi}\int_{-\infty}^{\infty}\dfrac{1}{2}G(\omega+\omega_{0})e^{-i\beta(\omega)z}\,e^{i\omega t}d\omega.
\end{equation}

The real electric field $\mathcal{E}$ is finally obtained from the
analytical field $\mathcal{\tilde{E}}$ by: 
\begin{align}
\mathcal{E}(t,z) & =\mathcal{E}^{+}(t,z)+\mathcal{E}^{-}(t,z)=\mathcal{E}^{+}(t,z)+[\mathcal{E}^{+}(t,z)]^{\ast}=2\operatorname{Re}[\mathcal{E}^{+}(t,z)]\nonumber \\
 & =\operatorname{Re}[\mathcal{\tilde{E}}(t,z)],
\end{align}

\subsection{Spectral width and transform-limited pulses}

\subsubsection{Uncertainty principle}

The propagation of an optical gaussian pulse along a dispersive single-mode
fiber is a paradigmatic example of optical pulse propagation that
can be found in any textbook on optical communications or general
photonics. The well-known result is that the temporal width of the
pulse envelope increases with the propagation distance. However, in
the frequency domain, the \emph{width} of the pulse spectrum at the
fiber output is found to be exactly the same as at the input. This
result often causes confusion as it apparently contradicts the idea
that a narrower time signal should have a broader frequency spectrum
and vice versa. We will clarify this point here.

The width of an arbitrary pulse-shaped waveform $f(t)$ is usually
characterized by its \emph{rms width} $\sigma_{t},$ given by 
\begin{equation}
\sigma_{t}^{2}\equiv\frac{\int_{-\infty}^{\infty}(t-\bar{t})^{2}|f(t)|^{2}dt}{\int_{-\infty}^{\infty}|f(t)|^{2}dt},\qquad\text{with\quad}\bar{t}\equiv\frac{\int_{-\infty}^{\infty}t|f(t)|^{2}dt}{\int_{-\infty}^{\infty}|f(t)|^{2}dt},\label{2}
\end{equation}
$\bar{t}$ being the \emph{mean arrival time} of the pulse. Note that
$\sigma_{t}$ has the same functional form as the standard deviation
(the square root of the variance) of a statistical continuous distribution,
with $\bar{t}$ playing the role of the the statistical mean. The
same definition is used in the frequency domain for the Fourier transforms:
\begin{equation}
\sigma_{\omega}^{2}\equiv\frac{\int_{-\infty}^{\infty}(\omega-\bar{\omega})^{2}|F(\omega)|^{2}d\omega}{\int_{-\infty}^{\infty}|F(\omega)|^{2}d\omega}\qquad\text{with\quad}\bar{\omega}\equiv\frac{\int_{-\infty}^{\infty}\omega|F(\omega)|^{2}d\omega}{\int_{-\infty}^{\infty}|F(\omega)|^{2}d\omega}.\label{wnum}
\end{equation}

The famous \emph{uncertainty principle}, first suggested by Heisenberg
in the context of Quantum mechanics, follows merely from the mathematical
properties of the Fourier transform and is proved with the aid of
the Schwartz inequality. It reads 
\begin{equation}
\sigma_{t}\,\sigma_{\omega}\geq\dfrac{1}{2}.\label{wheis}
\end{equation}

However, the \emph{equality} in (\ref{wheis}) holds \emph{only} for
a \emph{pure gaussian-shaped} Fourier transform pair: 
\begin{equation}
f(t)=\exp(-t^{2}/2T^{2})\quad\longleftrightarrow\quad F(\omega)=\sqrt{2\pi}Te^{-\frac{1}{2}T^{2}\omega^{2}}.\label{welem}
\end{equation}
In this case,\footnote{All the improper integrals needed in this subsection are obtained
as particular cases of some the following results:

\begin{align}
\int_{-\infty}^{\infty}\left(x-x_{0}\right)^{2}\exp(-x^{2}/\delta^{2}+bx)dx & =(\sqrt{\pi}/4)\left(\delta^{4}b^{2}+2\delta^{2}(1-2bx_{0})+4x_{0}^{2}\right)\delta\exp(\delta^{2}b^{2}/4)\\
\int_{-\infty}^{\infty}x\exp(-x^{2}/\delta^{2}+bx)dx & =\allowbreak\frac{1}{2}\sqrt{\pi}b\delta^{3}e^{\frac{1}{4}b^{2}\delta^{2}}\\
\int_{-\infty}^{\infty}\exp(-x^{2}/\delta^{2}+bx)dx & =\sqrt{\pi}\delta e^{\frac{1}{4}b^{2}\delta^{2}},
\end{align}
which are valid for $b$ complex too.} 
\begin{equation}
\sigma_{t}^{2}=T^{2}/2,\qquad\sigma_{\omega}^{2}=1/(2T^{2})\qquad\text{and\quad}\sigma_{t}\,\sigma_{\omega}=\frac{1}{2}.\label{wbuena}
\end{equation}
In any case other than a pure gaussian pulse, only ''greater than''\ applies.

We can never have a pure gaussian baseband pulse in an optical fiber
anyway, but, at most, a gaussian pulse-\emph{modulated} carrier (or
``gaussian bandpass pulse''). The electric field and its spectrum
at the input of the fiber, $z=0,$ are, in this case, 
\begin{align}
\mathcal{E}_{0}(t) & =e^{-t^{2}/2T^{2}}\cos\omega_{0}t\label{wp1}\\
E_{0}(\omega) & =\frac{1}{2}\sqrt{2\pi}Te^{-\frac{1}{2}T^{2}(\omega-\omega_{0})^{2}}+\frac{1}{2}\sqrt{2\pi}Te^{-\frac{1}{2}T^{2}(\omega+\omega_{0})^{2}}\nonumber \\
 & \equiv\frac{1}{2}(G(\omega-\omega_{0})+G(\omega+\omega_{0})).\label{wp2}
\end{align}
The spectrum is as sketched in Fig. \ref{wfig1}. Since (\ref{wp1})--(\ref{wp2})
is not a \emph{pure} gaussian (baseband) Fourier pair, we cannot expect
that it will satisfy the equality of relation (\ref{wheis}). The
calculations yield 
\begin{equation}
\sigma_{t}^{2}=\frac{T^{2}}{2}\left(1-\frac{2T^{2}\omega_{0}^{2}}{1+e^{T^{2}\omega_{0}^{2}}}\right)\simeq\frac{T^{2}}{2},\qquad\sigma_{\omega}^{2}=\frac{1}{2T^{2}}+\omega_{0}^{2}\simeq\omega_{0}^{2},\label{w2}
\end{equation}
$\allowbreak$leading to 
\begin{equation}
\sigma_{t}\,\sigma_{\omega}\simeq\frac{1}{2}(\omega_{0}T)^{2}>>>\frac{1}{2},\label{wpmod}
\end{equation}
where we have used the narrow-band approximation,\footnote{Having a strongly peaked, very narrow spectrum is equivalent to having
a slowly-varying time envelope (at the scale of an optical carrier
period), which, in turn, means that there are very many optical cicles
within the duration of the modulating pulse.} $T>>1/\omega_{0}.$ The result (\ref{wpmod}) indeed complies with
(\ref{wheis}), but it is not very informative! The problem obviously
lies on $\sigma_{\omega}.$ In view of the form of the spectrum, it
is not surprising that $\sigma_{\omega}\simeq\omega_{0}$ regardless
of the modulating pulse; since $\omega_{0}$ is so much larger than
the spectral width of the envelope ($\sim1/T$), the role of the latter
is irrelevant. It is then more profitable to use the analytical signal,
i.e., the positive spectrum only. In this case, the pair is 
\begin{equation}
e^{-t^{2}/2T^{2}}e^{i\omega_{0}t}\longleftrightarrow e^{-T^{2}(\omega-\omega_{0})^{2}}.\label{worig}
\end{equation}
The temporal rms width, which we will call $\tilde{\sigma}_{t},$
is, again, 
\begin{equation}
\tilde{\sigma}_{t}^{2}=\frac{T^{2}}{2}.\label{wrrr}
\end{equation}
In the spectral domain, we obtain $\bar{\omega}=\omega_{0}$ and 
\begin{align}
\tilde{\sigma}_{\omega}^{2} & =\frac{\int_{-\infty}^{\infty}(\omega-\omega_{0})^{2}e^{-T^{2}(\omega-\omega_{0})^{2}}d\omega}{\int_{-\infty}^{\infty}e^{-T^{2}(\omega-\omega_{0})^{2}}d\omega}\label{wints}\\
 & =\frac{1}{2T^{2}}\qquad\text{\textsf{{\small\{narrow band\}}}.}\label{wapr}
\end{align}
Results (\ref{wrrr})--(\ref{wapr}) appear identical to those for
the pure (baseband) gaussian shape, $\sigma_{t}^{2}$ and $\sigma_{\omega}^{2}$
in (\ref{wbuena}), but we have left the legend ``narrow band''\ as
a reminder that the narrow band approximation is implicit in the integrals
in (\ref{wints}), which, as explained in the previous subsection,
should really be extended between $0$ and $\infty,$ and include
the $G(\omega-\omega_{0})$ contribution. As discussed before, the
committed error is absolutely negligible. We then obtain 
\begin{equation}
\tilde{\sigma}_{t}\,\tilde{\sigma}_{\omega}\simeq\frac{1}{2}.\label{wsame}
\end{equation}

The product is, strictly speaking, a \emph{little bit} larger than
$1/2$ --- a value only reached by a true baseband gaussian pulse
---, but the difference is unnoticeable to all practical purposes
in the narrow band situation, as one could intuitively expect.

We are finally ready to address the original objective of this subsection.
Let us examine what happens to the optical gaussian pulse after propagation
along a distance $z.$ We will consider first-order dispersion only
and, following the considerations made above for bandpass pulses,
we will work with the analytical fields. The following result can
be found in any textbook on optical communications: 
\begin{align}
\mathcal{\tilde{E}}_{z}(t) & =\exp\left[-\frac{T^{2}(t-z/v_{g})^{2}}{2(T^{4}+\beta_{2}^{2}z^{2})}\right]\times\exp i\left[\omega_{0}t-\beta_{0}z+\frac{\beta_{2}z\,(t-z/v_{g})^{2}}{2(T^{4}+\beta_{2}^{2}z^{2})}\right]\label{wy1}\\
\tilde{E}_{z}(\omega) & =e^{-\frac{1}{2}T^{2}(\omega-\omega_{0})^{2}}e^{-i\beta(\omega-\omega_{0})z}.\label{wy2}
\end{align}
It follows that $\bar{t}=z/v_{g}$ and 
\begin{equation}
\tilde{\sigma}_{t}^{2}(z)=\frac{T^{2}}{2}+\frac{z^{2}\beta_{2}^{2}}{2T^{2}}.\label{wq}
\end{equation}
Therefore, the temporal width of the pulse does increase with $z$.
But the spectral width remains constant: 
\begin{equation}
\tilde{\sigma}_{\omega}^{2}(z)=\tilde{\sigma}_{\omega}^{2}(0)=1/(2T^{2}),\label{wqq}
\end{equation}
since the the dispersive propagation only changes the phase factor
of the spectrum, leaving its \emph{modulus} untouched: $|\exp[-(i\beta(\omega)z)|^{2}=1.$
If follows that 
\begin{equation}
\tilde{\sigma}_{t}\,(z)\tilde{\sigma}_{\omega}(z)=\sqrt{\left(\frac{T^{2}}{2}+\frac{z^{2}\beta_{2}^{2}}{2T^{2}}\right)\times\frac{1}{2T^{2}}}=\frac{1}{2}\sqrt{1+\frac{z^{2}\beta_{2}^{2}}{T^{4}}}\geq\frac{1}{2}.\label{wv}
\end{equation}

The result (\ref{wv}) does not violate the condition (\ref{wheis});
therefore, the uncertainty principle is indeed satisfied.

Note that the product $\tilde{\sigma}_{t}\,(z)\tilde{\sigma}_{\omega}(z),\ $apart
from being larger than $1/2$, is \emph{not constant} --- it depends
on $z.$ Perhaps, we would have expected a result of the type $\tilde{\sigma}_{t}\,(z)\tilde{\sigma}_{\omega}(z)=C$
with $C\geq1/2$ but fixed. In other words, we would have expected
a situation wherein the temporal pulse broadens more and more as it
propagates, while its spectrum gets more and more narrow, both effects
resulting in constant $\tilde{\sigma}_{t}\,(z)\tilde{\sigma}_{\omega}(z)$
product. But this is not the case. Why? The reader might object, regardless
of the mathematical evidence, that both the time signal and the spectrum
remain gaussian at any distance $z$, so their Fourier pair \emph{should}
be characterized by the same relation as the pair at $z=0,$ (\ref{wsame}).
The key here is that the premise is \emph{not} true --- the propagated
pulse $\mathcal{\tilde{E}}_{z}(t)$ is \emph{not} ``gaussian''\ anymore.
Even if its \emph{envelope} is gaussian-shaped, its carrier has now
\emph{chirp}, this is, (dispersion-induced) frequency modulation.
Furthermore, the spectrum is \emph{not} gaussian either; again, only
its spectral \emph{envelope} is, since $\tilde{E}_{z}(\omega)$ now
includes a new distortive \emph{phase} contribution, $\exp[-(i\beta(\omega)z)]$.
These characteristics were not present in the original pair (\ref{worig}),
so (\ref{wy1}) does not qualify as a true gaussian amplitude-modulated
carrier, and the result (\ref{wsame}) is no longer obtained. Expressed
in another way: it is not fair to compare the spectral widths of (\ref{wp1})
and (\ref{wy1}) because they are \emph{different} pulses. These remarks
are important for understanding the concept of \emph{transform-limited
pulse}, to be addressed in the next subsection.

In contrast to the case considered above, baseband pulses \emph{do}
have a constant product or $\sigma_{t}\,\sigma_{\omega}$, as we have
seen with the pure baseband gaussian pulse. Typically, the width of
the temporal pulse is proportional to some parameter $T$ and the
width of the FT turns out to be proportional to $1/T,$ so that their
product maintains a fixed value.\footnote{More specifically, it can be shown that any pulse having a functional
dependence $f(t)=g(t/T),$ where the parameter $T$ appears only scaling
$t$ (and, at most, as a multiplicative prefactor $T^{m}$), has an
rms width $\sigma_{t}^{2}=T^{2}\sigma_{t}^{\prime2},$ where $\sigma_{t}^{\prime}$
does not depend on $T.$ Its FT then has the form $F(\omega)\propto G(\omega T),$
according to (\ref{wprop}), with an rms width $\sigma_{\omega}^{2}=\sigma_{\omega}^{\prime2}/T^{2},$
where again $\sigma_{\omega}^{\prime}$ does not depend on $T.$ Consequently,
$\sigma_{t}\sigma_{\omega}=\sigma_{t}^{\prime}\sigma_{\omega}^{\prime}$
is independent of $T.$} Another simple example is the baseband symmetric rectangular pulse
of duration $T,$ which FT is a sine cardinal: $\pi_{T}(t)\longleftrightarrow\Pi_{T}(\omega)=\sin(T\omega/2)/(\omega/2).$
In this case, however, the rms measure is useless because\footnote{The integrand $\omega^{2}|(\sin T\omega/2)/(\omega/2)|^{2}$ does
not tend to $0$ as $\omega\rightarrow\infty,$ making the integral
divergent. This problem is not so rare, affecting any pulse which
TF decreases as $\omega^{-1}$ or slower.} $\sigma_{t}=T$ but $\sigma_{\omega}=\infty.$ Indeed $\infty>1/2,$
but this is not certainly a great piece of information, so a more
suitable definition for the width should be used. Since both functions
are symmetric, the \emph{HWHM} (half width at half maximum) width\footnote{The time and frequency HWHM parameters are defined through $|\pi_{T}(\sigma_{t}^{\text{HWHM}})|^{2}=\frac{1}{2}|\pi_{T}(0)|^{2}$
and $|\Pi_{T}(\sigma_{\omega}^{\text{HWHM}})|^{2}=\frac{1}{2}|\Pi_{T}(0)|^{2}$,
respectively.} can be a good choice. The obtained result is $(\sigma_{t}^{\text{HWHM}})(\sigma_{\omega}^{\text{HWHM}})\simeq(T/2)(2.78/T)=1.39.$

As already said, all these considerations lead us to the concept of
\emph{``Fourier transform-limited pulse''}, which we explain next.

\subsubsection{Transform-limited pulses}

We have stated that baseband pulses satisfy the Heisenberg relation
with a fixed value for the product $\sigma_{t}\,\sigma_{\omega}$
(which depends on the specific pulse shape). This automatically imposes
an inverse time-frequency width relation. On the contrary, we have
seen two different examples of bandpass pulses, the first, but not
the second, having a fixed value of the $\tilde{\sigma}_{t}\,\tilde{\sigma}_{\omega}$
product. What is their difference?

In short, the bandpass pulses with fixed $\tilde{\sigma}_{t}\,\tilde{\sigma}_{\omega}$
product are the so-called ``transform-limited''\ pulses. A \emph{transform-limited}
(or bandwidth-limited) optical pulse is defined as a bandpass pulse
``as short as its spectral bandwidth permits''. This statement certainly
requires elaboration, but we can explain it with the example of the
gaussian pulse. At $z=0$ we have the pulse and its FT given by (\ref{worig}).
At $z>0,$ we have (\ref{wy1})--(\ref{wy2}). In both cases the
frequency span of the FT is the same, but in the second case the pulse
is longer. The pulse (\ref{worig}) is transform-limited, but the
pulse (\ref{wy1}) is not, because, given its temporal \emph{width
}--- and this implies looking at its \emph{envelope }---, \emph{it
could have had a narrower spectrum.} Actually, the pulse (\ref{worig})
is the reference example.

\begin{figure}[h]
\centering{}\includegraphics[scale=0.1]{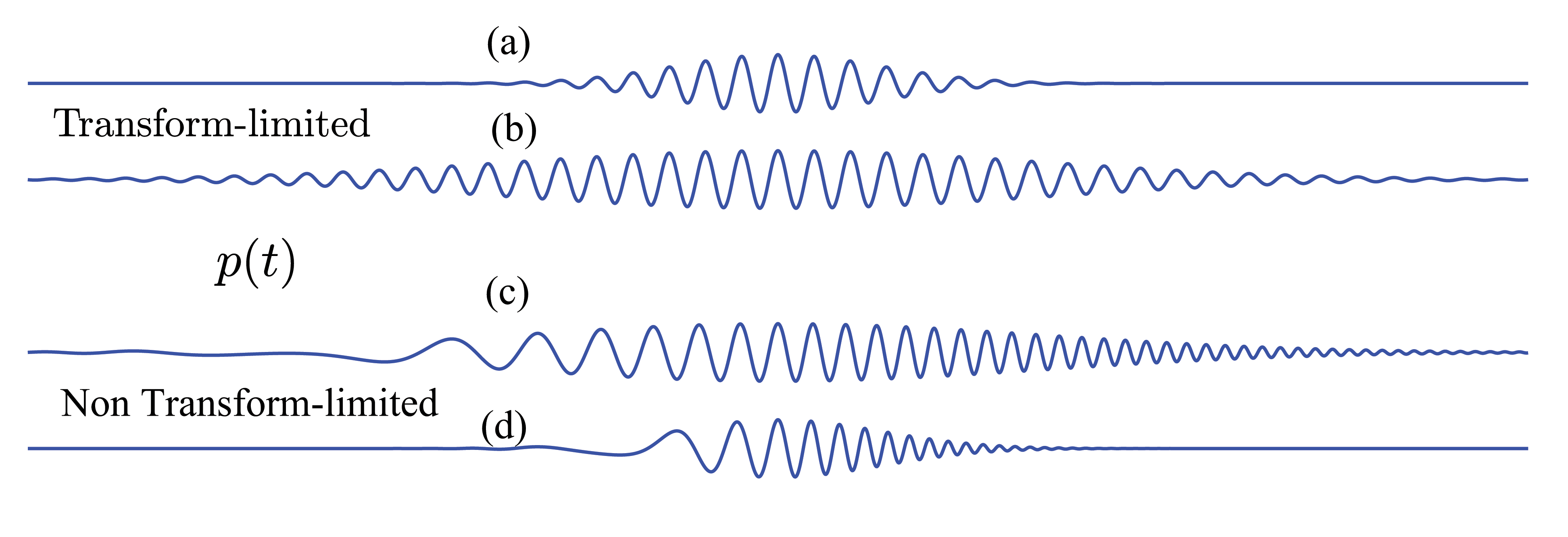}\caption{Illustration of the concept of transform-limited pulses. Temporal
pulses. See text.}
\label{wfig2} 
\end{figure}

\begin{figure}[h]
\centering{}\includegraphics[scale=0.1]{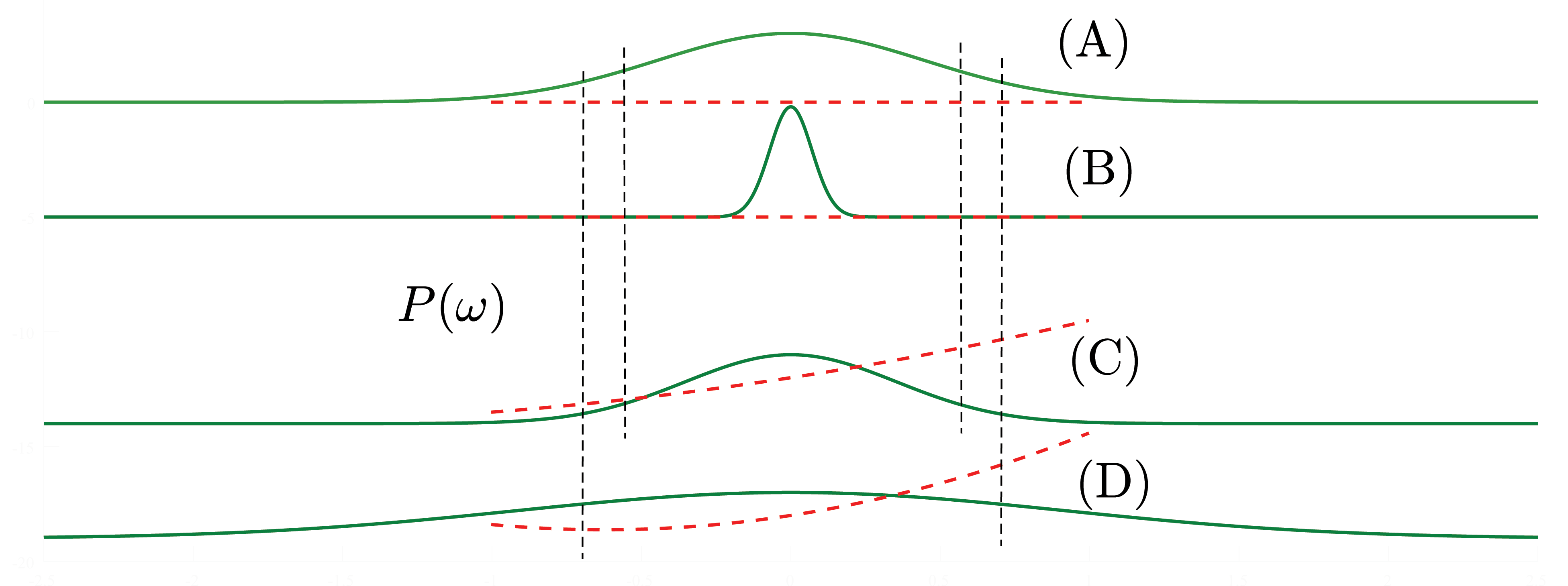}\caption{Fourier transforms corresponding to the pulses in Fig. \ref{wfig2}.
See text.}
\label{wfig22} 
\end{figure}

At this stage we know that the temporal ``extra''\ width of (\ref{wy1})
stems from the modulation of its carrier, and this is precisely the
idea. Essentially, the transform-limited pulses have a perfectly monochromatic
carrier, without any kind of phase/frequency modulation. Such ``substructure'',\ when
present, causes the pulse envelope to broaden beyond the minimum duration
it could have attained given its given spectral width. This result
is summarized in Figs. \ref{wfig2} and \ref{wfig22}.

Pulse (a) is an unchirped bandpass gaussian pulse, thus transform-limited.
Its spectrum (A) has gaussian modulus and zero phase (dashed line).
If the width of the pulse is increased, that of the spectrum decreases,
as shown by (b)--(B), and the product $\tilde{\sigma}_{t}\,\tilde{\sigma}_{\omega}$
remains constant (with a value of $1/2$ in this case).

The pair (c)--(C) illustrates the behavior of the pulse (a) after
propagating over a distance $z$ along a dispersive optical fiber.
The width of the new time pulse (c) is larger than that of pulse (a)
---and the carrier becomes chirped---, yet the spectral width of
(C) is the same as that of (A). This makes us suspect that the spectrum
(C) is ``too wide'', and it indeed is: pulse (b), having the same
\emph{envelope} as pulse (c), is spectrally narrower. Consequently,
(c) is not transform-limited. (As we have discussed above, the time-chirping
of (c) only affects the \emph{phase} of its spectrum, which is now
quadratic, as given by (\ref{wy2}): $\beta(\omega-\omega_{0})z\simeq\beta(\omega_{0})z+(\omega-\omega_{0})v_{g}^{-1}x+\frac{1}{2}(\omega-\omega_{0})^{2}\beta_{2}z.$

Finally, (d) is a narrower version of (c), i.e., both pulses have
the same expression (\ref{wy1}) but (d) has a shorter $T$. As one
would expect, (\ref{wqq}) predicts a spectrum (D) which is wider
than (C), but not in an exactly inverse fashion, with the result that
$\tilde{\sigma}_{t}\,\tilde{\sigma}_{\omega}\neq\,$constant, as (\ref{wv})
evidences. Note that, as sketched in the figure, the spectrum (D)
is necessarily broader than (A) as well, because, while the corresponding
time envelopes of (a) and (d) being identical, (d) is chirped and
(a) is not. Naturally, all these considerations are not exclusive
for gaussian-shaped pulses; they apply to any waveform.

 \bibliographystyle{IEEEtran}
\bibliography{acompat,IEEEfull,dond}

\end{document}